\documentclass[entropy,article,submit,moreauthors,10pt,a4paper,pdftex]{mdpi}
\pdfoutput=1
\firstpage{1} 
\articlenumber{x}
\doinum{10.3390/------}
\pubvolume{xx}
\pubyear{2017}
\copyrightyear{2017}
\externaleditor{Academic Editor: name}
\history{Received: date; Accepted: date; Published: date}

\preto{\abstractkeywords}{\nolinenumbers}
\usepackage{verbatim}
\usepackage{bm} 
\usepackage{bbm}
\usepackage[normalem]{ulem}
\usepackage{dsfont}
\usepackage{amsmath}
\usepackage{amsthm}
\usepackage{amsfonts}
\usepackage{amssymb}
\usepackage{mathptmx}
\usepackage{subfigure}

\Title{Thermoelectrics of Interacting Nanosystems
	-- Exploiting Superselection instead of Time-Reversal Symmetry}

\Author{Jens Schulenborg$^{1}$, Angelo Di Marco$^{1}$, Joren Vanherck$^{2,3}$, Maarten R. Wegewijs$^{4,5}$, and Janine Splettstoesser$^{1}$}

\AuthorNames{Jens Schulenborg, Angelo Di Marco, Joren Vanherck, Maarten R. Wegewijs, Janine Splettstoesser}

\address{%
$^{1}$ \quad Department of Microtechnology and Nanoscience (MC2), Chalmers University of Technology, SE-41298 G{\"o}teborg, Sweden\\
$^{2}$ \quad Physics Department, Universiteit Antwerpen, B-2020 Antwerpen, Belgium \\
$^{3}$ \quad Imec, B-3001 Leuven, Belgium\\
$^{4}$ \quad Peter Gr{\"u}nberg Institut, Forschungszentrum J{\"u}lich, 52425 J{\"u}lich,  Germany \& JARA-FIT\\
$^{5}$ \quad Institute for Theory of Statistical Physics, RWTH Aachen, 52056 Aachen,  Germany}

\abstract{% A single paragraph of about 200 words maximum.
Thermoelectric transport is traditionally analyzed using relations imposed by time-reversal symmetry, ranging from Onsager's results to fluctuation relations in counting statistics. In this paper, we show that a recently discovered duality relation for fermionic systems -- deriving from the fundamental fermion-parity superselection principle of quantum many-particle systems -- provides new insights into thermoelectric transport.
Using a master equation, we analyze the stationary charge and heat currents through a weakly coupled, but strongly interacting single-level quantum dot subject to electrical and thermal bias.
In linear transport, the fermion-parity duality shows that features of thermoelectric response coefficients are actually dominated by the average and fluctuations of the charge in a \emph{dual} quantum dot system, governed by \emph{attractive} instead of repulsive electron-electron interaction.
In the nonlinear regime, the duality furthermore relates most transport coefficients to much better understood \emph{equilibrium} quantities.
Finally, we naturally identify the fermion-parity as the part of the Coulomb interaction relevant for both the linear and nonlinear Fourier heat.
Altogether, our findings hence reveal that next to time-reversal, the duality imposes equally important symmetry restrictions on thermoelectric transport. As such, it is also expected to simplify computations and clarify the physical understanding for more complex systems than the simplest relevant interacting nanostructure model studied here.}


\robustify{\uparrow}
\robustify{\downarrow}
\robustify{\sum}
\robustify{\int}
\robustify{\nonumber}
\robustify{\cite}
\robustify{\footnote}
\newrobustcmd{\Figure}[2]{
  \begin{figure}[ht]
    \includegraphics[width=1.0\linewidth]{#1}
    \caption{#2}
  \end{figure}
}
\expandafter\newrobustcmd\csname Figure2\endcsname[3]{
\begin{figure}[ht]
  \includegraphics[width=0.9\linewidth]{#1}
  \\
  \includegraphics[width=0.9\linewidth]{#2}
  \caption{#3}
\end{figure}
}
\renewrobustcmd{\Re}{{\text{Re}}}
\renewrobustcmd{\Im}{{\text{Im}}}
\newrobustcmd{\eff}{\text{eff}} 
\newrobustcmd{\dagtot}{{\dag_\tot}}
\newrobustcmd{\dagres}{{\dag_\res}}
\newrobustcmd{\Ttot}{{\text{T}_\tot}}
\newrobustcmd{\Tres}{{\text{T}_\res}}
\newrobustcmd{\T}{{\text{T}}}
\newrobustcmd{\tot}{\text{tot}}
\newrobustcmd{\tun}{\text{T}}
\newrobustcmd{\res}{\text{R}}
\newrobustcmd{\un}{\text{i}}   
\newrobustcmd{\In}{\text{0}}   
\newrobustcmd{\state}{inverted stationary state\xspace}   
\newrobustcmd{\K}{\mathcal{K}}
\newrobustcmd{\D}{\mathcal{I}}
\renewrobustcmd{\P}{\mathcal{P}}   
\newrobustcmd{\W}{\tilde{W}}
\newrobustcmd{\Temp}{T}
\newrobustcmd{\dual}[1]{\bar{#1}}
\newrobustcmd{\one}{\mathds{1}}
\newrobustcmd{\ket}[1]{|#1\rangle}
\newrobustcmd{\bra}[1]{\langle#1|}
\newrobustcmd{\brkt}[1]{\langle #1 \rangle}
\newrobustcmd{\braket}[2]{\langle #1 | #2 \rangle}
\newrobustcmd{\Ket}[1]{\bm{|}#1\bm{)}}
\newrobustcmd{\Bra}[1]{\bm{(}#1\bm{|}}
\newrobustcmd{\Braket}[2]{\bm{(}#1\bm{|}#2\bm{)}}
\newrobustcmd{\Brkt}[1]{\bm{(} #1 \bm{)}}
\newrobustcmd{\op}[1]{\hat{#1}}
\newrobustcmd{\psiIn}{\psi_{\mathrm{I}}}
\newrobustcmd{\kvecIn}{\boldsymbol{k}_{\mathrm{I}}}
\newrobustcmd{\kvecInParr}{\boldsymbol{k}^{\|}_{\mathrm{I}}}
\DeclareMathOperator{\Tr}{Tr}
\newrobustcmd{\tr}{\underset{\res}{\Tr}}
\newrobustcmd{\tri}{\Tr_\res}
\newrobustcmd{\col}[4]{{\begin{bmatrix}#1 \\ #2 \\ #3 \\ #4 \end{bmatrix}}}
\newrobustcmd{\row}[4]{{\begin{bmatrix}#1 &  #2  & #3  & #4 \end{bmatrix}}}
\newrobustcmd{\suppmat}{\cite{Schulenborg15Suppmat}}
\newrobustcmd{\Eq}[1]{Eq.~(\ref{#1})}
\newrobustcmd{\BrackEq}[1]{[Eq.~(\ref{#1})]}
\newrobustcmd{\eq}[1]{(\ref{#1})}
\newrobustcmd{\Fig}[1]{Fig.~\ref{#1}}
\newrobustcmd{\BrackFig}[1]{[Fig.~\ref{#1}]}
\newrobustcmd{\fig}[1]{\ref{#1}}
\newrobustcmd{\Figs}[1]{Figs.~\ref{#1}}
\newrobustcmd{\Sec}[1]{Sec.~\ref{#1}}
\newrobustcmd{\Ref}[1]{Ref.~\cite{#1}}
\newrobustcmd{\Refs}[1]{Refs.~\cite{#1}}
\newrobustcmd{\App}[1]{App.~\ref{#1}}
\newrobustcmd{\fpop}{(-\one)^N}
\newrobustcmd{\fpOp}{(-\one)^N}
\newrobustcmd{\fpOpHat}{(-\one)^{\hat{N}}}
\newrobustcmd{\hatfpop}{(-\one)^{\hat{N}}}
\newrobustcmd{\hatfpOp}{(-\one)^{\hat{N}}}
\newrobustcmd{\zin}{z_{\text{i}}}
\newrobustcmd{\hatzin}{\hat{z}_{\text{i}}}
\newrobustcmd{\Phif}{\Phi_\mathrm{f}}
\newrobustcmd{\Phil}{\Phi_\mathrm{l}}
\newrobustcmd{\kvec}{\boldsymbol{k}}
\newrobustcmd{\kvecperp}{\boldsymbol{k}^{\perp}}
\newrobustcmd{\kvecparr}{\boldsymbol{k}^{\|}}
\newrobustcmd{\xvec}{\boldsymbol{x}}
\newrobustcmd{\pvec}{\boldsymbol{p}}
\newrobustcmd{\Vvec}{\boldsymbol{V}}
\newrobustcmd{\qZvec}{\boldsymbol{q}_0}
\newrobustcmd{\qZvecParr}{\boldsymbol{q}^{\|}_0}
\newrobustcmd{\kZvec}{\boldsymbol{k}_0}
\newrobustcmd{\uvec}{\boldsymbol{u}}
\newrobustcmd{\vvec}{\boldsymbol{v}}
\newrobustcmd{\kZvecperp}{\boldsymbol{k}^{\perp}_0}
\newrobustcmd{\kZvecalpbet}{\kvec_{\alpha\beta}}
\newrobustcmd{\kZvecalpbetParr}{\kvec^{\|}_{\alpha\beta}}
\newrobustcmd{\kZvecalpbetperp}{\kvec^{\perp}_{\alpha\beta}}
\newrobustcmd{\hOp}{\hat{h}}
\newrobustcmd{\hSC}{\hat{h}_{S}}
\newrobustcmd{\HSC}{H_{S}}
\newrobustcmd{\hW}{h_{\mathrm{W}}}
\newrobustcmd{\hWalpbet}{h_{\mathrm{W},\alpha\beta}}
\newrobustcmd{\PsiBCS}{\Psi_{BCS}}
\newrobustcmd{\HBCS}{H_{BCS}}
\newrobustcmd{\Htot}{H_{\text{tot}}}
\newrobustcmd{\hatHtot}{\hat{H}_{\text{tot}}}
\newrobustcmd{\Htun}{H_{\text{tun}}}
\newrobustcmd{\hatHtun}{\hat{H}_{\text{tun}}}
\newrobustcmd{\Hpair}{H_{\mathrm{pair}}}
\newrobustcmd{\Jens}[1]{ {\color{red}(JENS: #1)\color{black}} }
\newrobustcmd{\EBCS}{E_{BCS}}
\newrobustcmd{\EWSM}{E_{W}}
\newrobustcmd{\Csb}{C_{\mathrm{sb}}}
\newrobustcmd{\sgnfn}[1]{\mathrm{sgn}\left(#1\right)}
\newrobustcmd{\markinred}[1]{\color{red}{#1}\color{black}}
\newrobustcmd{\rsepa}{\,,\quad}
\newrobustcmd{\lsepa}{\quad ,\,}
\newrobustcmd{\lrsepa}{\quad , \quad}
\newrobustcmd{\colvec}[1]{\begin{pmatrix}#1\end{pmatrix}}
\newrobustcmd{\nbrack}[1]{\left(#1\right)}
\newrobustcmd{\sqbrack}[1]{\left[#1\right]}
\newrobustcmd{\cbrack}[1]{\left\{#1\right\}}
\newrobustcmd{\nlbrack}[1]{\left(#1\right.}
\newrobustcmd{\sqlbrack}[1]{\left[#1\right.}
\newrobustcmd{\clbrack}[1]{\left\{#1\right.}
\newrobustcmd{\nrbrack}[1]{\left.#1\right)}
\newrobustcmd{\sqrbrack}[1]{\left.#1\right]}
\newrobustcmd{\crbrack}[1]{\left.#1\right\}}
\newrobustcmd{\mean}[1]{\left\langle#1\right\rangle}
\newrobustcmd{\MeanO}[1]{\mean{#1}_0}
\newrobustcmd{\Mset}[1]{\left\{#1\right\}}
\newrobustcmd{\Mserset}[2]{\left\{#1,\dotsc,#2\right\}}
\newrobustcmd{\Forall}{\quad\forall}
\newrobustcmd{\Exists}{\quad\exists}
\newrobustcmd{\ExistsOne}{\quad\exists_1}
\newrobustcmd{\existsOne}{\exists_1}
\newrobustcmd{\MTupel}[1]{\left(#1\right)}
\newrobustcmd{\MserTup}[2]{\left(#1,\cdots,#2\right)}
\newrobustcmd{\Thetafn}[1]{\varTheta(#1)}
\newrobustcmd{\Deltafn}[1]{\delta(#1)}
\newrobustcmd{\expfn}[1]{\exp\left(#1\right)}
\newrobustcmd{\lnfn}[1]{\ln\left(#1\right)}
\newrobustcmd{\Fermfn}[2]{\Over{\expfn{\beta\left(#1-#2\right)}+1}}
\newrobustcmd{\fermfn}[2]{f_{#2}(#1)}
\newrobustcmd{\bosefn}[2]{b_{#2}(#1)}
\newrobustcmd{\abs}[1]{\left|#1\right|}
\renewrobustcmd{\Re}{\mathrm{Re}}
\renewrobustcmd{\Im}{\mathrm{Im}}
\newrobustcmd{\rhofn}[2]{\rho_{#1}(#2)}
\newrobustcmd{\cosfn}[1]{\cos{\left(#1\right)}}
\newrobustcmd{\arcsinfn}[1]{\mathrm{arcsin}\left(#1\right)}
\newrobustcmd{\sinfn}[1]{\sin{\left(#1\right)}}
\newrobustcmd{\tanfn}[1]{\tan{\left(#1\right)}}
\newrobustcmd{\cotfn}[1]{\cot{\left(#1\right)}}
\newrobustcmd{\cothfn}[1]{\coth{\left(#1\right)}}
\newrobustcmd{\tanhfn}[1]{\tanh\!\left(#1\right)}
\newrobustcmd{\sumsub}[2]{\sum_{\substack{{#1}\\{#2}}}}
\newrobustcmd{\sumsubsub}[3]{\sum_{\substack{{#1}\\{#2}\\{#3}}}}
\newrobustcmd{\prodsub}[2]{\prod_{\substack{{#1}\\{#2}}}}
\newrobustcmd{\cc}[1]{\overline{#1}}
\newrobustcmd{\TODO}[1]{\color{red}TODO: #1\color{black}}
\newrobustcmd{\equalby}[1]{\overset{#1}{=}}
\newrobustcmd{\equalbys}[2]{\underset{#2}{\overset{#1}{=}}}
\newrobustcmd{\equalbyeqn}[1]{\overset{\eqref{#1}}{=}}
\newrobustcmd{\equalbyeqns}[2]{\underset{\eqref{#2}}{\overset{\eqref{#1}}{=}}}
\newrobustcmd{\evalAtEqui}[1]{\left.#1\right|_{\text{eq}}}
\newrobustcmd{\evalAtBal}[1]{\left.#1\right|_{\text{bal}}}
\newrobustcmd{\zeq}{z_{\text{eq}}}
\newrobustcmd{\hatzeq}{\hat{z}_{\text{eq}}}
\newrobustcmd{\zieq}{z_{\text{i,eq}}}
\newrobustcmd{\hatzieq}{\hat{z}_{\text{i,eq}}}
\newrobustcmd{\zia}{z_{\text{i}\alpha}}
\newrobustcmd{\hatzia}{\hat{z}_{\text{i}\alpha}}
\newrobustcmd{\nzia}{n_{\text{i}\alpha}}
\newrobustcmd{\pzia}{p_{\text{i}\alpha}}
\newrobustcmd{\Eeq}{E_{\text{eq}}}
\newrobustcmd{\gamc}{\gamma_c}
\newrobustcmd{\gamp}{\gamma_p}
\newrobustcmd{\gamca}{\gamma_{c\alpha}}
\newrobustcmd{\gampa}{\gamma_{p\alpha}}
\newrobustcmd{\nza}{n_{z\alpha}}
\newrobustcmd{\pza}{p_{z\alpha}}
\newrobustcmd{\nzap}{n_{z\alpha'}}
\newrobustcmd{\nz}{n_{z}}
\newrobustcmd{\nzeq}{n_{z,\text{eq}}}
\newrobustcmd{\delnsqzeq}{\delta n^2_{z,\text{eq}}}
\newrobustcmd{\pzeq}{p_{z,\text{eq}}}
\newrobustcmd{\pzieq}{p_{i,\text{eq}}}
\newrobustcmd{\gamceq}{\gamma_{c,\text{eq}}}
\newrobustcmd{\gampeq}{\gamma_{p,\text{eq}}}
\newrobustcmd{\nzieq}{n_{\text{i,eq}}}
\newrobustcmd{\delnsqzieq}{\delta n^2_{\text{i,eq}}}
\newrobustcmd{\nzi}{n_{\text{i}}}
\newrobustcmd{\pzi}{p_{\text{i}}}
\newrobustcmd{\kBT}{k_{\text{B}}T}
\newrobustcmd{\kB}{k_{\text{B}}}
\newrobustcmd{\Hdot}{H}
\newrobustcmd{\Hleads}{H_{\text{leads}}}
\newrobustcmd{\hatHleads}{\hat{H}_{\text{leads}}}
\newrobustcmd{\INa}{I_{\text{N}}^{\alpha}}
\newrobustcmd{\INap}{I_{\text{N}}^{\alpha'}}
\newrobustcmd{\IEa}{I_{\text{E}}^{\alpha}}
\newrobustcmd{\GamR}{\Gamma_{\text{R}}}
\newrobustcmd{\GamL}{\Gamma_{\text{L}}}
\newrobustcmd{\muR}{\mu_{\text{R}}}
\newrobustcmd{\muL}{\mu_{\text{L}}}
\newrobustcmd{\mueq}{\mu}

\begin{document}
\allowdisplaybreaks

\setcounter{section}{0} %% Remove this when starting to work on the template.

%%%%%%%%%%%%%%%%%%%%%%%%%%%%%%%%%%%%%%%%%%
\section{Introduction}
%%%%%%%%%%%%%%%%%%%%%%%%%%%%%%%%%%%%%%%%%%

%%%%%%%%%%%%%%%%%%%%%%%%%%%%%%%%%%%%%%%%%%
\subsection{Motivation and general outline}
%%%%%%%%%%%%%%%%%%%%%%%%%%%%%%%%%%%%%%%%%%
A thorough understanding of the thermoelectric operation of basic circuit elements such as quantum dots is important for future energy- and information-technologies, see, for example, the review articles~\cite{Sothmann2015,Benenti2017Jun} and references therein.
Their \emph{nonlinear} operation, due to large temperature and voltage gradients, is nontrivial and has only recently
received more attention~\cite{Scheibner2008,Esposito2009,Wierzbicki2010Oct,Kennes2013,Sanchez2013Jan,Meair2013,Svensson:2013,Sothmann2015,Sanchez2016Dec,Erdman2017Jun,Benenti2017Jun,Shiraishi2017May,Josefsson2017Oct},
the results indicating new possibilities for thermoelectric applications.
To get a better grip on this nonequilibrium regime, the implications of time-reversal symmetry -- often exploited in thermoelectrics -- have been expressed in fluctuation-relations using the powerful tools of counting statistics~\cite{Saito2008Sep,Forster2008Sep,Esposito2009Dec,Silaev2014Aug,Utsumi2014May}.
This is helpful since in strongly confined devices the coupled nonlinear transport of charge and heat is particularly rich and correspondingly difficult to compute. The reason for this is the energy-dependent transmission caused by the system itself, having discrete quantum states and strong Coulomb interaction, rather than by its coupling.
In particular, the effect of Coulomb interaction is multifaceted, since local interactions can both store energy and modify the transfer of energy by individual particles. For applications this can be both damaging and helpful: interaction can limit the efficiency since it provides a way to transfer heat without inducing a charge current, thereby leading to leakage; yet, it can also modify the effective level structure in an advantageous way~\cite{Karlstrom11,Kennes2013} or be used for energy harvesting in multi-terminal devices~\cite{Sanchez2011Feb,Sothmann2012May,Whitney2016Jan}
as shown experimentally~\cite{Thierschmann2015Oct,Hartmann2015Apr,Roche2015Apr}.

One reason why the discussion of efficiencies and performance of thermoelectric devices
has been mostly confined to the linear regime of operation is the powerful Onsager reciprocity~\cite{Onsager1931Feb,Butcher1990,Jacquod2012Oct}. The latter is implied by time-reversal symmetry and it allows to make straightforward statements about efficiencies.
%%%%%%% General intro duality %%%%%%%%%
However, recently a new general symmetry relation has been discovered for electronic systems~\cite{Schulenborg16} that is independent of time-reversal symmetry.
Its formulation starts from the very general observation that the dynamics of any system obeying linear superposition of its states is characterized by its time-evolution \emph{modes}, the right eigenvectors of the evolution generator. The \emph{amplitude} of excitation of these modes is governed by the corresponding left eigenvectors.
For closed systems, these mode and amplitude vectors of the Schr\"odinger dynamics $\partial_t \ket{\phi}= -i H \ket{\psi}$ are the energy eigenkets and bras, respectively, which are trivially related to each other by taking the Hermitian-adjoint (since $H=H^\dagger$).
However, for open systems no such simple general relation exists due to the more complicated nature of nonunitary time evolution.
Nevertheless, some of us have shown~\cite{Schulenborg16} that for any fermionic open system in the wide-band limit, a quite general duality exists between the modes and amplitudes of the time-evolution kernel. This holds even when the reduced density operator of the open system obeys a kinetic equation with the \textit{time-nonlocal} form $\partial_t \rho(t) = -i[H,\rho(t)] + \int_0^t dt' \mathcal{W}(t,t') \rho(t')$. Importantly, the duality relies on a fundamental principle \emph{other} than time-reversal, namely the superselection rule based on fermion parity~\cite{Wick1952Oct,Aharonov1967Mar,Streater2000,Bogolubov1990}.
This principle is obeyed by all fermionic systems, and therefore the duality is expected to have far-reaching physical consequences.

For thermoelectric transport, the implications of this additional general principle are to date largely unexplored. 
Fully exploiting the new duality relation, the present paper revisits the thermoelectric response of a weakly coupled, but strongly interacting single-level quantum dot subject to both electrical and thermal biases,
both in the linear and nonlinear regime.
We derive new results, significantly clarify known results, and thereby -- most notably -- outline a completely new approach for the analytical computation of thermoelectrics of interacting nanoscale systems.
%
%%%%%%% from transient behavior to advantages in stationary transport %%%%%%%
In \Refs{Schulenborg16,Vanherck17}, it was already realized that the duality relation between decay modes and their excitation amplitudes would naturally express itself in the \textit{transient} charge and heat currents
as the system decays to a new stationary state after a sudden switch (quench).
This paper shows that the duality is a similarly powerful tool for the experimentally more accessible \emph{stationary} thermoelectric transport.
Indeed, we demonstrate that both in the linear and nonlinear stationary regime -- after some judicious choices -- an expansion in open-system evolution eigenmodes is advantageous. The compact analytical formulas we obtain in this way offer several new interesting insights into important measurable thermoelectric transport quantities, and provide a truly nonstandard physical view point.
Most importantly and as further outlined below, they expose the remarkable fact that a strongly \textit{repulsive} system may prominently feature behavior reminiscent of strong \textit{attractive} interaction in the thermoelectric response coefficients.\footnote{This should be clearly distinguished from the effects studied in \cite{Andergassen2011Dec} for systems which from the start have attractive interaction.} Since this was not  recognized as such due to lack of a proper framework, our paper does not merely present trivial or subjective rewritings, even though we re-derive and re-express some known results.
Instead, our approach is uniquely dictated by the physically-motivated evolution-mode decomposition, and the outcomes are virtually impossible to ``guess'' without the systematics and new intuition provided by the duality.
%
%%% Technical advantages and new insight %%%%

We finally note that the very general origin and formulation~\cite{Saptsov2012Dec,Saptsov2014Jul,Schulenborg16} of the duality allows us to extend our method beyond the weakly coupled quantum dot setup treated here. As our study illustrates, the technical simplifications can be such that one can avoid numerical calculations altogether.
We therefore expect that the ideas presented here can also simplify the analysis of more complex systems studied most recently~\cite{Sierra2016Jun,Jaramillo2017Feb,Erdman2017Jun,Rossello2017Jun}. This includes situations with broken time-reversal symmetry~\cite{Vanherck17}, non-diagonal density operators due to, e.g., noncollinear magnetic fields, and also strongly coupled systems exhibiting time non-local effects.

%%%%%%% Structure of the paper %%%%%%%
The paper is structured as follows:
after outlining the main ideas and results in general terms [\Sec{sec_ideas}]
and introducing the microscopic model and the weak-coupling master equation [\Sec{sec_model}],
we review the minimal details of the duality relation
that we will exploit for a gated quantum dot coupled to two electronic reservoirs.
The mode-decomposition of the charge and energy current formulas [\Eq{eq_INz}-\eq{eq_IEz}] are then analyzed in the linear [\Sec{sec_linear}]
and nonlinear transport regimes [\Sec{sec_nonlinear}].
The manuscript also contains an appendix which is substantial,
not because the new derivations are complicated,
but because the steps are nonstandard and deserve to be outlined.

%%%%%%%%%%%%%%%%%%%%%%%%%%%%%%%%%%%%%%%%%%
\subsection{Overview of main ideas and results\label{sec_ideas}}
%%%%%%%%%%%%%%%%%%%%%%%%%%%%%%%%%%%%%%%%%%

\begin{figure}[!t]
	\center
	\includegraphics[width=0.8\textwidth]{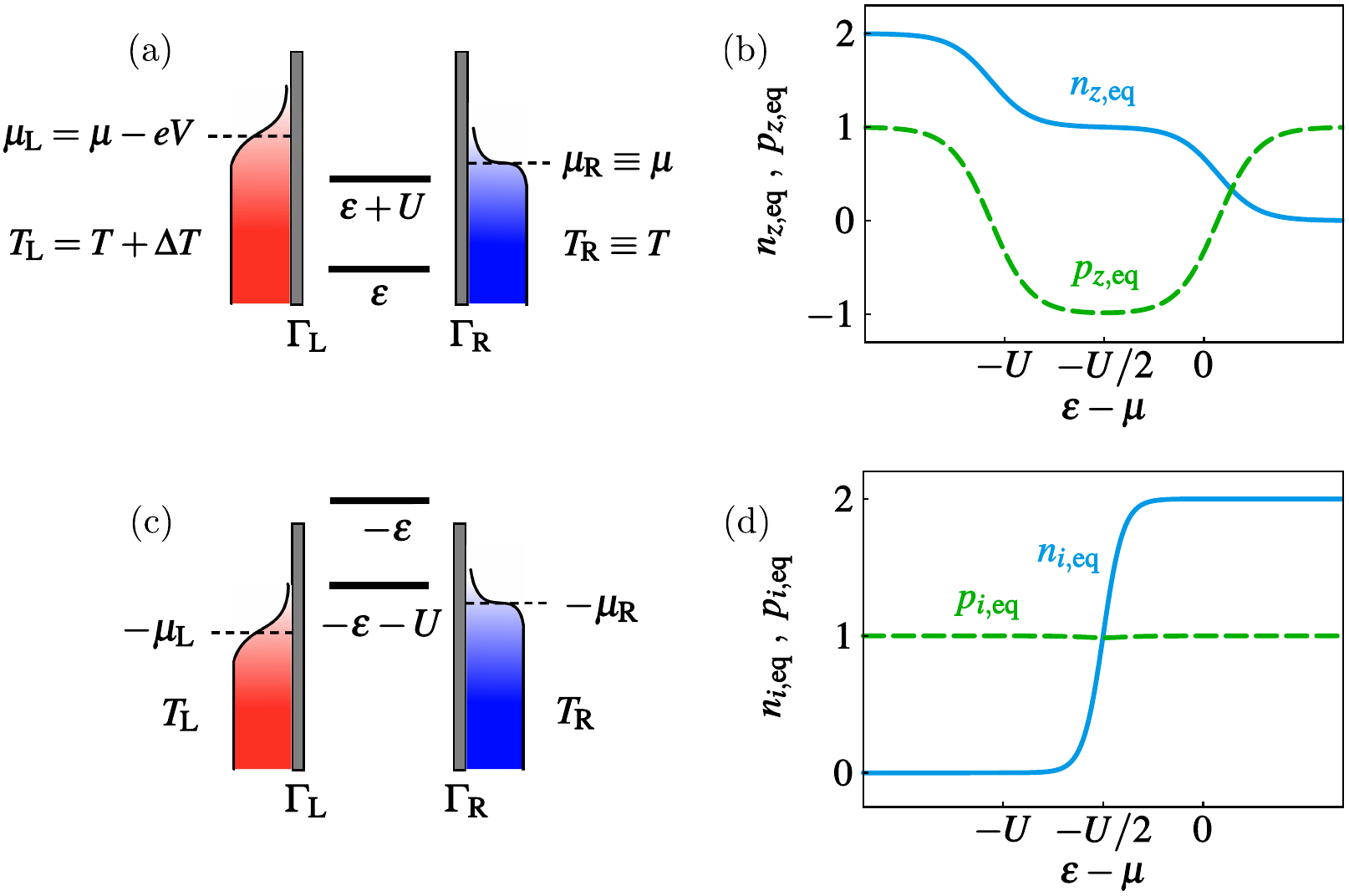}
	\caption{
		\label{fig_setup}
		(a) Energy landscape of a single-level quantum dot with repulsive interaction $U>0$ in contact with a hot and a cold electronic reservoir with an applied bias voltage.
				(b)  Equilibrium occupation number and parity of the single-level quantum dot as a function of the level position [taken with respect to the right reservoir only, or equivalently obtained for the full system at equilibrium, $\Delta T=V=0$, as indicated by the ``eq'' subscript]. The occupation $\nzeq$ shows $T$-broadened steps at the two Coulomb resonances $\mueq = \epsilon$ and $\epsilon+U$, respectively, and the parity $\pzeq$ alternates  correspondingly.
		(c) Energy landscape of the \emph{inverted} system that is dual to the system of interest in (a).
		(d) Equilibrium occupation number and parity of the single-level quantum dot in its \textit{inverted} stationary state. In the dual, attractive system, the charge shows a \emph{single} double-sized step at $\mueq = \epsilon + \tfrac{1}{2}U$, with \emph{half} the temperature broadening $T/2$
		where two electrons enter or leave the dot by two first-order processes sequential in time,
		and the parity is essentially always even.
		Parameters for (b) and (d): $U=10T$, $V=\Delta T=0$, and $\GamL=\GamR\ll T$. 	}
\end{figure}

By discussing a simple, yet relevant model of a nanoscale system, this paper aims to illustrate how the understanding of the role of Coulomb interaction in thermoelectric transport is advanced by the fermion-parity duality.
Before going into further details in \Sec{sec_fermion_parity} it is important to appreciate three of its main ideas in general terms:

(1) The duality relation maps the eigenmodes of the system of interest to the amplitudes for a \emph{different physical system} -- the dual system.
We will refer to the latter as the \emph{inverted} system because the duality transformation \emph{inverts the interaction} (as well as other energies),
going from repulsive to attractive and \emph{vice versa}.
For quantum dots, this can be easily visualized by inverting the energy landscape in \Fig{fig_setup}(a) to that of \Fig{fig_setup}(c), whose details will be discussed later on.
This mapping already explains the seemingly strange occurrence of features of attractive interaction in quantities computed for repulsive systems, as first noted in \cite{Schulenborg16,Vanherck17}. The straightforward interpretation of such puzzling properties is done resorting to the \emph{inverted stationary state}, which allows to understand the nontrivial dependence on the original system's parameters from the -- often simple and well-understood~\cite{Anderson1975Apr} -- physics of the attractive dual model as in  \Fig{fig_setup}(c).

(2) Another reason why the duality clarifies interaction effects is that the ``essential'' correlating parts of the Coulomb interactions,
say, between two orbitals $i$ and $j$
with occupation operators $\hat{n}_i$ and $\hat{n}_j$, respectively, is simply given by parity operators,
$(1-2\hat{n}_i)(1-2\hat{n}_j)=(-1)^{\hat{n}_i+\hat{n}_j}$.
In fact, correlated electron model Hamiltonians are often directly formulated in terms of the operators on the left hand side.
The duality reveals that the total parity operator always corresponds to a special eigenmode of open fermion-system dynamics~\cite{Saptsov2012Dec,Saptsov2014Jul}, and is hence protected. In simple yet relevant situations, one thereby cleanly separates, \emph{throughout} the entire calculation, the contributions of the Coulomb interaction into an ``essential'' correlating part and a nontrivial ``average'' contribution carried by a charge mode.
Since Coulomb interaction is an important source of energy dependence and energy storage in quantum dots, thermoelectrics is thus seen to be intimately tied up with fermion parity and the corresponding duality.

(3) Finally, in the context of thermoelectricity, it is important to emphasize that the duality -- in the simple form used here -- requires energy-independent \emph{coupling} between system and reservoir (wide-band limit). This does not mean that it is irrelevant to thermoelectric transport, where properly engineered energy-dependence of the coupling can be of interest for the device operation, see e.g. Refs.~\cite{Sanchez2011Feb,Thierschmann2015Oct,Sothmann2012May,Hartmann2015Apr,Roche2015Apr}. Here, the nanoscale system \emph{itself} provides the strong energy dependence required for thermoelectric effects,
both through strong size-quantization and Coulomb interaction.
Models of this kind are relevant in many thermoelectric studies~\cite{Crepieux2011Apr,Juergens2013Jun,Jordan2013Feb,Zhou2015Oct,Sierra2016Jun}
and the duality applies to their description, even when the energy-independent coupling is strong and the temperature is low~\cite{Schulenborg16}, cf. \cite{Saptsov2012Dec,Saptsov2014Jul}. Also, effective energy-dependent couplings as realized in multi-dot systems~\cite{Juergens2013Jun,Jordan2013Feb} can be treated in terms of the duality relation presented here.
Finally, the duality considerations can be extended~\cite{Schulenborg_tobepublished} systematically to account for the energy-dependence of the coupling.

%%%%%% New results in this paper %%%%%
As a guide to the present paper, we now outline how the above general points turn up in the specific insights of our study,
most of which remain hidden when approaching the thermoelectric problem in the standard way:

%%%%% Linear regime %%%%
In linear response [\Sec{sec_linear}],
we combine the duality with Onsager's reciprocity derived from time-reversal symmetry.
%%%%% other known dualities and time-reversal %%%%%%%%
The discussion of the linear response coefficients benefits from the combined insights of both relations.\footnote{Previously, our duality was similarly combined with the independent well-known Iche-duality~\cite{Iche1972Jun,Taraphder1991May,Koch2007May} based on an electron-hole transformation, yielding new relations for quantum dots in a magnetic field~\cite{Vanherck17}.}
Even more so, in our particular example, we can show how the duality enforces the Onsager relations in linear response, and restricts \emph{how} these relations break down beyond the linear regime.
To achieve this, we first formulate the linear response in a way that is compatible with the duality.
Thereby, we find a simple, explicit expression for the average energy carried by electrons, the tight-coupling part of the heat current.
Remarkably, it depends on the mean occupation of the dot in the \emph{inverted} stationary state: this formula exposes the unexpectedly \emph{sharp crossover} behavior of the thermopower between well-separated resonances [\Fig{fig_seebeck}] as a two-particle \emph{resonance} in the dual attractive model.
Also, the obvious formal similarity between the linear Ohm and Fourier laws, 
\begin{align}
	I = G V,\qquad
	J =  K \Delta T
	\label{eq_ohm_fourier}
\end{align}
gets an unexpected twist:
Whereas it is well-known that the stationary-state fluctuations of the dot occupation govern the electrical conductance $G$,
the duality reveals that the linear Fourier heat coefficient\footnote{The Seebeck coefficient $S$ and the Peltier coefficient $\Pi$ are introduced in Sec.~\ref{sec_linear}.} $\kappa=K-\Pi S G$, for the heat current in the absence of a charge current, $J|_{I=0}=\kappa\Delta T$, is dominated instead by occupation fluctuations in the \textit{inverted} stationary state. As such, the $\epsilon$-dependence of $\kappa$ is also governed by attraction, exhibiting the same two-particle resonance as the thermopower.

%%%%%%%%%%%%%%%%%%%%%%%%%%%%%%%%%%%%%%%%%%
%%%%%%%%%%%%%%%%%%%%%%%%%%%%%%%%%%%%%%%%%%

%%%%% Nonlinear regime %%%%
The generally more complicated regime of nonlinear response  [\Sec{sec_nonlinear}]
can also be addressed with the duality.
For the calculation of the nonlinear Seebeck and Fourier coefficient, we show that an evaluation of \textit{equilibrium} dot observables -- both in the original and dual, inverted system -- is often sufficient to obtain the nonequilibrium heat currents.
This leads to compact analytical expressions
and major simplifications in their explicit calculation,
while clarifying the underlying physical picture.
For example, the nonlinear Fourier heat is essentially the difference between the parity when the quantum dot is in equilibrium with a single lead on the left or on the right.
Although it is well-known that the Fourier heat is carried by electron-electron interaction -- even for macroscopic devices~\cite{Ashcroft1976Jan} --
the duality pinpoints precisely \emph{which part} of the interaction is crucial:
it is the parity operator that is entirely responsible for the transferred Fourier heat.
Finally, the strong difference between Peltier and Seebeck effects in the nonlinear regime, indicating the breakdown of Onsager reciprocity, can be rationalized completely. The nonlinear Peltier coefficient can be decomposed into an equilibrium Seebeck contribution that stems from the heat transport tightly coupled to the charge transport and 
a parity-mode contribution. Both of these contributions are sensitive to the attractive-interaction physics of the dual system.

%%%%%%%%%%%%%%%%%%%%%%%%%%%%%%%%%%%%%%%%%%
\section{Model, master equation, and duality}\label{sec_model}
%%%%%%%%%%%%%%%%%%%%%%%%%%%%%%%%%%%%%%%%%%

%%%%%%%%%%%%%%%%%%%%%%%%%%%%%%%%%%%%%%%%%%
\subsection{Model, assumptions, and notation}
%%%%%%%%%%%%%%%%%%%%%%%%%%%%%%%%%%%%%%%%%%

We are interested in thermoelectric transport  through a single-level quantum dot between two\footnote{Many statements can be generalized in a straightforward manner to multiple reservoirs.} reservoirs labeled $\alpha=\text{L,R}$,
as sketched in \Fig{fig_setup}(a).
The full Hamiltonian 
\begin{equation}
 \hatHtot = \hat{\Hdot} + \hatHleads + \hatHtun\label{eq_hamiltonian_full} 
\end{equation}
decomposes into three parts.
The dot Hamiltonian
\begin{equation}
\hat{\Hdot}=\sum_{\sigma=\uparrow,\downarrow}\epsilon \hat{n}_\sigma +U\hat{n}_\uparrow \hat{n}_\downarrow\label{eq_hamiltonian_dot}
\end{equation}
describes the correlated dynamics of a single spin-degenerate level $\epsilon$
with Coulomb charging energy $U$ for doubly occupying the level.
Here $\hat{n}_{\sigma}=\hat{d}^\dagger_\sigma \hat{d}^{}_\sigma$
are number operators for dot electrons of spin $\sigma=\uparrow,\downarrow$
with creation (annihilation) operators $\hat{d}^\dagger_\sigma$ ($\hat{d}^{}_\sigma$).
The quantum dot is coupled to noninteracting electronic reservoirs $\alpha$
 described by
\begin{equation}
\hatHleads = \sum_{\alpha}\hat{\Hdot}_{\alpha}
=\sum_{\alpha,k,\sigma}\epsilon_{\alpha k} \hat{c}_{\alpha k \sigma}^\dagger \hat{c}^{}_{\alpha k \sigma}
,
\qquad
\hat{N}_\alpha:=\sum_{k,\sigma}\hat{c}_{\alpha k \sigma}^\dagger \hat{c}^{}_{\alpha k \sigma}
\label{eq_Nalpha}
\end{equation}
with operators  $\hat{c}^{}_{\alpha k \sigma}$ and $\hat{c}_{\alpha k \sigma}^\dagger$ for electrons with spin $\sigma$ and orbital quantum number $k$. The reservoir energies $\epsilon_{\alpha k}$ and tunneling amplitudes $t_{\alpha k}$ in the tunnel coupling
\begin{equation}
\hatHtun = \sum_{\alpha,k,\sigma}t_{\alpha k} \hat{d}^\dagger_\sigma \hat{c}^{}_{\alpha k \sigma}+\text{h.c.}
\end{equation}
are assumed to be spin-independent.
We moreover assume that the relevant coupling strengths $\Gamma_\alpha(E)=2\pi\sum_{k\sigma}\delta(\epsilon_{\alpha k} - E)|t_{\alpha k}|^2$ determining the typical scale of the tunneling rates are energy independent:
 $\Gamma_\alpha(E) \approx \Gamma_\alpha$.
This wide-band limit assumption is crucial for the derivation of the fermion-parity duality in the simple form given below in Eq.~(\ref{eq_duality}),
see also our earlier remarks in \Sec{sec_ideas} under point (3).
Finally, the individual grand-canonical leads $\alpha$ are characterized by the electrochemical potential $\mu_\alpha$, and the inverse temperature $\beta_\alpha = 1/(\kBT_\alpha)$
entering through the particle / hole Fermi distribution functions $f_\alpha^\pm(x)=[e^{\pm \beta_\alpha(x-\mu_\alpha)}+1]^{-1}$.
We let $T=T_\text{R}$ and $\mu=\mu_\text{R}$ denote the global-equilibrium values of the reservoirs
and $\Delta T$ and $V$ the applied biases through
$T_\text{L}=T+\Delta T$ and $\mu_\text{L}=\mu-eV$.

In the next two sections, we set up the calculation of the stationary charge and heat currents, $I^\alpha$ and $J^\alpha$, through the quantum dot which can be written in terms of \emph{particle} and \emph{energy} currents, $\INa$ and $\IEa$:
\begin{eqnarray}
I^\alpha=-e \INa \qquad \textrm{and} \qquad  J^\alpha=\IEa-\mu_\alpha \INa \ . \label{eq_charge_heat}
\end{eqnarray}
We will outline how to compute the currents 
$\INa
:=
-\frac{\partial}{\partial t}\langle \hat{N}_\alpha\rangle 
$  and $\IEa
:=
-\frac{\partial}{\partial t}\langle \hat{H}_\alpha\rangle 
$ [cf. \Eq{eq_Nalpha}]
from lead $\alpha$ into the dot
in a way that exploits the duality.
From here on, we set $e=\kB=\hbar=1$.

%%%%%%%%%%%%%%%%%%%%%%%%%%%%%%%%%%%%%%%%%%
\subsection{Master equation and non-equilibrium currents\label{sec:master}}
%%%%%%%%%%%%%%%%%%%%%%%%%%%%%%%%%%%%%%%%%%

Following previous work~\cite{Schulenborg16,Vanherck17,SchulenborgLic,VanherckMas}, we outline how to compute the particle and energy currents via the reduced density operator $\hat{\rho}$ of the quantum dot
obtained by tracing out the reservoir degrees of freedom.
For weak coupling and high temperatures, $\Gamma_\alpha \ll T_\alpha$,
the mixed state obeys a Born-Markov master equation:
\begin{equation}\label{eq_master}
\frac{d}{dt}\Ket{\rho}=W\Ket{\rho}\ .
\end{equation}
In \Eq{eq_master}, operators are written as supervectors, indicated by round kets: $\hat{\rho}=\Ket{\rho}$.
In the following we also need to consider the ``superhermitian'' conjugate of such a supervector $\Ket{x}$ which is conveniently written as bra $\Bra{x}$.
This denotes a linear function acting on any $\Ket{y}$, i.e., another operator $\hat{y}$, as follows:
 $\Braket{x}{y} := \text{Tr} \, [ \hat{x}^\dagger \hat{y} ]$.
In this notation, the kernel $W$ is a \emph{superoperator}
whose specific matrix elements are given in \App{app:explicit} for completeness, 
but which are not required in the following [cf. \Eq{eq_table}].
Such a matrix element $\Bra{f} W \Ket{i}$ physically relates to the rate for a tunnel-induced transition, $\Ket{i} \to \Ket{f}$, between two of the dot states
\begin{equation}
 \Ket{0} = \ket{0}\bra{0} \lrsepa \Ket{1} = \frac{1}{2}\sqbrack{\ket{\uparrow}\bra{\uparrow} + \ket{\downarrow}\bra{\downarrow}} \lrsepa \Ket{2} = \ket{2}\bra{2}.\label{eq_states}
\end{equation}
The empty state $\ket{0}$ has energy $E = 0$, the \emph{mixture} of the spin states $\ket{\uparrow}$ and $\ket{\downarrow}$ has energy $E = \epsilon$, and the doubly occupied state $\ket{2}=\ket{\uparrow\downarrow}$ has energy $E = 2\epsilon + U$. Since the system is fully rotationally invariant\footnote{
	The rotational symmetry of the full Hamiltonian \eq{eq_hamiltonian_full} causes the master equation for the probabilities in the states \eq{eq_states} to completely decouple from the local spin-dynamics, including the coherences.}
and because we focus on the particle- and energy currents,
we only need the kernel $W$ in a linear subspace spanned by the three trace-normalized operators given in Eq.~\eq{eq_states},
see the supplement to \cite{Schulenborg16} for details.
Importantly, due to the weak coupling, the kernel is the sum of kernels $W_\alpha$ that would be obtained if the system was coupled \emph{only}\footnote{Regarding the use of the reservoir index $\alpha$, we note that the particle and energy current in reservoir $\alpha$ depend on the properties of \textit{all} reservoirs. This is in contrast to, e.g., the kernel $W_\alpha$, whose contributions are due to a single reservoir $\alpha$ only. We emphasize this difference by choosing superscripts for the former case and subscripts for the latter.}
to reservoir $\alpha$:
\begin{equation}
W = \sum_{\alpha} W_{\alpha} \ .   \label{eq_kernel_decomposition}
\end{equation}
In the same approximation, the currents $\INa$ and $\IEa$  can be expressed \cite{SchulenborgLic,Vanherck17} in terms of the reservoir-resolved kernel $W_\alpha$, the number operator $\hat{N} = \sum_\sigma \hat{n}_\sigma$,
as well as the Hamiltonian $\hat{\Hdot}$ \emph{of the dot}, and the solution  of \Eq{eq_master}:
\begin{eqnarray}
\INa=-\frac{\partial}{\partial t}\langle \hat{N}_\alpha\rangle = \Bra{N}W_{\alpha}\Ket{\rho}\label{eq_IN}\\
\IEa=-\frac{\partial}{\partial t}\langle \hat{\Hdot}_{\alpha}\rangle = \Bra{\Hdot}W_{\alpha}\Ket{\rho}.\label{eq_IE}
\end{eqnarray}
In writing the second equality in Eqs.~(\ref{eq_IN}) and (\ref{eq_IE}), particle and energy conservation have been used.\footnote{The second step in the energy current Eq.~(\ref{eq_IE}) is only valid in the weak-coupling limit, see remarks in \Refs{Schulenborg16,SchulenborgLic} and \Ref{Gergs_tobepublished}.}

From this point onward, the \textit{standard} way to obtain the currents in the stationary limit [namely, where the left hand side of Eq.~(\ref{eq_master}) equals zero] seems straightforward, knowing the Fermi Golden-Rule expressions for the rates $\Bra{f} W_\alpha \Ket{i}/\Braket{f}{f}$:
(i) Find the non-equilibrium state $\Ket{z}$ by solving the stationary limit of \Eq{eq_master}, $W \Ket{z} = \sum_\alpha W_\alpha \Ket{z} = 0$ (``zero mode''), and normalize it to $\Braket{\one}{z} = \Tr\sqbrack{\hat{z}} = 1$.
(ii) Plug
$\Ket{\rho}=\Ket{z}$ into the current equations \eq{eq_IN} and \eq{eq_IE}.
While this procedure yields explicit expressions for the non-equilibrium currents, it often provides only limited analytical and physical insights, in particular for thermoelectric quantities.
Moreover, it unnecessarily complicates the evaluation of bias-derivatives required for the linear-response coefficients, even when using Beenakker's linearization~\cite{Beenakker1991Jul,Beenakker1992Oct}.
However, inspection of equations (\ref{eq_master}), (\ref{eq_IN}), and (\ref{eq_IE}) suggests an alternative route based on our recent results~\cite{Schulenborg16,Vanherck17}
which we outline next.

%%%%%%%%%%%%%%%%%%%%%%%%%%%%%%%%%%%%%%%%%%
\subsection{Fermion-parity duality and its use in thermoelectrics.}
\label{sec_fermion_parity}
%%%%%%%%%%%%%%%%%%%%%%%%%%%%%%%%%%%%%%%%%%

The calculation of the stationary state and the stationary currents
\begin{align}
\INa=\Bra{N}W_{\alpha}\Ket{z}
,\qquad
\IEa=\Bra{\Hdot}W_{\alpha}\Ket{z}
\label{eq_Is}
\end{align}
are expected to be drastically simplified when expressed in a basis of time-evolution modes,
a standard practice when solving linear systems of equations.
Looking at \Eq{eq_Is}, three different eigenvector bases suggest themselves:
those of the two separate kernels $W_\alpha$ ($\alpha=\text{L,R}$)
and of the total kernel $W=W_{\textrm{L}}+W_{\textrm{R}}$.
A key technical point of the present paper is that even though the evaluation of the currents \eqref{eq_Is} depends on the total kernel $W$ through its zero mode $\Ket{z}$,
it often turns out that a mode decomposition of the separate kernels $W_\alpha$  suffices to compute $\Ket{z}$.
This provides an important advantage: the kernels $W_\alpha$ of the separate leads are by definition \textit{always} equilibrium kernels. We can thus express and understand nonequilibrium currents in terms of equilibrium quantities which have much simpler dependencies on parameters.

There are three reasons why this decomposition is possible.
First, the weak-coupling approximation allows the decomposition given in Eq.~\eq{eq_kernel_decomposition}.
Second, our study focuses in most parts either on the linear response regime [\Eq{eq_linear_response_state}],
or on the nonlinear regime but with a vanishing \emph{average} charge current. We will return to this point at the end of this section.
The third, most crucial ingredient is 
the \emph{fermion-parity duality}, which strongly restricts the form of the kernels $W$ and $W_\alpha$ and thereby identifies the optimal variables in which to express the currents.

We first discuss the duality for a quantum dot coupled to only one lead $\alpha$,
allowing us to take over the expression for $W_\alpha$ given in \Ref{Schulenborg16} for this case.
There, it was shown that there is a duality relation between the kernel $W_\alpha$ and its Hermitian conjugate at inverted energies:
\begin{equation}
 \sqbrack{W_{\alpha}(\hat{\Hdot},\mu_\alpha)}^\dagger = -\Gamma_\alpha -\hat{\P} \ W_\alpha(-\hat{\Hdot},-\mu_\alpha) \ \hat{\P} \lrsepa \hat{\P}\Ket{\bullet} = \Ket{\fpOp \bullet} \ .\label{eq_duality}
\end{equation}
Here, $W_{\alpha}(-\hat{\Hdot},-\mu_\alpha)$ denotes the kernel of the \emph{dual} system, in which the signs of all energies in the Hamiltonian $\hat{\Hdot}$, as well as the electrochemical potential $\mu_\alpha$ in lead $\alpha$, have been inverted.
This is sketched in \Fig{fig_setup}(a) and (c). The operator $\hatfpOp$ is the fermion parity of the dot, giving $+1$ for even and $-1$ for an odd occupation number, and its appearance reflects the fundamental origin of the duality, Eq.~\eq{eq_duality}.
We refer to Refs.~\cite{Schulenborg16,SchulenborgLic} for an introduction to fermion-parity superselection, the general duality relation and its special form used here, as given in Eq.~\eq{eq_duality}.

The duality relation imposes strong restrictions on the fermionic master equation \eq{eq_master}.
In fact, these restrictions are so strong that they completely determine the rate matrix $W_\alpha$. For illustration, this construction is carried out explicitly in \App{app:kernel} and directly produces the three eigenvalues $-\gamma_{m \alpha}$  of $W_\alpha$
and the corresponding left and right eigenvectors$\Bra{m'_\alpha}$ respectively $\Ket{m_\alpha}$ labeled using $m=z,c,p$:
\begin{align}
	W_\alpha= - \sum_{m=c,p} \gamma_{m \alpha} \Ket{m_\alpha} \Bra{m'_\alpha}
	,
	\label{eq_Walphaexp}
\end{align}
where the zero mode $\hat{z}_\alpha$ does not show up because $\gamma_{z \alpha} = 0$.
Alternatively, one may directly compute these quantities by diagonalizing the matrix $W_\alpha$ employing the duality~\eq{eq_duality} as done in \Ref{Schulenborg16} and obtain:
\begin{equation}
\begin{array}{c|c|c|c}
	\text{Label}           &               \text{Amplitude}                &                           \text{$-$ Eigenvalue $=$ decay rate}                             &                           \text{Mode}                                \\\hline\hline
	 \text{Zero }  &         \Bra{z'_\alpha} = \Bra{\one}          &                            \gamma_{z \alpha} = 0                             &                           \Ket{z_\alpha}                              \\
	    (z)          &            \text{\small{[trace]}}             &                                                                              &                     \text{\small{[stationary state]}}                     \\\hline
	\text{Charge } & \Bra{{c_\alpha}'} = \Bra{N} - \nza \Bra{\one} & \gamca = \frac{\Gamma_\alpha}{2} \Big[ f^{+}_\alpha(\epsilon)+f^-_\alpha(\epsilon+U) \Big] & \Ket{c_\alpha} = \frac12 \fpOpHat \Big[\Ket{N}-\nzia \Ket{\one}\Big]  \\
	   (c)           &    \text{\small{[$\sim$ charge operator]}}    &                                                                              &               \text{\small{[$\sim$ charge operator]}}                  \\\hline
	\text{Parity } &      \Bra{p'_\alpha} = \Bra{\zia \fpOp}       &                           \gampa =  \Gamma_\alpha                            &                     \Ket{p_\alpha} = \Ket{\fpOp}                       \\
	      (p)        &   \text{\small{[$\sim$ inverted stationary state]}}    &                                                                              &                   \text{\small{[parity operator]}}                   \\\hline
\end{array}
\label{eq_table}
\end{equation}
Inspecting this table, one explicitly sees that the mode vectors and amplitude covectors
are \emph{cross-related} through the parity operator $(-\one)^{\hat{N}}$ and the energy inversion (indicated by subscript ``i'' on $\hatzia$ and $\nzia$) that appear in the duality \eq{eq_duality}.
The stationary state -- given by the equilibrium grandcanonical ensemble $\Ket{z_\alpha} = \hat{z}_\alpha = e^{-\beta_\alpha(\hat{H} - \mu_\alpha\hat{N})}/\Tr\sqbrack{e^{-\beta_\alpha(\hat{H} - \mu_\alpha\hat{N})}}$ \BrackEq{eq_derive_step3} in the here considered weak tunnel-coupling regime  -- has the eigenvalue $0$. Hence, duality relates $\Ket{z_\alpha}$ to $\Bra{p'_\alpha}$  with eigenvalue $-\gampa = -\Gamma_\alpha$, where $\Bra{p'_\alpha}$ is the trace with the \emph{inverted} stationary state $\times$ parity.
Similarly, the trace operation $\Bra{z'_\alpha}$ is cross-related to the parity operator $\Ket{p_\alpha}$.
Finally, since the Fermi function with respect to any chemical potential $\mu_\alpha$ obeys $f_{\mu_\alpha}^{+}(x) = f^{-}_{-\mu_\alpha}(-x)$, the charge eigenvalue $-\gamca$ is self-dual and $\Bra{c'_\alpha}$ is mapped to $\Ket{c_\alpha}$.

The crucial point made by the table in \eq{eq_table} is that it reveals the natural variables in which the currents \eq{eq_Is} are expressed when inserting the mode expansion \eq{eq_Walphaexp}.
These include the expectation values of the charge and parity in the stationary state,
\begin{align}
\nz = \Braket{N}{z} = \Tr \left[\hat{N} \hat{z}\right]
,
\qquad
p_{z} =\Braket{\fpOp}{z} = \Tr \left[\fpOpHat \hat{z}\right]\ ,
\label{eq_nz}
\end{align}
entering through the factors $\Braket{m'_\alpha}{z}$.
However, due to the factors $\Braket{N}{m_\alpha}$ and $\Braket{H}{m_\alpha}$, additional expectation values with respect to the reservoir-$\alpha$ resolved equilibrium state \BrackEq{eq_derive_step3} appear:
\begin{align}
	\nza = \Braket{N}{z_\alpha}
	\lrsepa
	\pza = \Braket{\fpOp}{z_\alpha} 
	\label{eq_nza},
\end{align}
explicitly given as a functions of $\epsilon,U,\mu_\alpha,T_\alpha$ in Eqs.~\eq{eq_derive_step9} and \eq{eq_parity_expectation}, and most importantly, expectation values
\begin{align}
\nzia &= \Braket{N}{\zia}
 = \nza\sqbrack{\epsilon, U, \mu_\alpha \rightarrow -\epsilon, -U, -\mu_\alpha}
\lrsepa
\pzia &= \Braket{\fpOp}{\zia}
 = \pza\sqbrack{\epsilon, U, \mu_\alpha \rightarrow-\epsilon, -U, -\mu_\alpha}
\label{eq_nia},
\end{align}
where $\hat{z}_{\text{i}\alpha}$ is the corresponding \emph{inverted} stationary state of the \textit{dual system} \BrackEq{eq_derive_step6},
\begin{align}\label{eq_def_zi}
\hat{z}_{\text{i}\alpha} = \hat{z}_{\alpha}\sqbrack{\hat{\Hdot}, \mu_\alpha \rightarrow -\hat{\Hdot}, -\mu_\alpha} = e^{+\beta_\alpha(\hat{H} - \mu_\alpha\hat{N})}/\Tr\sqbrack{e^{+\beta_\alpha(\hat{H} - \mu_\alpha\hat{N})}}\ .
\end{align}
In \Fig{fig_setup} (b) and (d), we plot the averages
\eq{eq_nza} and \eq{eq_nia} for $\nza$ and $\pza$ in the right lead, i.e., $\mu_\alpha = \mu$ and $T_\alpha = T$:
They are simple, stepped functions of the physical parameters
whose shapes are straightforwardly rationalized based on the physics of a repulsive and attractive quantum dot in \emph{equilibrium} with a single lead $\alpha$, respectively, as explained in the caption to \Fig{fig_setup}.
Inverting the interaction $U$ -- going from \Eq{eq_nza} to \Eq{eq_nia} -- \emph{qualitatively} changes the parameter dependence of these quantities, as shown in the following.
We point out that the standard treatment of the problem completely overlooks that expressions \eq{eq_nia} are part of the natural variables in which to express the currents \eqref{eq_Is}, and in which to understand the parameter dependence of the transport.

The importance of this insight has been demonstrated for the transient time-dependent response of a quantum dot coupled to a single reservoir in \cite{Schulenborg16}.
The final issue that we face in extending this to the \emph{stationary} nonequilibrium thermoelectric currents \eqref{eq_Is} is that these currents depend on the stationary state $\Ket{z}$ of the system coupled to \emph{both} leads.
This state is the zero mode of the \textit{total} kernel $W$, for which the duality holds independently, as shown in \Ref{Schulenborg16}:
\begin{equation}
\sqbrack{W(\hat{\Hdot},\mu_\text{L},\mu_\text{R})}^\dagger =
-\Gamma-\hat{\P} \ W(-\hat{\Hdot},-\mu_\text{L},-\mu_\text{R}) \hat{\P} \label{eq_duality_tot}
,
\end{equation}
introducing the lump sum of couplings  $\Gamma=\sum_{\alpha} \Gamma_\alpha$. 

Importantly, while equation \eq{eq_duality_tot} itself follows from the lead-resolved duality \eq{eq_duality} with $W = \sum_\alpha W_\alpha$, its implications for the eigenvectors of $W= - \sum_{m=c,p} \gamma_{m} \Ket{m'} \Bra{m}$ and their eigenvalues are nevertheless non-trivial, since $W_\text{L}$ and $W_\text{R}$ do not commute.
The nonequilibrium stationary state that we need
is, in particular, not generally a sum of lead-resolved stationary states with some lead-resolved\footnote{If one considered more than two leads and obtained 3 linearly-independent vectors already from all $\Ket{z_\alpha}$, one could formally expand $\Ket{z}$ in these vectors. However, this would not be beneficial, since it would still require knowledge of the explicit form of $\Ket{z}$, and the weights $\lambda_\alpha$ would be functions of all system parameters, including the temperature and chemical potential of all other reservoirs.} weights $\lambda_\alpha$,
\begin{align}
\Ket{z}  \neq \sum_{\alpha} \lambda_\alpha \Ket{z_\alpha}
.
\end{align}
Section \ref{sec_linear} will, however, show that such a simple decomposition does hold when taking the first temperature-bias or voltage-bias derivative [\Eq{eq_linear_response_state}] at equilibrium.
More remarkably, as exploited in \Sec{sec_nonlinear}, this decomposition continuous to hold even in the \emph{nonlinear} regime \BrackEq{eq_bal_state} whenever the system is balanced to maintain zero particle current, $I_\alpha = 0$ \BrackEq{eq_balanced}.
Both cases thus allow us to by-pass the computation of $\Ket{z}$ and fully exploit the duality \eq{eq_duality} for $W_\alpha$ for each lead separately in the way outlined above.
This procedure directly ties the \emph{non-equilibrium} thermoelectric transport to \emph{equilibrium} physics by introducing the optimal variables \eq{eq_nza} and \eq{eq_nia} from the start.

%%%%%%%%%%%%%%%%%%%%%%%%%%%%%%%%%%%%%%%%%%
\subsection{Charge and energy currents\label{eq_currents}}
%%%%%%%%%%%%%%%%%%%%%%%%%%%%%%%%%%%%%%%%%%
Combining all above outlined ideas, we insert the eigenmode decomposition $W_\alpha= - \sum_{m=z,c,p} \gamma_{m \alpha} \Ket{m'_\alpha} \Bra{m_\alpha}$ into the current equations \eq{eq_Is} and use the table in \eq{eq_table} to obtain the central current formulas of the paper:
\begin{subequations}
\begin{align}
\INa &= - \gamca \Braket{c_\alpha'}{z}  =  \gamca \left[ \nza - n_z \right]\label{eq_INz}\\
\IEa &=  \left[\epsilon+\frac{U}{2}\left(2-\nzia\right)\right] \INa - \gampa \ U \  \Braket{\zia\fpOp}{z}\ .
\label{eq_IEz}
\end{align}
The particle current \eq{eq_INz} is simply the charge relaxation rate $\times$ the excess dot charge relative to its equilibrium value with respect to lead $\alpha$, where the current is measured.
The energy current \eq{eq_IEz} shows a more interesting nonstandard decomposition: it has a tight-coupling contribution directly proportional to the charge current, and a contribution which is independent of the charge current \eq{eq_INz}, and hence associated to the parity mode. The prefactor of the tight-coupling term involves the energy scale
\begin{eqnarray}
E_\alpha =
 \epsilon+\frac{U}{2}\left(2-\nzia\right) \ .
 \label{eq_average_E}
\end{eqnarray}
Remarkably, the interaction contribution to this energy is \emph{not} naturally expressed in the stationary state charge $n_{z\alpha}$, but instead in the \emph{inverted} stationary charge $\nzia$ with respect to lead $\alpha$.
As expected, $E_\alpha$ is approximately $\epsilon$ in the vicinity of transitions between zero- and singly occupied state, and approximately $\epsilon+U$ in the vicinity of transitions between singly and doubly occupied state.
However, in the crossover regime between these two resonances it is the physics of attractive interaction in $\nzia$ that dominates its behavior,
cf. \Fig{fig_setup}.

The overlap $\Braket{\zia\fpOp}{z}\approx\Braket{\zia}{z}$, contributing to the non-tightly coupled term of Eq.~\eq{eq_IEz}, approximately equals a state overlap since the parity in the inverted stationary state is almost always even due to the attractive interaction of the dual model, see \Fig{fig_setup}(d). This relation will turn out to be useful for the understanding of the characteristic features of the nonlinear Peltier coefficient discussed in Sec.~\ref{sec_peltier_nonlinear}. For explicit numerical results and plots, it is helpful to write
\begin{equation}\label{eq_overlap_explicit}
\Braket{\zia\fpOp}{z}=\frac{1}{4} \left(p_z + p_{i\alpha} \right) + \frac{1}{2} \left( n_z -1 \right) \left( n_{i\alpha} -1 \right),
\end{equation}\label{eq_Iz}
\end{subequations}
fully expressing it in terms of the measurable quantities introduced in Eqs.~\eq{eq_nz} to \eq{eq_nia}. See, e.g., Ref.~\cite{VanherckMas} for the derivation of this relation, and appendices~\ref{app:kernel}-\ref{app:linear_response} for the explicit calculation of all appearing quantities.

The current formulas \eq{eq_Iz} are valid both in the linear and in the nonlinear response regime and form the starting point of the remainder of the paper.
The following sections discuss their implications for the stationary thermoelectric response of the quantum dot from the new perspective offered by the duality \eq{eq_duality}.

%%%%%%%%%%%%%%%%%%%%%%%%%%%%%%%%%%%%%%%%%%
\section{Linear response regime\label{sec_linear}}
%%%%%%%%%%%%%%%%%%%%%%%%%%%%%%%%%%%%%%%%%%

\begin{figure}[!t]
	\center
	\includegraphics[width=0.45\textwidth]{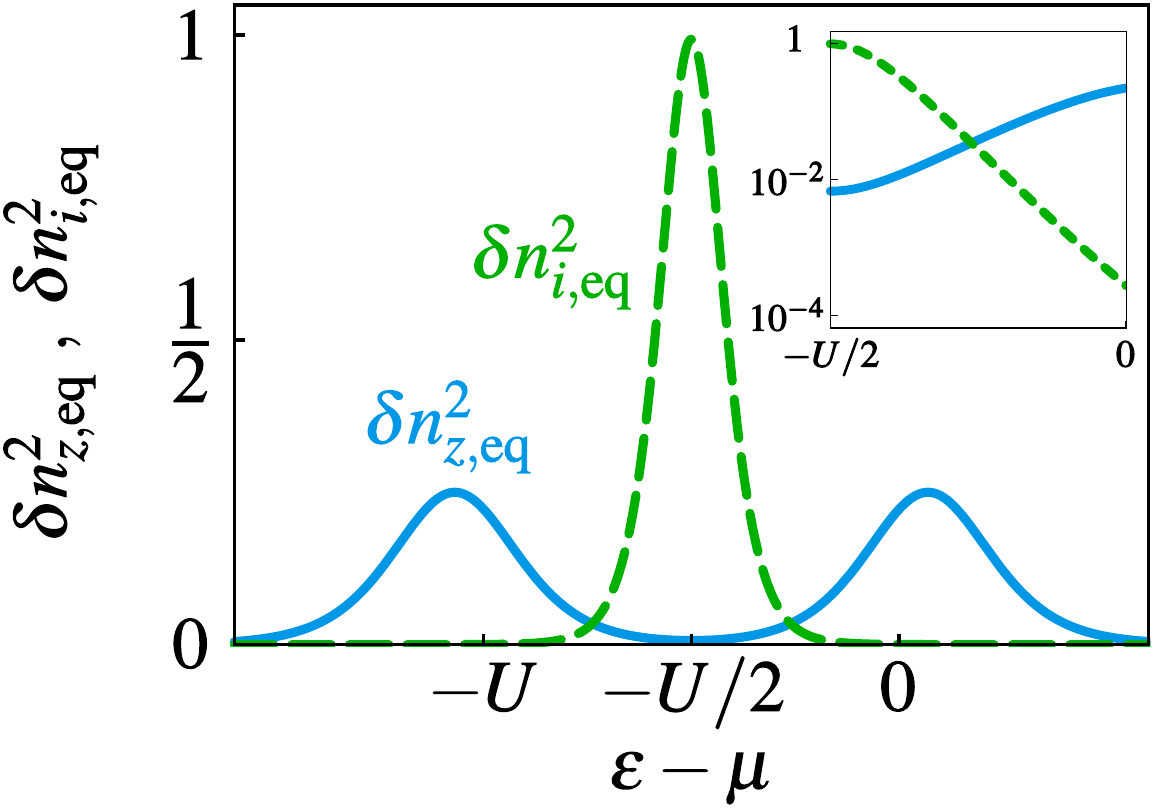}
	\caption{
		\label{fig_fluctuations}
		Fluctuations of the dot occupation number in the equilibrium state of the original, repulsive model (blue, full line) and of the inverted dual model (green, dashed line), showing features of attractive interaction. The inset shows the section between $\epsilon-\mueq=-U/2$ and $\epsilon-\mueq= 0$ on a log scale to show the different  decay of the two types of fluctuations due to thermal smearing. The parameters are $U=10T$, $V=\Delta T=0$, and $\GamL=\GamR\ll T$.
	}
\end{figure}

We start with the investigation of the linear thermoelectric response of the quantum dot
to voltage and temperature gradients,
$|V|,|\Delta T| \ll T$.
In this case, we can consider the symmetrized charge and heat currents, $I=(I^\text{L}-I^\text{R})/2$ and $J=(J^\text{L}-J^\text{R})/2$
which are standardly written in terms of the Onsager matrix,
\begin{eqnarray}
\left(\begin{array}{c}
I\\J
\end{array}\right)
\approx 
\left(\begin{array}{cc}
L_{11} & L_{12}\\L_{21} & L_{22}
\end{array}\right) \left(\begin{array}{c}
V/T \\ \Delta T/T^2
\end{array}\right) \ .\label{eq_onsager_matrix}
\end{eqnarray}
Here, the diagonal elements
$L_{11}=TG$
and
$L_{22}=T^2K$
present the electric and thermal conductances,
$G=dI/dV|_\textrm{eq}$
and
$K=dJ/d\Delta T |_\textrm{eq}$
respectively, whereas the off-diagonal elements $L_{12}=T^2\partial I/\partial\Delta T|_\text{eq}$ and $L_{21}=T\partial J/\partial\Delta T|_\text{eq}$ characterize the thermoelectric and the electrothermal responses. Here and below, we denote the evaluation of any quantity $q$ in the equilibrium limit by either of the expressions
$q |_\text{eq} = q_\text{eq} := q |_{V=\Delta T=0}$. The linearization of our general result \eq{eq_Iz} for the currents around equilibrium
is conveniently found from 
\begin{eqnarray}
&& \evalAtEqui{\tfrac{d}{dx}\Ket{z}} = \sum_\alpha\frac{\Gamma_\alpha}{\Gamma}\evalAtEqui{\tfrac{d}{dx}\Ket{z_\alpha}} = -\sum_\alpha\frac{\Gamma_\alpha}{\Gamma}
	{\sqbrack{
		{\tfrac{d\hat{A}_\alpha}{dx}} - \Braket{{\tfrac{d A_\alpha}{dx} }}{z_\alpha}\cdot\one}}\cdot\evalAtEqui{\Ket{z_\alpha}} \label{eq_linear_response_state}
\end{eqnarray}
for derivatives with respect to $x=\mu_\alpha,\beta_\alpha$, where we have introduced the affinities
$\hat{A}_\alpha := \beta_\alpha\nbrack{\hat{\Hdot} - \mu_\alpha \hat{N}}$. The derivation of this result [\App{app:near_equi}] only involves the linearization around equilibrium and nothing else, i.e., no detailed balance or other balancing relations used in previous derivations~\cite{Beenakker1991Jul,Beenakker1992Oct} and their extensions~\cite{Erdman2017Jun}.
Formulating the linear response on the level of the stationary \textit{state} has the crucial advantage that it can be combined with the duality.
Thereby we can
circumvent the cumbersome process of taking derivatives of expectation values and subsequently simplifying them.
We instead exploit \Eq{eq_linear_response_state}, the mode expansion given in \eq{eq_table} and two simple orthogonality relations involving the
equilibrium state of the original and the inverted model,
\begin{equation}
 \Braket{\zieq\fpOp}{\zeq} = 0 \lrsepa
 \Braket{\zieq\fpOp\cdot N}{\zeq} = 0\ .\label{eq_orthogonality_relations}
\end{equation}
The first is simply the orthogonality of left eigenvector $\Bra{p'}$ and right stationary vector $\Ket{z}$, here applied at equilibrium, which by duality even holds beyond linear response. The second relation follows from
$\hat{z}_\text{eq}=e^{-(\hat{H}-\mu)/T} / \Tr[e^{-(\hat{H}-\mu)/T}]$ which for the inverted equilibrium state implies $\hatzieq(\hat{H},\mu) =\hat{z}_\text{eq} (-\hat{H},-\mu)\propto e^{(\hat{H}-\mu)/T} \propto (\hat{z}_\text{eq})^{-1}$. Therefore the second relation is found from 
\begin{align}
	\Braket{\zieq\fpOp\cdot N}{\zeq}
	\propto  \Tr[\hat{N}\fpOpHat] = 0\ .
\end{align}
Using the above equations, we finally obtain the linear response coefficients of the charge and heat currents~(\ref{eq_Iz}) [see \App{app:linear_response} for the details of this non-standard derivation],
\begin{subequations}
\begin{eqnarray}
L_{11} & = & \frac{\GamL\GamR}{\Gamma^2}\ \gamceq\ \delnsqzeq \ \ 
\label{eq_G} \\
L_{12}=L_{21}& = &-L_{11} \left[ \left( \epsilon - \mueq \right) + \frac{U}{2}\left(2-\nzieq\right)\right] \  
\label{eq_M}\\
L_{22} & = & L_{11}\left[\left( \epsilon - \mueq \right) + \frac{U}{2}\left(2-\nzieq\right)\right]^2+ \frac{\GamL\GamR}{\Gamma^2} \ \gampeq\left(\frac{U}{2}\right)^2
\delnsqzeq \ \delnsqzieq \ ,
\label{eq_K}
\end{eqnarray}\label{eq_linear_response_coefficients}\end{subequations}
with equilibrium decay rates
$\gamceq= \frac{\Gamma}{2} \left[f^{+}(\epsilon)+ f^{-}(\epsilon+U)\right]$
and
$\gampeq = \Gamma = \gamp$.
These results warrant a detailed discussion given in the following sections.
However, what is immediately clear is that even in the linear response regime, the natural quantities in which the response coefficients are expressed are not only the coupling asymmetry $\GamL\GamR/\Gamma^2$, the expected energies ($\epsilon$, $\mu$, $U$) and the occupation number $\nzeq$ and its fluctuations in the original dot model,
\begin{eqnarray}
	\delnsqzeq & = & \brkt{\hat{N}^2}_{\text{eq}}  - \brkt{\hat{N}}_{\text{eq}}^2  = \Braket{N^2}{\zeq}  - \Braket{N}{\zeq}^2\ .\label{eq_fluct_nz}
\end{eqnarray}
In addition,
the occupation number $\nzieq$ [cf. \Eq{eq_average_E}] and its fluctuations
\begin{eqnarray}
\delnsqzieq & = & \evalAtEqui{\sqbrack{\brkt{\hat{N}^2}_{\zin}  - \brkt{\hat{N}}_{\zin}^2}} = \Braket{N^2}{\zieq}  - \Braket{N}{\zieq}^2
= \delnsqzeq\sqbrack{\epsilon,U,\mu \rightarrow -\epsilon,-U,-\mu}
\label{eq_fluct_ni}
\end{eqnarray}
in the equilibrium state of the \textit{dual, attractive} system $\hat{z}_\text{i,eq}$ appear as well. They enter via the overlap of the equilibrium state and dual equilibrium state, $\Braket{\zieq}{\zeq} = \delnsqzieq \cdot \delnsqzeq $, as proven in \App{app:linear_response_peltier_fourier}.
In Fig.~\ref{fig_fluctuations}, we plot the $\epsilon$-dependence of both fluctuations by evaluating the explicit expression of $\delnsqzeq$ as a function of $\epsilon,U,\mu,T$, stated in \Eq{eq_fluctuation_expectation_explicit}.
These fluctuations show peaks at $\epsilon=\mu$ and $\epsilon+U=\mu$ for the original model
and at $\epsilon + U/2 = \mu$ for the dual attractive model.
Using that the fluctuations also follow from an $\epsilon$-derivative of the equilibrium occupation numbers \BrackEq{eq_fluctuation_expectation_derivative}, $\delnsqzeq=-T\partial \nzeq/\partial\epsilon$ and $\delnsqzieq=T\partial \nzieq/\partial\epsilon$ due to \Eq{eq_fluct_ni}, the observed resonances in $\delnsqzeq$ and $\delnsqzieq$ are readily understood from $\nzeq$ and $\nzieq$ as shown earlier in \Fig{fig_setup}(b) and (d).

%%%%%%%%%%%%%%%%%%%%%%%%%%%%%%%%%%%%%%%%%%
%%%%%%%%%%%%%%%%%%%%%%%%%%%%%%%%%%%%%%%%%%

%%%%%%%%%%%%%%%%%%%%%%%%%%%%%%%%%%%%%%%%%%
\subsection{Electric response}
\label{sec_cond}
%%%%%%%%%%%%%%%%%%%%%%%%%%%%%%%%%%%%%%%%%%
For reference, we note that the linear electric conductance
\begin{align}
G = \frac{L_{11}}{T} = \frac{1}{T} \frac{\GamL\GamR}{\Gamma^2}\ \gamceq\ \delnsqzeq 
\label{eq_G_trivial}
\end{align}
is essentially the charge-relaxation rate $\gamma_\text{c,eq}$ $\times$
the equilibrium charge fluctuations $\delnsqzeq$ plotted in Fig.~\ref{fig_fluctuations}.
It shows the well-known Coulomb peaks~\cite{Beenakker1991Jul} whenever fluctuations of the charge by one electron are energetically allowed, see \Fig{fig_setup}(b).
Starting from the charge current formula \eq{eq_INz}, we note that the conductance \eq{eq_G_trivial} derives from the charge-mode of the evolution, as physically expected.

%%%%%%%%%%%%%%%%%%%%%%%%%%%%%%%%%%%%%%%%%%
\subsection{Thermo-electric response and Seebeck thermopower}
\label{sec_seebeck_linear}
%%%%%%%%%%%%%%%%%%%%%%%%%%%%%%%%%%%%%%%%%%

The charge mode  also solely determines the response \eq{eq_M} of the charge current  to a small temperature difference $\Delta T$.
To tie this to an energy scale, we consider
the  thermopower or Seebeck coefficient $S=-\frac{1}{T} \frac{L_{12}}{L_{11}}$
corresponding to the induced thermo-voltage $V|_{I=0}$ across the two leads in the absence of a charge current, $V|_{I=0}=S\Delta T$ (Seebeck effect).
The expression $\Eeq :=  T\,S$ has the unit of an energy, related to the voltage required to counterbalance the thermally induced charge current~\cite{Mott:1969,Matveev99}.
Our linearized result \eq{eq_M} gives for this energy
\begin{eqnarray}
\Eeq
=  T\, S
= \left( \epsilon - \mueq \right) + \frac{U}{2}\left[ 2-\nzieq(\epsilon - \mu,U,T) \right]
	= E_\alpha|_\text{eq} \ .
 \label{eq_seebeck}
\end{eqnarray}
Before discussing this result, we immediately note that $\Eeq$ equals the tight-coupling energy \eq{eq_average_E} [for $\nzia|_\text{eq}=\nzieq$ with $\mu_\alpha = \mu, T_\alpha = T$] which by duality naturally appears in the decomposition of the \emph{heat} current \eq{eq_IEz}.
As we will detail when introducing the Peltier coefficient in \Sec{sec_peltier_linear}, this fact indicates that the Onsager relation $L_{12}=L_{21}$ is obeyed.
However, since we purposely refrain from using time-reversal symmetry, it is first of interest to see how, by 
using only linear response of the state \eq{eq_linear_response_state}
and the mode-amplitude duality~\eq{eq_table}, the \emph{charge} current formula \eq{eq_INz} -- apparently of very different form from the heat current -- produces nonetheless the same energy scale. This follows by noting that for the linear response coefficient $L_{12}$, only equilibrium correlators are relevant\footnote{
The linearization of the particle current \eq{eq_INz} in either $x = \mu_\alpha,T_\alpha$ is proportional to terms of the form $\Bra{N}\evalAtEqui{\tfrac{d}{dx}\Ket{z}}$.
	By \Eq{eq_linear_response_state} the $\Delta T$-derivative of $\Ket{z}$ translates to a temperature derivative of the Boltzmann factor $e^{-(\hat{\Hdot} - \mu \hat{N})/T}$ which pulls down both $\hat{N}$ and $\hat{\Hdot}$,
	giving correlators $\brkt{\hat{N}^2}_{\text{eq}}$ and $\brkt{\hat{\Hdot}\hat{N}}_{\text{eq}}$.
}, in particular $\brkt{\hat{\Hdot}\hat{N}}_{\text{eq}}$.
The latter\footnote{
	\Ref{Crepieux2016} also links the thermoelectric conversion efficiency to mixed charge-heat noise, which in essence relates to the time-nonlocal version of the mixed particle-energy correlator appearing here.} introduces the characteristic energy $\Eeq$:
as a consequence of  the orthogonality relations \eq{eq_orthogonality_relations},
the only contribution to $\brkt{\hat{\Hdot}\hat{N}}_{\text{eq}}$
comes from the part of the energy $\hat{\Hdot}$ that does \emph{not} couple to the parity $\hatfpOp$ [see \App{app:linear_response_seebeck}]. 
The additional appearing correlator $\brkt{\hat{N}^2}_{\text{eq}}$ is responsible for the $\mu$-shift occurring in $\Eeq$.

The energy \eq{eq_seebeck} thus naturally appears by mode decomposition of these correlators, showing that the duality is essential for the understanding of the thermopower.
As we noted above [\Eq{eq_Iz}], the interaction-induced part in $\Eeq$, respectively $S$
is governed by the behavior of the average occupation $\nzieq$ in the \emph{inverted} stationary state.
Importantly, this energy is not simply the noninteracting part
[$\nzieq$ depends nontrivially on $U$, cf. \Eq{eq_derive_step9}] or a mean-field-like energy which would involve the average $\nzeq/2$ rather than $(2 - \nzieq)/2$.
%%%%%%%%%%%%%%%%%%%%%%%%%%%%%%%%%%%%%%%%%%%%%%%%%%%%%%%%
\begin{figure}[!t]
	\center
	\includegraphics[width=4.5in]{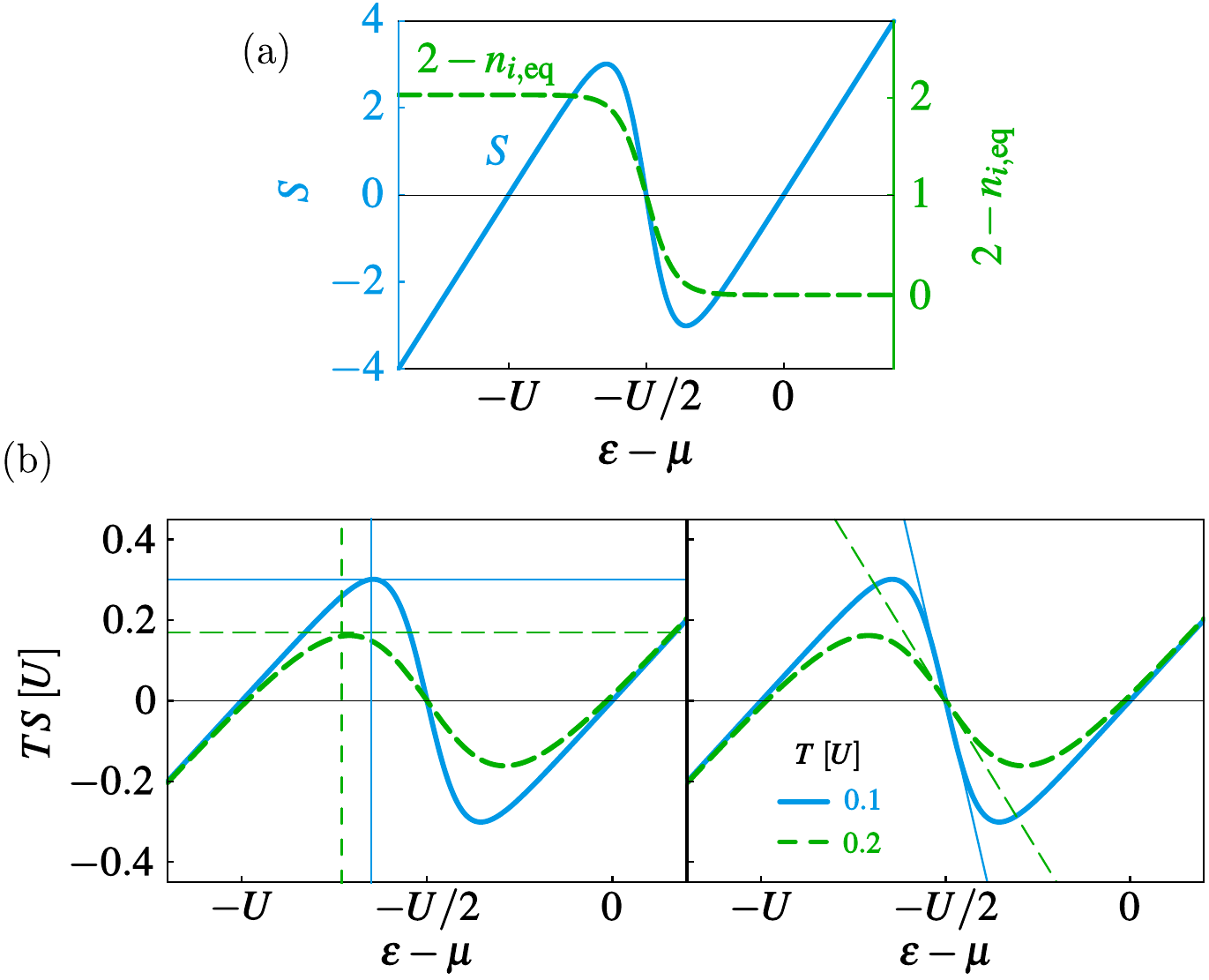}
	\caption{
		(a) Linear Seebeck thermopower $S$, Eq.~(\ref{eq_seebeck}), (solid-blue line), and $2-\nzieq$ (dashed-green line) as a function of the level position of the dot $\epsilon - \mueq$, for $U=10T$. Here, $\nzieq$ is the occupation number of the inverted stationary state at equilibrium. In the units chosen here ($e=\kB=1$), the Seebeck coefficient is dimensionless.\\
		(b) Comparison with the approximate position and height of the local maxima of $S$, given by Eq.~(\ref{eq_S_minmax}), (left panel), and of the slopes at the particle-hole symmetry point, given by Eq.~(\ref{eq_S_slope}), (right panel).
		For all panels, the couplings are chosen $\GamL=\GamR\ll T$.
		\label{fig_seebeck}
	}
\end{figure}
%%%%%%%%%%%%%%%%%%%%%%%%%%%%%%%%%%%%%%%%%%%%%%%%%%%%%%%%

In Fig.~\ref{fig_seebeck}(a), we plot the Seebeck coefficient, Eq.~(\ref{eq_seebeck}), as a function of the energy level position $\epsilon - \mueq$.
The sawtooth behavior has been known for a long time~\cite{Beenakker1992Oct} and it is traditionally understood as follows:
Each linear branch of the curve is explained by considering just one of the two Coulomb resonances and ignoring the other.
For each branch, the magnitude of the voltage that can be sustained increases linearly
\footnote{In the vicinity of a single resonance, the linear behavior of the Seebeck coefficient can simply be understood from the exact solution of a single noninteracting level with transmission probability $\mathcal{T}(\omega)$ peaked around $\omega=\epsilon$. Then, $I=\int d\omega \mathcal{T}(\omega)\left(f_\text{L}(\omega)-f_\text{R}(\omega)\right)\approx\int d\omega \mathcal{T}(\omega)\left.(\partial f/\partial x)\right|_{(x=\epsilon-\mu)/T}\left[-(\omega-\mu)\Delta T/T^2+V/T\right]$, when linearizing with respect to $\Delta T$ and $V$. Approximating the transmission probability to be proportional to a $\delta$ function, this yields $I\approx (G/T)\left[(\epsilon-\mu)\Delta T/T-V\right]$. Getting a vanishing current requires $V=S\Delta T=(\epsilon-\mu)\Delta T/T$, such that $S$ is linear in $\epsilon$.}
with $\epsilon - \mueq$.
The shape of the whole curve is then understood as a crossover from the particle-dominated transport of one resonance to the hole-dominated transport of the next one when changing $\epsilon - \mueq$:
by a continuity argument the curve must cross zero between the two resonances to connect the two branches.
However, it remains unclear from this line of arguing whether this is a sharp jump -- as here in the linear regime -- or a smooth crossover -- as in the nonlinear regime discussed later.
Moreover, close inspection shows that this sharp jump has an anomalous thermal broadening by \emph{half} the temperature, $T/2$.

Both issues are readily understood from our result \eq{eq_seebeck} deriving from duality:
the mean energy $\Eeq$ probed by the thermopower depends on the average charge $\nzieq$ of the quantum dot with \emph{inverted} energies at equilibrium.
As \Fig{fig_setup}(c) and (d) illustrate, the occupation in the inverted stationary state $\nzieq$ abruptly drops from 2 to 0 once $\epsilon - \mueq$ exceeds $-U/2$, as is well-known for impurities with attractive Coulomb interaction~\cite{Anderson1975Apr}.
The duality thus explains the unexpectedly sharp jump in the thermopower in \Fig{fig_seebeck}.
Moreover, since for $U\gg T$ the occupation for the strongly attractive dual model is well-approximated by 
\begin{equation}
	\nzieq \approx 2f^-(2\epsilon+U) =
	2
	\left.
	f^-\nbrack{\epsilon + \frac{U}{2}}
	\right|_{T \rightarrow T/2}
	,
	\label{eq_dual_charge_approx}
\end{equation}
we see that the anomalous thermal scale of \emph{half} the reservoir temperature $T/2$ appears. This has an intuitive interpretation in the dual model:
since the attractive interaction favors a single transition involving an electron pair, each electron feels half of the thermal noise\footnote{
	Note that here electron pairs are \emph{not coherently} transported: such coherent pair tunneling appears only in processes of order $\Gamma^2$. In the analysis~\cite{Leijnse2009Oct} of the rates for such effects a Bose function appears with $2\epsilon + U$ in its argument, indicating coherent transport of fermion pairs. Interestingly, we note that this rate as function of $\epsilon$ shows the same halving of the thermal noise as discussed in the present paper.}.
Note that in our calculation for the attractive dual model, two electrons enter sequentially (in time) by two separate processes with rates described by Fermi functions\footnote{The rate matrix \eqref{eq_W_matrix} contains no nonzero rate connecting the empty and doubly occupied state.}. Their net effect is that the transition becomes allowed at a single value of $\epsilon-\mu=-U/2$.

Since our result \eq{eq_seebeck} expresses the thermopower compactly in the natural and well-understood variable $\nzieq$,
we can easily find a simple yet accurate formula for the level positions at which the thermopower achieves a local minimum $(\epsilon_-)$ and maximum $(\epsilon_+)$.
This requires finding $\epsilon$ from the extremal condition
\begin{align}
	\frac{\partial \nzieq }{\partial (\epsilon/T)}  = \frac{2T}{U}\ .\label{eq_S_minmax_step}
\end{align}
The strong attractive interaction in the dual system relative to the anomalous lower temperature scale $T/2$  makes \Eq{eq_dual_charge_approx} an excellent approximation to the left hand side of \Eq{eq_S_minmax_step}, already for $U \gtrsim 5T$.
This gives
\begin{equation}
\frac{\epsilon_\pm - \mueq}{U}  \approx   -\frac{1}{2} \mp \frac{T/2}{U}\lnfn{\frac{U}{T/2}}
,\qquad
S(\epsilon_\pm)
 \approx  \pm \sqbrack{\frac{U}{2T} - \frac{1}{2}\nbrack{1 + \lnfn{\frac{U}{T/2}}}}.\label{eq_S_minmax}
\end{equation}
Furthermore, the sensitivity of the thermopower to a gate change in the vicinity of these extrema is relevant for applications\footnote{Note that, when increasing $\Gamma/T$, the slope at the crossover is expected to be modified by renormalization due to higher-order effects~\cite{Kubala2006May}.}.
We see that in the crossover regime, the negative slope is dominated by the interactions $U \gtrsim T$,
\begin{eqnarray}\label{eq_S_slope}
\left. \frac{dS}{d\epsilon} \right|_{\epsilon - \mueq = - U/2}
\approx
- \frac{1}{T} \left( \frac{U}{2T} -1  \right)
,
\end{eqnarray}
in contrast to the positive slopes $1/T$ of the two branches associated with isolated resonances.
Fig.~\ref{fig_seebeck}(b) shows that Eqs.~(\ref{eq_S_minmax}) and (\ref{eq_S_slope}) indeed approximate very well the features of the Seebeck coefficient for different temperatures.
Measurements of the Seebeck coefficient of a quantum dot have confirmed this behavior, see for example Refs.~\cite{Dzurak1997Apr,Staring2007Jul,Svensson:2012,Svensson:2013}.

%%%%%%%%%%%%%%%%%%%%%%%%%%%%%%%%%%%%%%%%%%%%%%%%%%%%%%%%
\subsection{Peltier coefficient\label{sec_peltier_linear}}
%%%%%%%%%%%%%%%%%%%%%%%%%%%%%%%%%%%%%%%%%%%%%%%%%%%%%%%%

The Peltier coefficient
$\Pi := \partial J/\partial V |_\textrm{eq} \big/ \partial I/\partial V |_\textrm{eq}$
determines the \textit{heat} current generated per transferred particle by a small voltage bias $|V| \ll  T$ alone ($\Delta T=0$).
It thus also defines an energy scale and our heat-current formula \eqref{eq_IEz} gives
\begin{align}
	\Pi = \Eeq\ .
\end{align}
This provides a physically different picture of the energy scale \eq{eq_average_E} as compared to the thermopower: it is the characteristic energy carried by electrons across a biased junction. This is reflected by how it is obtained: it is the prefactor of the first, tight-coupling part in the \emph{heat} current \eq{eq_IEz} [cf. \Eq{eq_seebeck}].

Of course, Onsager reciprocity dictates $L_{12}=L_{21}$, i. e., 
that the \emph{values} of the energy scales obtained from the electro-thermal and thermoelectric response are the same,
where
$L_{21}=T\partial J/\partial V |_\textrm{eq}$
and $\Pi = -L_{21} / L_{11} $.
However, since we explicitly avoid arguing from time-reversal symmetry, it is of interest to see how the reciprocity relation $L_{21} = L_{12}$ emerges when only using our duality \eq{eq_duality}.
The crucial point in our approach is that the fermion-parity part of the energy current \eq{eq_IEz}, carrying \textit{part} of the interaction energy, is to linear  order insensitive to the bias $V$ for any value of the remaining parameters:
\begin{eqnarray}
 	\evalAtEqui{\frac{d}{d\mu_\alpha}\Braket{\zia \fpOp}{z}} = 0 \ .\label{eq_vanishing_FP}
\end{eqnarray}
This follows from the linear-response of the stationary state $\hat{z}$ \BrackEq{eq_linear_response_state} together with the orthogonalities \eq{eq_orthogonality_relations} [see  \App{app:linear_response_peltier_fourier}].
The result \eq{eq_vanishing_FP} thus implies that in linear response to the voltage bias, the heat current coefficient $L_{21}$ stems from the charge mode only, see also Ref.~\cite{Contreras-Pulido2012Feb}. Since this is equally true for the thermopower in $L_{12}$, we indeed reobtain Onsager's reciprocity relation.
We stress in particular that the tight-coupling of the heat- and charge current is not an approximation in our treatment, but follows from duality.

In \Sec{sec_nonlinear} we will see that the parity contribution to heat current \eq{eq_IEz} \emph{does} play a role for finite bias, pinpointing how the interesting deviations between $\Pi$ and $S$ emerge beyond linear response.

%%%%%%%%%%%%%%%%%%%%%%%%%%%%%%%%%%%%%%%%%%%%%%%%%%%%%%%%
\subsection{Thermal response and Fourier heat conductance}
\label{sec_fourier}
%%%%%%%%%%%%%%%%%%%%%%%%%%%%%%%%%%%%%%%%%%%%%%%%%%%%%%%%

%%%%%%%%%%%%%%%%%%%%%%%%%%%%%%%%%%%%%%%%%%%%%%%%%%%%%%%%
\begin{figure}[!t]
	\center
	\includegraphics[width=2.5in]{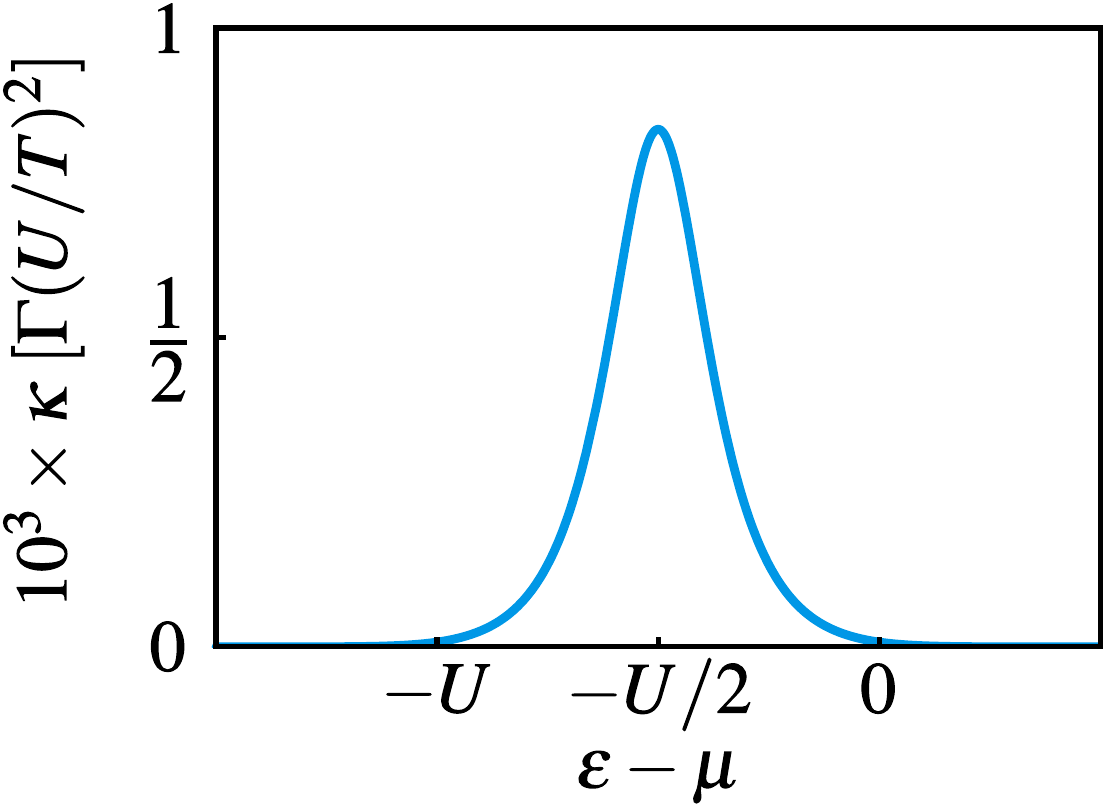}
	\caption{
		 Linear Fourier coefficient $\kappa$, Eq.~(\ref{eq_fourier_coeff}) as a function of $\epsilon - \mueq$. The occurrence of a \emph{single} peak in $\kappa$ at $\epsilon - \mu = -U/2$ directly reflects the fluctuations $\delnsqzieq$ of the \emph{inverted} stationary state at the electron-pair resonance of the attractive dual quantum dot system, see Fig.~\ref{fig_fluctuations} and \Fig{fig_setup}(d). The smallness of the Fourier coefficient on the scale determined by $U/T$ and $\Gamma$ stems from the product of occupation number fluctuations $\delnsqzieq \ \delnsqzeq$.
		 Parameters: $U=10T$, $\GamL=\GamR\ll T$.
		\label{fig_fourier}
			}
\end{figure}
%%%%%%%%%%%%%%%%%%%%%%%%%%%%%%%%%%%%%%%%%%%%%%%%%%%%%%%%

Finally, the thermal coefficient $L_{22}=T^2\partial J / \partial \Delta T |_\textrm{eq}$, given by Eq.~(\ref{eq_K}),
acquires contributions from both
terms in \Eq{eq_IEz}.
This suggests that there are two types of energy fluctuations that correspond to linear thermal transport
when a thermal gradient is applied.
Indeed, the fluctuation of the grand-canonical energy
$\hat{\mathcal{H}} := \hat{H}-\mu \hat{N}$
determining the equilibrium state $\hat{z}_\text{eq} \propto e^{-\hat{\mathcal{H}}/T}$
decomposes into two corresponding parts:
%[\Eq{eq_average_E}, \eq{eq_fluct_nz}, \eq{eq_fluct_ni}]:
\begin{eqnarray}
\delta \hat{\mathcal{H}}_{\textrm{eq}}^2
=\brkt{\hat{\mathcal{H}}^2}_{\text{eq}}  - \brkt{\hat{\mathcal{H}}}_{\text{eq}}^2  = \Braket{\mathcal{H}^2}{\zeq}  - \sqbrack{\Braket{\mathcal{H}}{\zeq}}^2
=\Eeq^2 \delnsqzeq + \left( \frac{U}{2} \right)^2  \delnsqzieq \ \delnsqzeq , \label{eq_E_fluct}
\end{eqnarray}
see \App{app:linear_response_peltier_fourier}.
Thus, the charge (parity) contribution to the linear heat transport coefficient $L_{22}$ \BrackEq{eq_K} are the charge- (parity-) related energy fluctuations $\times$ the charge (parity) rate $\gamma_\text{c,eq}$ ($\gamma_\text{p,eq}$).

In thermoelectric applications, one often wants to know the heat transferred from the hot to the cold reservoir when no electrical power is generated. This purely thermal current
$J|_{I=0} \approx \kappa \Delta T$ for a small thermal bias $\Delta T$  at zero charge current is characterized by the Fourier heat conductance, $\kappa=\partial J|_{I=0}/\partial \Delta T |_\text{eq} $. It is given by $T^2\kappa = L_{22} - L_{12}L_{21}/L_{11}$, for which we obtain
\begin{eqnarray}
\kappa & = & \frac{1}{T^2} \frac{\GamL\GamR}{\Gamma^2} \gampeq\left(\frac{U}{2}\right)^2 \delnsqzeq \ \delnsqzieq  \ \ \label{eq_fourier_coeff}
\end{eqnarray}
using \Eq{eq_linear_response_coefficients}. This coefficient stems entirely from the parity-mode contribution to the heat current.
As in the Drude theory for metals~\cite{Ashcroft1976Jan}, the pure thermal conductance $\kappa$ is due to the electron-electron interaction.\footnote{
	For $U=0$ there are no further contribution mechanisms to thermal transport in the absence of charge transport to the leading order in tunnel coupling $\Gamma$. Higher orders in the tunnel coupling would, however, allow for processes in which energy is transferred in a multi-particle tunneling process, while no net charge transport occurs.}
However, we note that the interaction also nontrivially enters into the other term of the heat current that does not contribute to $\kappa$.
The heat-current decomposition dictated beforehand by duality thus naturally pinpoints \emph{which part} of the interaction enters the \emph{purely thermal} Fourier heat conductance -- namely the fermion-parity part.
We again stress that both a naive perturbative decomposition of the heat current (noninteracting part $+$ interaction corrections)
as well as a mean-field decomposition fail to achieve this.

\begin{figure}[!t]
	\center
	\includegraphics[width=0.7\textwidth]{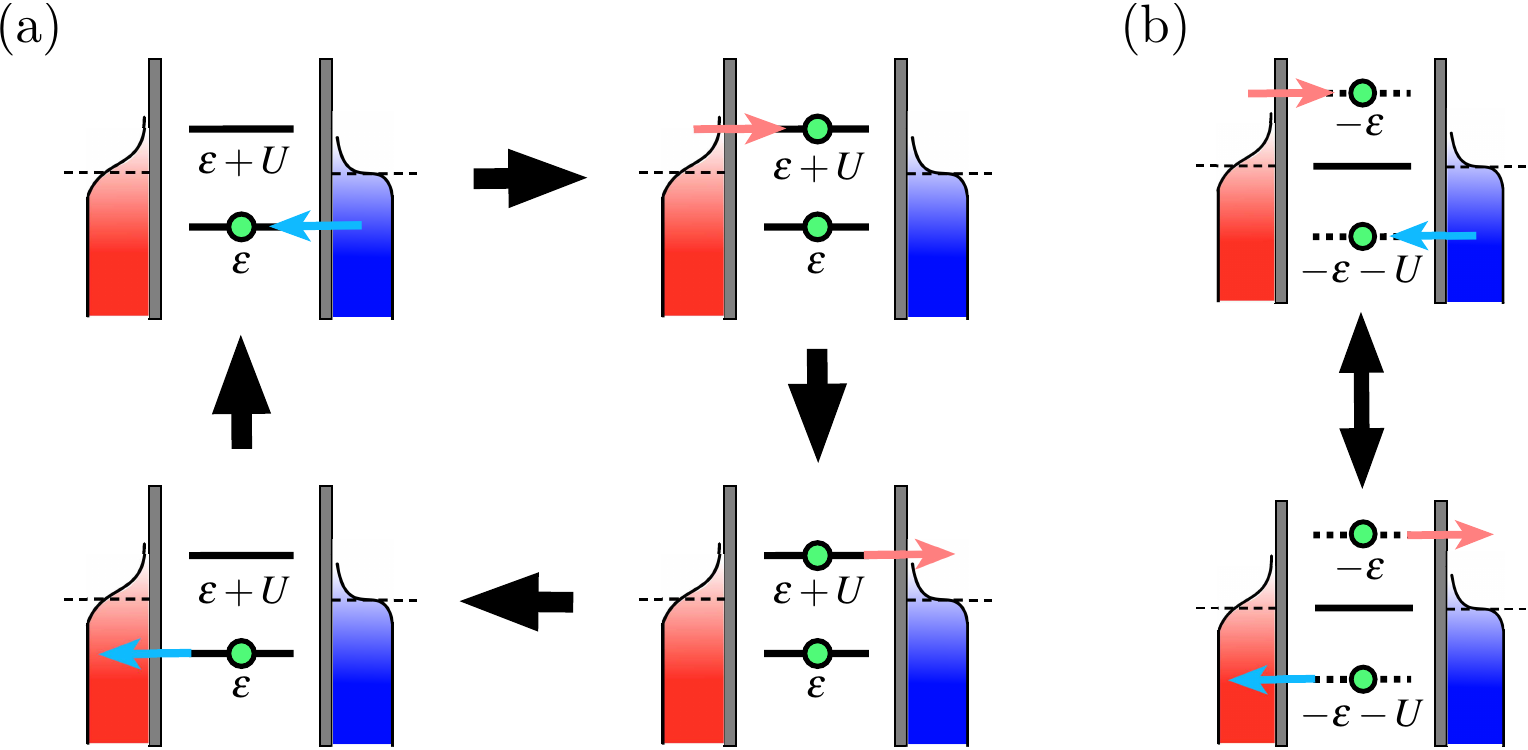}
	\caption{Sketch of the processes leading to Fourier heat transfer in the thermally broadened region around the particle-hole symmetry point, $\epsilon - \mueq = - U/2$, for (a) the normal and (b) the inverted quantum dot models. In both cases shown, the energy $U$ flows from the hot lead to the cold one without net charge transfer from the left to the right lead.
	\label{fig_inverted}
	}
\end{figure}

The duality furthermore clarifies the parameter-dependence of $\kappa$.
Being the product of the charge fluctuations in the stationary \emph{and} in the inverted stationary state, the Fourier coefficient might be expected to reflect the fluctuations $\delnsqzeq$ of the actual system, for example in its $\epsilon$-dependence.
However, these fluctuations that we plotted in \Fig{fig_fluctuations} are unable to account for 
the single resonance that $\kappa$ shows at $\epsilon-\mueq = -U/2$ in \Fig{fig_fourier}.
Instead, this peak entirely comes from the fluctuations $\delnsqzieq$
in the dual model, where due to the attractive interaction only a transition from zero and double dot occupation occurs precisely for this level position, cf. \Fig{fig_inverted}.
Interestingly, the anomalous thermal broadening $T/2$ in the attractive model [\Eq{eq_dual_charge_approx} ff.] is crucial to make these fluctuations dominate:
it leads to a difference in the exponential thermal suppressions of the two fluctuations when $\epsilon$ is varied, as shown in the inset of Fig.~\ref{fig_fluctuations}.
As a result, when multiplying the two curves in \Fig{fig_fluctuations}, the peak (tails) of $\delnsqzieq$ can overcome (suppress) the tails (peaks) of $\delnsqzeq$
and thereby determine the shape of the result for $\kappa$ in \Fig{fig_fourier}.
The very good approximation to the peak height for $U \gg T$,
\begin{align}\label{eq_plateau_linearized}
\kappa \approx 
-
\frac{\GamL \GamR}{\Gamma}
\frac{\partial}{\partial x}
\left.\tanh\nbrack{\frac{1}{x}}\right|_{x=\tfrac{4T}{U}}
= \frac{\GamL \GamR}{\Gamma}\left(\frac{U}{4T}\frac{1}{\text{cosh}\left(\frac{U}{4T}\right)}\right)^2
,
\end{align}
will be derived later on [\Eq{eq_plateau}] by exploiting the duality directly in the nonlinear regime and linearizing afterwards.

Thus, whereas $\delta n_z^2$ determines the electric dissipation of the electric conductance,
the fluctuation of the \emph{inverted} dot population, $\delta n_{\text{i}}^2$ dominates the thermal dissipation described by Fourier heat.
This is the unexpected twist to the formal analogy between Ohm's law and Fourier law that we announced earlier [\Eq{eq_ohm_fourier}]. It appears only when one examines the origin of their coefficients, \Eq{eq_G_trivial} and \eq{eq_fourier_coeff}, taking into account the fundamental restrictions imposed by duality.

%%%%%%%%%%%%%%%%%%%%%%%%%%%%%%%%%%%%%%%%%%%%%%%%%%%%%%%%
%%%%%%%%%%%%%%%%%%%%%%%%%%%%%%%%%%%%%%%%%%%%%%%%%%%%%%%%

Finally, we show how the duality simplifies the Fourier heat transfer even on a pictorial level.
\Fig{fig_inverted}(a) shows how this energy flow can be pictured in the original, repulsive model as a cycle of single-electron tunneling processes,
the cycle being restricted to have zero net charge current, $I = 0$.
Two cycles contribute to the Fourier heat,
in which the energy $\epsilon$ ($\epsilon + U$) is transferred from the cold (hot) lead to the hot (cold) one through the lower (higher) resonance\footnote{corresponding to the addition energy for the transition between zero and single occupation (between single and double occupation)} of the dot.
In either cycle, an amount of energy $U$ is removed from the hot reservoir and released into the cold one. What remains unclear in this picture is how a single sharp peak at $\epsilon -\mueq = - U/2$ with a thermal broadening given by $T/2$ can emerge due to a delicate cancellation between particle and hole processes which are all \emph{off resonant}.
In \Fig{fig_inverted}(b) we show how (in linear response) the same energy transfer can be understood in the inverted dual model in terms of the single resonance available in an attractive quantum dot which can be either empty or doubly occupied.
This process clearly shuts off for $|\epsilon + U/2 -\mueq | \gg T/2$, implying the fluctuations $\delnsqzieq$ vanish and $\kappa$ with it.

%%%%%%%%%%%%%%%%%%%%%%%%%%%%%%%%%%%%%%%%%%
\section{Nonlinear regime}\label{sec_nonlinear}
%%%%%%%%%%%%%%%%%%%%%%%%%%%%%%%%%%%%%%%%%%

In the regime where either of the biases $V$ and $\Delta T$ is large enough to invoke a nonlinear current response, Onsager's relations are of no help anymore. Although nonlinear fluctuation relations may provide interesting insights~\cite{Esposito2009,Lopez2012Jun},
this would require additional machinery of counting statistics.
The fermion-parity duality \Eq{eq_duality}, instead, can be exploited in the nonlinear regime
without further ado: it allows us to 
simplify the derivation and improve our understanding of the \emph{nonlinear} Seebeck, Peltier, and Fourier coefficients obtained from the full non-equilibrium currents \eq{eq_Iz}.
We note that since now heat currents are not anymore conserved due to the finite Joule heating, it is only meaningful to consider lead-resolved heat currents $J^\alpha$ for $\alpha = \text{L}$ or $\text{R}$.

%%%%%%%%%%%%%%%%%%%%%%%%%%%%%%%%%%%%%%%%%%%%%%%%%%%%%%%%
\subsection{Thermo-electric response and Seebeck thermopower}\label{sec_seebeck_nonlinear}
%%%%%%%%%%%%%%%%%%%%%%%%%%%%%%%%%%%%%%%%%%%%%%%%%%%%%%%%

We start by analyzing the nonlinear thermopower / Seebeck coefficient $S_\text{nl} = V|_{I=0} / \Delta T$.
The required thermo-voltage $V|_{I=0}$ is obtained by solving
$I= 0$
for $V$ using the currents \eq{eq_charge_heat}.
The form of the charge current \eq{eq_INz} shows that this is equivalent to maintaining equal lead-resolved and, therefore, equilibrium occupations on the dot:
\begin{align}
	n_{z \textrm{L}} = n_{z \textrm{R}}\ .\label{eq_balanced}
\end{align}
This simplification is ultimately a consequence of the fact that for the spin-degenerate quantum dot model considered here, duality \eq{eq_duality} dictates the charge mode to be an exact eigenmode both in and out of equilibrium, see \eq{eq_table}.
Since we consider $\muR =\mueq$ to be fixed, we can readily solve\footnote{For symmetric bias this is a transcendental equation for $\muL-\muR$ which cannot be solved analytically any longer.} the balance equation \eq{eq_balanced}
for $\muL\equiv\mueq-V$ and obtain [\App{app:cur_bal}] the nonlinear thermopower, using the explicit expressions for $n_{z \textrm{L}}$ and $n_{z \textrm{R}}$ given in Eq.~(\ref{eq_derive_step9}). While the result could be written in terms of Fermi functions directly, a particularly insightful form is
\begin{align}
	S_\text{nl}
=&
\frac{1}{T}
\nbrack{ \epsilon - \mu + U} - \frac{T+\Delta T}{\Delta T}
\ln{\left[
	\frac{
		1 - n_{\text{iR}}
		+
		\sqrt{\nbrack{1 - n_{\text{iR}}}^2
			+
			\expfn{ \frac{U\cdot\Delta T}{T\cdot(T+\Delta T)}} 
		\cdot n_{\text{iR}}(2-n_{\text{iR}})
		}
	}
	{
		2 - n_{\text{iR}}
	}
\right]}\ ,
\label{eq_thermovoltage_nl}
\end{align}
where $n_{\text{iR}}$ is the $\epsilon$-dependent inverted stationary charge number with respect to the reference chemical potential $\mu$ and temperature $T$ in the right lead. 
The representation in terms of the dual occupation number is motivated by its connection to the linear limit, in which $\nzieq = n_{\text{iR}}$ governs the interaction-related contribution to the Seebeck coefficient for a fundamental reason. This allows us, in the following, to exploit the simple single-step behavior and anomalous thermal scale $T/2$ of the dual charge \BrackEq{eq_dual_charge_approx} in order to further analyze $S_\text{nl}$.\par

%%%%%%%%%%%%%%%%%%%%%%%%%%%%%%%%%%%%%%%%%%%%%%%%%%%%%%%%
\begin{figure}[!t]
	\center
	\includegraphics[width=4.5in]{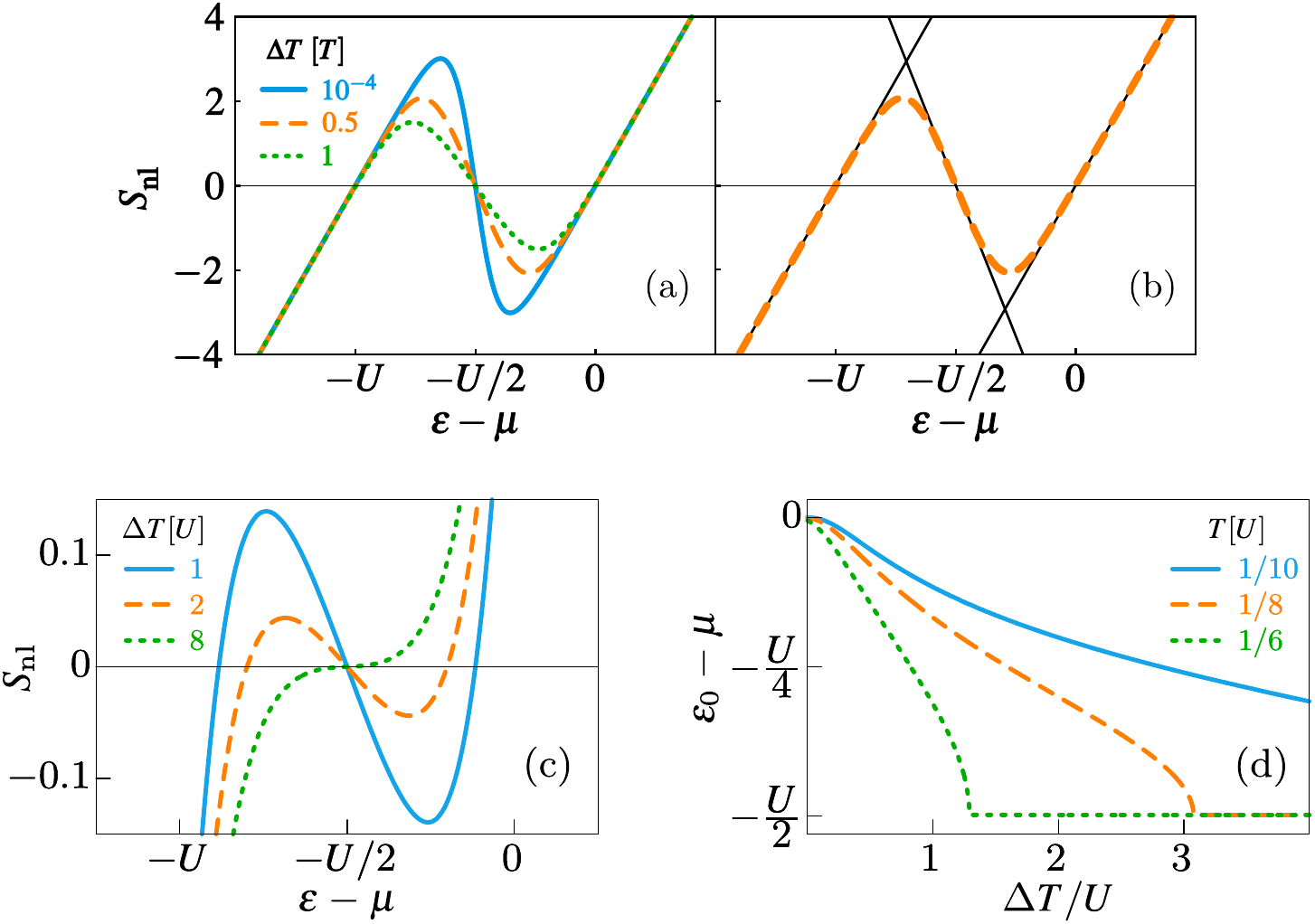}
	\caption{(a) Nonlinear Seebeck coefficient $S_\text{nl}$ as a function of the level position of the dot for different values of the temperature difference $\Delta T$. The solid-blue line reproduces the linear response result shown in Fig.~\ref{fig_seebeck}, to contrast with $\Delta T=0.5T$ (dashed-orange line), and $\Delta T=T$ (dotted-green line). (b) Curve for $\Delta T=0.5T$ (dashed-orange line), together with the approximate slopes, $1/T$ and $-1/T_\text{eff}$ as discussed in the text. (c) $S_\text{nl}$ for large temperature gradients $\Delta T \gg T$. (d) Level position $\epsilon_0 - \mu > -U/2$ at which $S_\text{nl}(\epsilon_0) = 0$ as a function of the temperature gradient $\Delta T/U$ and for different $T/U$. Here, the blue curve represents the situation in (c). Note that we always assume $\GamL=\GamR\ll T$ and, except for (d), $U=10T$.
	\label{fig_seebeck_nonlinear}}
\end{figure}
%%%%%%%%%%%%%%%%%%%%%%%%%%%%%%%%%%%%%%%%%%%%%%%%%%%%%%%%

Let us first consider small to moderate temperature gradients $0 < \Delta T \lesssim T$. Plotting $S_\text{nl}$ 
as a function of $\epsilon - \mueq$ in \Fig{fig_seebeck_nonlinear}, we see in the here studied regime $U \gg T$
that the \textit{nonlinear} thermopower maintains the characteristic $\epsilon$-dependent shape from the linear regime, exhibiting in particular the qualitative signature of $n_{\text{iR}}$.
The only important impact of the increasing thermal bias $\Delta T$ is that the slope of $S_\text{nl}$ at $\epsilon+U/2=\mu$ becomes increasingly less negative,
and correspondingly, local extrema are less pronounced in Fig.~\ref{fig_seebeck_nonlinear}(b).
To analytically understand this behavior, we realize that due to $\Delta T \lesssim T$ and due to the sharp step of $n_{\text{iR}}(\epsilon)$ around $\epsilon - \mu = -U/2$, we can efficiently carry out a linear expansion of \Eq{eq_thermovoltage_nl} in $\epsilon - \mu$ by first expanding $S_\text{nl}$ in $1 - n_{\text{iR}}$ around  $1 - n_{\text{iR}} = -1, 0, 1$ for $\epsilon - \mu$ close to $0,-U/2, -U$ respectively:
\begin{align}
T S_\text{nl} \approx
\begin{cases}
	\displaystyle \epsilon - \mu + \frac{U}{2}\sqbrack{\frac{T(T+\Delta T)}{U\cdot\Delta T}\nbrack{e^{\frac{U\cdot\Delta T}{T(T+\Delta T)}} - 1}(2 - n_{\text{iR}})} \approx \epsilon-\mu
	& ,\quad \epsilon \approx \mu
	\\\vspace{-0.3cm}
	\\
	\displaystyle \epsilon - \mu + \frac{U}{2}\sqbrack{1 + \frac{T(T+\Delta T)}{U\cdot\Delta T/2}\nbrack{1 - e^{-\frac{U\cdot\Delta T/2}{T(T+\Delta T)}}}(1 - n_{\text{iR}})} \approx -\frac{T}{T_{\text{eff}}}(\epsilon-\mu+\frac{U}{2})
	 & ,\quad \epsilon + \frac{U}{2} \approx \mu
	\\\vspace{-0.3cm}
	\\
	\displaystyle \epsilon - \mu + \frac{U}{2}\sqbrack{2 - \frac{T(T+\Delta T)}{U\cdot\Delta T}\nbrack{e^{\frac{U\cdot\Delta T}{T(T+\Delta T)}} - 1}n_{\text{iR}}} \approx \epsilon-\mu + U
	& ,\quad \epsilon + U \approx \mu\ .
	\end{cases}
	\label{eq_thermovoltage_nl_lines}
\end{align}
For $\epsilon - \mu$ close to $0, -U$, the flat $\epsilon/U$-dependence and the anomalously small thermal broadening of $n_{\text{iR}}$ by $T/2$ \BrackEq{eq_dual_charge_approx} lead to a complete suppression of the $\Delta T$-dependent non-equilibrium terms. By contrast, the sizable linear $\epsilon/U$-dependence of $1 - n_{\text{iR}}$ with slope $-U/T$ around $\epsilon - \mu = -U/2$ \BrackEq{eq_S_slope} does introduce a relevant $\Delta T$-dependence of the slope of $S_\text{nl}$ linearized in $\epsilon$, given by the inverse effective temperature
\begin{equation}
 \frac{1}{T_{\text{eff}}} \overset{U \gg T}{\approx} \frac{1}{T}\cdot\sqbrack{\frac{T}{\Delta T} - \frac{T + \Delta T}{\Delta T}\cdot\expfn{-\frac{U}{2T}\cdot\frac{\Delta T}{T + \Delta T}}} \overset{\frac{U\Delta T}{T+\Delta T} \gg T}{\approx} \frac{1}{\Delta T}\label{eq_effective_temp}.
\end{equation}
As expected, taking the limit $\Delta T \rightarrow 0$ in \Eq{eq_thermovoltage_nl_lines} immediately gives back the linear Seebeck coefficient \eq{eq_seebeck}. We also note that a useful demarcation of the regime where the second line in  \Eq{eq_thermovoltage_nl_lines} is a good approximation is obtained by finding the crossing points of the linear $\epsilon$-expansions in all three regimes:
\begin{align}
	|\epsilon +U/2-\mu| &\lesssim \frac{\Delta T}{T+\Delta T}\cdot\sqbrack{1 + \coth\nbrack{\frac{U}{4T}\cdot\frac{\Delta T}{T + \Delta T}}}\cdot \frac{U}{4}\overset{\frac{U\Delta T}{T+\Delta T} \gg T}{\approx} \frac{\Delta T}{T+\Delta T}\cdot\frac{U}{2}.
	\label{eq_regime}
\end{align}
The approximations in Eqs.~\eq{eq_thermovoltage_nl_lines} and \eq{eq_regime} will be helpful for the understanding of the nonlinear Fourier coefficient in Sec.~\ref{sec_fourier_nonlinear}.\par

Let us now address larger temperature gradients $\Delta T \gg T$, approaching and exceeding the interaction strength $U$. An interesting feature of this regime is that the roots of $S_\text{nl}(\epsilon)$ close to the Coulomb resonances are shifted towards the root at $\epsilon - \mu = -U/2$, and that all roots can even merge entirely at this single level-position for large enough $\Delta T$. This both experimentally~\cite{Svensson:2013} and theoretically~\cite{Sierra2014Sep} studied effect is captured by the analytic expression \eq{eq_thermovoltage_nl} of the thermovoltage, as clearly visible in \Fig{fig_seebeck_nonlinear}(c). 
Our duality makes explicit that there is no reason that the thermopower roots at $\epsilon - \mu = 0,-U$ should remain fixed; only the position of resonance at $\epsilon - \mu = -U/2$ is dictated by the dual model.
Indeed, the linear thermopower \eq{eq_seebeck} is not exactly $0$ at $\epsilon - \mu = 0,-U$, although this effect is exponentially suppressed with large interaction $[ 0 < (2-\nzieq)/2 \sim e^{-U/T}$ at $\epsilon = \mu]$.

For large temperature gradients $\Delta T /T \gg 1$, the effect can, however, clearly be seen: figure \ref{fig_seebeck_nonlinear}(d) displays a sizable root shift on the scale $U/2$ as a function of $\Delta T/U$, obtained numerically from \Eq{eq_thermovoltage_nl} for different $T/U$. We also observe that all roots merge at $\epsilon - \mu = -U/2$ at a relatively large temperature gradient, which we define as $\Delta T_0/U$.  
To get an approximate analytical description of this quantity, we start from the condition $d S_{\text{nl}}/d\epsilon = 0$ at $\epsilon - \mu = -U/2$, which sets the point beyond which only one root can exist. 
Since this typically happens at large $\Delta T/T$, we can Taylor expand the derivative in $T/\Delta T$ at fixed $T/U$ to analytically solve for
\begin{equation}
 \frac{\Delta T_0}{U} \approx \frac{1}{2}\sqbrack{\frac{T}{U}\expfn{\frac{U}{2T}} - \frac{T}{U} - \frac{1}{2}}.\label{eq_thermovoltage_nl_crossover}
\end{equation}
Equation \eq{eq_thermovoltage_nl_crossover} is in good agreement with the numerics shown in \Fig{fig_seebeck_nonlinear}(d). Altogether, this confirms that the root shift is observable for achievable temperature gradients as long as the the interaction strength $U$ does not exceed the base temperature $T$ by more than an order of magnitude.

%%%%%%%%%%%%%%%%%%%%%%%%%%%%%%%%%%%%%%%%%%%%%%%%%%%%%%%%
\subsection{Electro-thermal response and Peltier coefficient}\label{sec_peltier_nonlinear}
%%%%%%%%%%%%%%%%%%%%%%%%%%%%%%%%%%%%%%%%%%%%%%%%%%%%%%%%

The nonlinear Peltier coefficient $\Pi^\alpha_\text{nl} = (J^\alpha/I^\alpha)|_{\Delta T = 0}$ determines 
the heat transferred per charge in response to a finite  voltage bias $V$ ($\Delta T = 0$). It therefore quantifies one of the desired thermoelectric effects: the ability of the quantum dot system to electrically cool one of the contacts (via a heat current \textit{out of} this contact).
Due to the absence of heat current conservation under nonequilibrium conditions, the nonlinear Peltier coefficient depends on the lead $\alpha$ that it refers to,
$\Pi^\textrm{L}_\text{nl} \neq \Pi^\textrm{R}_\text{nl}$, and furthermore differs substantially from the Seebeck coefficient, as we now discuss.
 
Since the Peltier coefficient is not restricted by charge-current balancing~\footnote{This represents a contrast to the Seebeck and the Fourier coefficient.}, we can \emph{not} express it entirely in lead-resolved equilibrium variables. Nevertheless, we can exploit our duality to a large extent.
First, we decompose the Peltier coefficient into a part that stems from the heat current contribution that is tightly coupled to the charge current, $\Pi^\alpha_\text{nl,tc}$ [first term in \Eq{eq_IEz}], and the non-tightly coupled contribution, $\Pi^\alpha_\text{nl,ntc}$, stemming from the fermion-parity mode [second term in \Eq{eq_IEz}],
\begin{align}
	\Pi^\alpha_\text{nl}
	& = 
	\Pi^\alpha_\textrm{nl,tc} + \Pi^\alpha_\textrm{nl,ntc}\ .
	\label{eq_nonlinear_peltier}
\end{align}
%%%%%%%%%%%%%%%%%%%%%%%%%%%%%%%%%%%%%%%%%%%%%%%%%%%%%%%%
\begin{figure}[!t]
	\center
		\includegraphics[width=5in]{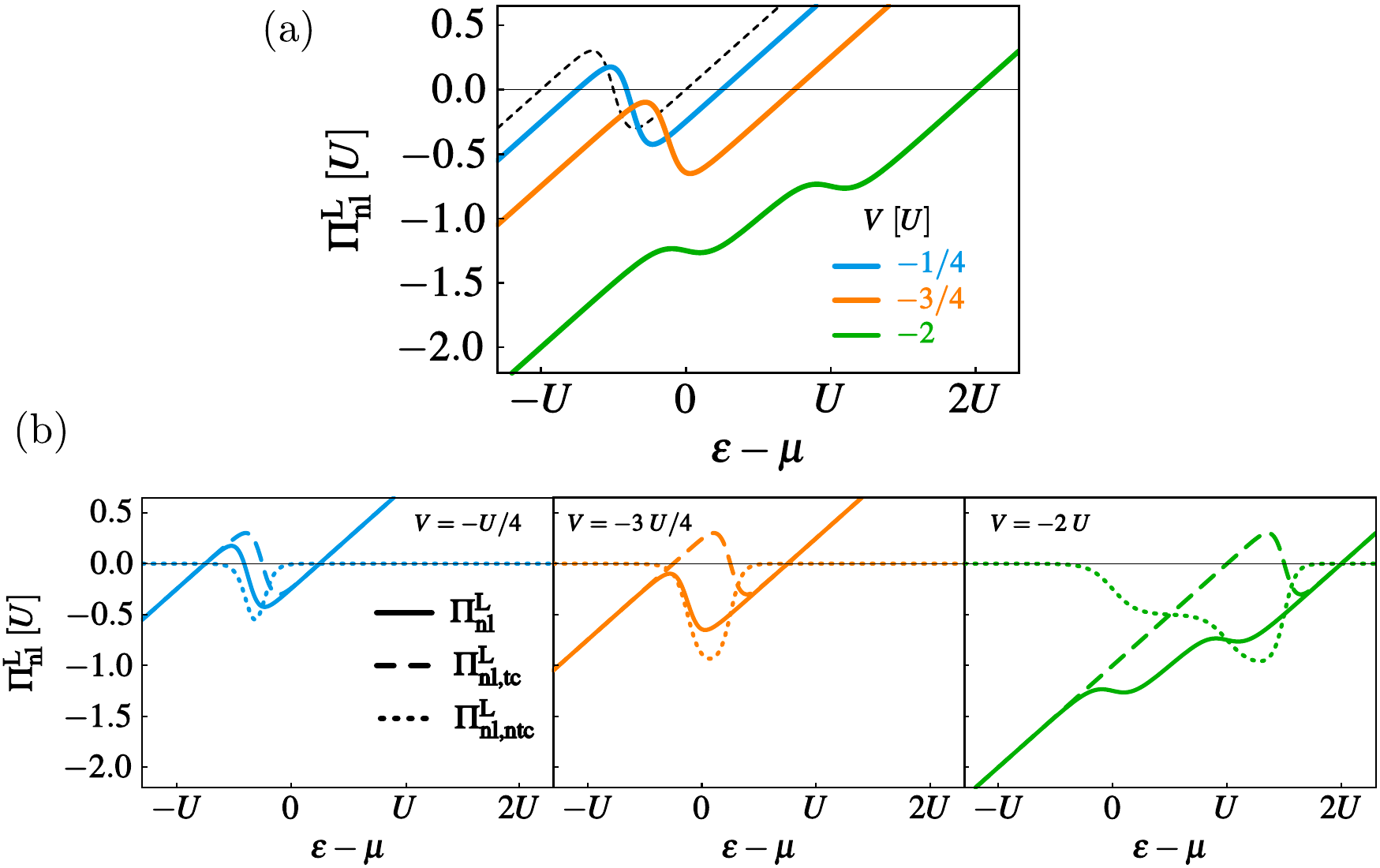}
	\caption{
		(a) Nonlinear Peltier coefficient $\Pi^\textrm{L}_\text{nl}$ as a function of $\epsilon - \mueq$ for different values of the voltage difference $V$. 		
		(b) Decomposition of the nonlinear Peltier coefficient for the parameters chosen for (a) into the tight-coupled $\Pi^\textrm{L}_\textrm{nl,tc}$ (dashed lines) and non-tight-coupled $\Pi^\textrm{L}_\textrm{nl,ntc}$ (dotted lines) contributions.
				 The parameters $U=10~T$ and $\Gamma_\text{L} = \Gamma_\text{R}\ll T$ are fixed for all panels.
		\label{fig_peltier_nl}
	}
\end{figure}
%%%%%%%%%%%%%%%%%%%%%%%%%%%%%%%%%%%%%%%%%%%%%%%%%%%%%%%%
The first term of \Eq{eq_IEz} couples the charge current to  $E_\alpha$, the average energy \eq{eq_average_E}. This is the same energy that determines the \emph{linear} thermopower (Seebeck coefficient), but with respect to $\mu_\alpha$. Therefore,
\begin{align}
\Pi^\alpha_\textrm{nl,tc} = T\, S |_{\mu\rightarrow\mu_\alpha }, 
\end{align}
and $\Pi^\text{L}_\textrm{nl,tc}$ is just the linear thermopower curve shifted by the applied bias $V$, in contrast to $\Pi^\text{R}_\textrm{nl,tc}$, which is not shifted.
This can be seen in \Fig{fig_peltier_nl}(a) and (b), where we plot the Peltier coefficient and its decomposition, the dashed curves showing the tight-coupling part. It means that, interestingly, the nonlinear Peltier coefficient, although \textit{not} related to the nonlinear Seebeck coefficient, is actually related to the \textit{linear} Seebeck coefficient.

The remaining non-tight-coupling contribution $\Pi^\alpha_\textrm{nl,ntc}$ is completely due to the parity mode, 
\begin{align}
	\Pi^\alpha_\textrm{nl,ntc} =-\frac{\gamma_{p\alpha}U}{I^\alpha}\Braket{z_{\text{i}\alpha} \fpOp}{z}|_{\Delta T=0}.
	\label{eq_overlap}
\end{align}
Using Eq.~\eq{eq_overlap_explicit}, this can be expressed in the average occupation and parity in the stationary state $\Ket{z}$ and $\alpha$-resolved inverted stationary state $\Ket{\zia}$. In order to make further quantitative progress, the evaluation of the expectation values $n_z$ and $p_z$ with respect to the stationary state $\Ket{z}$ cannot be avoided (see \App{app:explicit}). However, we stress that Eq.~\eq{eq_overlap} together with Eq.~\eq{eq_overlap_explicit} already present a significant simplification relative to the standard way of writing the same result.

Moreover, we can still further exploit the duality in a qualitative discussion.
For example, the results in \Fig{fig_peltier_nl} show that $\Pi^{\textrm{L}}_\text{nl,ntc}$ contributes only in a limited interval of $\epsilon$ values.
This can be explained using only the observation that this contribution depends on the overlap of the stationary state $\hat{z}$ and the $\alpha$-resolved \emph{inverted equilibrium} state $\hatzia$, as given in Eq.~\eq{eq_overlap}. Intuitively, this overlap is a measure of how much the two mixed states $\hat{z}$ and $\hatzia$ resemble each other, modulo parity signs.
Taking $\mu_\text{L}=\mu-V$, the inverted equilibrium state $\hat{z}_\text{iL}$
switches from an empty state for $\epsilon-\mu\leq-U/2-V$ to a doubly occupied state for $\epsilon-\mu\geq-U/2-V$.
In contrast, the nonequilibrium stationary state $\hat{z}$ has a less simple bias dependence and determines the evolution of the panels of Fig.~\ref{fig_peltier_nl}(b), where $V$ was chosen to be negative, $|V|=-V$:
(Left):
For $|V|\ll U/2$ the overlap \eq{eq_overlap} is nonzero only at $\epsilon+U/2=\mu$ up to thermal broadening.
(Center): For $U/2<|V|<U$, there exists a region $\epsilon-\mu \in [0,|V|-U/2]$, where the state $\hat{z}$ is a mixture of an empty and singly-occupied state
while the inverted state $\hat{z}_\text{iL}$ is empty, leading to a nonzero overlap \eq{eq_overlap}.
(Right) For $|V|>U$  there is a double step.
Here, the inverted state $\hat{z}_\text{iL}$ is empty for all $\epsilon-\mu\leq |V|-U/2$.
The double step in the overlap \eq{eq_overlap} arises due to the state $\hat{z}$ being empty with probability $p_0\approx 1/4$ and $1/3$ for $\epsilon -\mu \in [0,|V|-U]$ and $[|V|-U,|V|-U/2]$, respectively.

The duality thus allows to understand the stepwise contributions of $\Pi^\text{L}_\textrm{nl,ntc}$ and thereby the details of \Fig{fig_peltier_nl}(b):
its effect is to shift the tight-coupling part (determined by the Seebeck coefficient, i.e., the ``Onsager part'') to lower values for moderate bias $|V|<U$, and to double the saw-tooth behavior for high bias $|V| > U$.
Overall, a large bias gives more negative values for $\Pi^{\textrm{L}}_\text{nl}$, restricting the regime of effective cooling of the left reservoir to $\epsilon-\mu>|V|$, for which heat is carried \textit{out of} the left reservoir  (equivalently, the right reservoir for this setting can only be cooled if $\epsilon-\mu<-U$).

%%%%%%%%%%%%%%%%%%%%%%%%%%%%%%%%%%%%%%%%%%%%%%%%%%%%%%%%
\subsection{Thermal response and Fourier coefficient}\label{sec_fourier_nonlinear}
%%%%%%%%%%%%%%%%%%%%%%%%%%%%%%%%%%%%%%%%%%%%%%%%%%%%%%%%

\begin{figure}[!t]
		\center
			\includegraphics[width=0.8\textwidth]{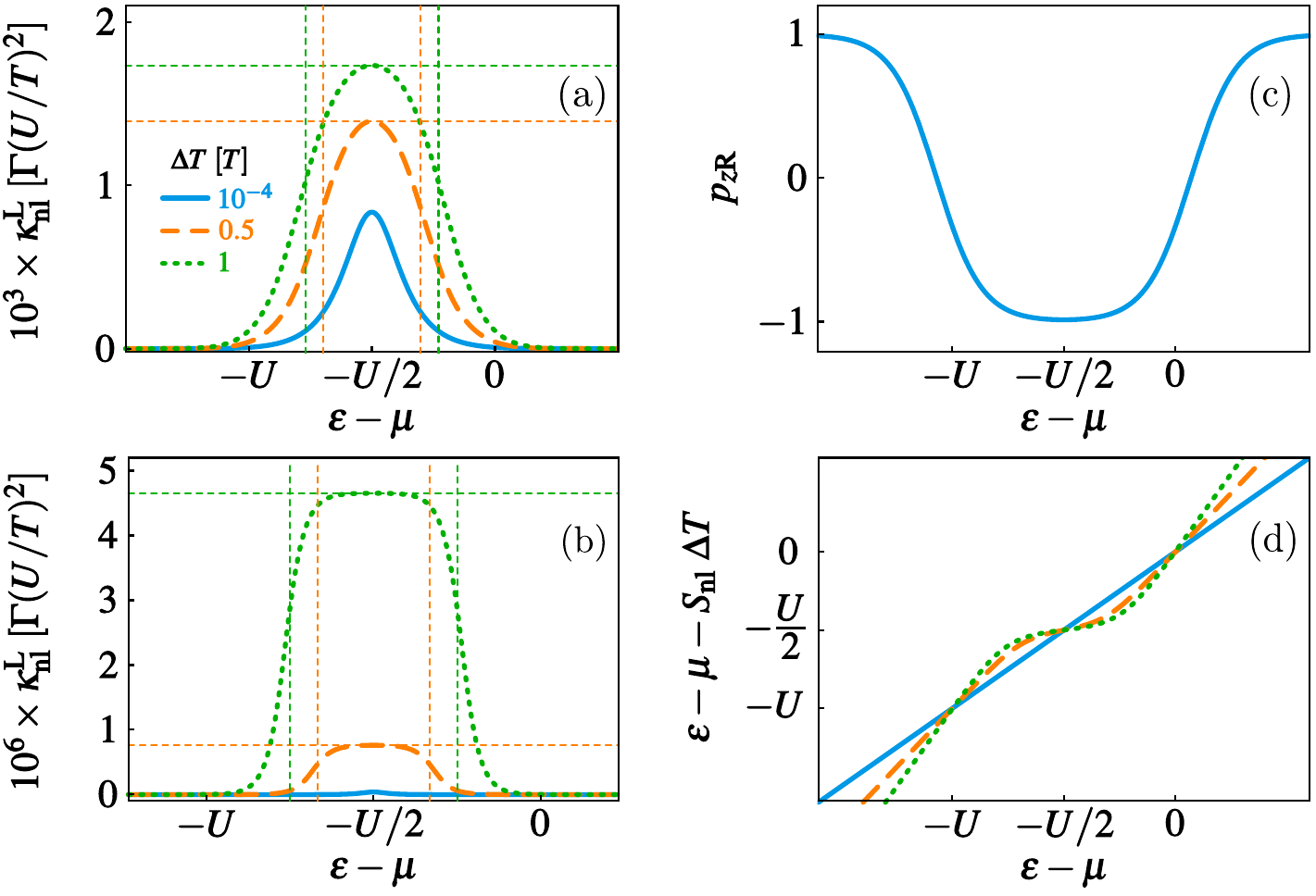}
		\caption{(a) Nonlinear Fourier coefficient $\kappa^\textrm{L}_\text{nl} \equiv - \kappa^\textrm{R}_\text{nl}$ as a function of the level position of the dot with respect to the electrochemical potential of the right lead, for different values of the temperature difference and $U=10T$. The blue lines reproduce the linear response result shown in Fig.~\ref{fig_fourier}. Dashed vertical lines indicate the plateau values, approximated in Eq.~(\ref{eq_plateau}), while fine vertical lines indicate the width \BrackEq{eq_regime} of the broadened peak.
			(b) Same as in (a), but for $U=30T$, where the Fourier coefficient develops a plateau for large values of $\Delta T$.
			(c) Parity $p_{z\text{R}}$, which is independent of $V,\Delta T$. 
			(d) Energy level shifted by the thermovoltage, $\epsilon-\mu-S_{\text{nl}}\Delta T$ as function of $\epsilon-\mu$ for the temperature differences of (a).
					The parameters $U=10T$, $\GamL=\GamR\ll T$ are fixed in all panels unless indicated otherwise.
			\label{fig_fourier_nl}
		}
\end{figure}

Finally, we examine the nonlinear Fourier heat coefficient,
$\kappa^\alpha_\text{nl} := J^\alpha|_{I = 0} / \Delta T$,
for lead $\alpha = \text{L,R}$,
the ratio of the heat current and the finite temperature bias
in the absence of a charge current.
By energy conservation\footnote{
	Since the Fourier heat is independent of the Joule heating, we have $\kappa^\text{L}_\text{nl}\equiv-\kappa^\text{R}_\text{nl}$ even though in the nonlinear regime heat currents are in general not conserved.},
we have $\kappa^\text{L}_\text{nl}\equiv-\kappa^\text{R}_\text{nl}$.
Our heat current formula, Eq.~\eq{eq_IEz}, immediately shows that the nonlinear Fourier heat current is produced solely by the parity mode contribution:
\begin{align}\label{eq_fourier_overlap}
	\kappa^{\textrm{L}}_\text{nl}
	& = 
	 - \gamma_{p \text{L}} \frac{U}{\Delta T} \ \Braket{z_{i \text{L}} \fpOp}{z} \Big|_{I = 0}\ .
\end{align}
In contrast to the expression for the non-tight-coupling contribution to the Peltier coefficient, Eq.~\eq{eq_overlap}, the above expression \eq{eq_fourier_overlap} can be fully analyzed in terms of equilibrium quantities, dictated by the duality. 
We therefore make use of the remarkable fact that under current balanced conditions \eq{eq_balanced}, the full non-equilibrium stationary state $\Ket{z}$ simplifies to a sum of the lead-resolved equilibrium states $\Ket{z_\alpha}$:
\begin{equation}
\left.\Ket{z}\right|_{I = 0} =\left.\sum_\alpha\frac{\Gamma_\alpha}{\Gamma}\Ket{z_\alpha}\right|_{I = 0} 
\label{eq_bal_state} \ .
\end{equation}
This is shown \App{app:cur_bal}, in fact for any number of leads $\alpha$.
With this, we find
\begin{align}
\kappa^{\textrm{L}}_\text{nl}
&	 =
	 \frac{1}{4}
	  \frac{\GamL \GamR}{\Gamma}
	  \frac{U}{\Delta T}
	  \left. \left(   p_{z \text{L}} - p_{z \text{R}} \right) \right|_{I = 0}
	     \ . \label{eq_fourier_nonlineara}
	    \end{align} 
Remarkably, it shows that the nonlinear Fourier heat coefficient can be rationalized in terms of two relatively simple equilibrium observables of the quantum dot, the lead-resolved parities $p_{z \text{L}}$ and $p_{z \text{R}}$.
As previously pointed out, the parity $p_{z \text{R}}$ with respect to the right lead, plotted Fig.~\ref{fig_fourier_nl}(c), simply equals the equilibrium parity $p_{z, \text{eq}}$, since $\muR = \mu$ and $T_\text{R} = T$ are kept fixed in the right lead. This equilibrium parity simply changes sign at both resonances, $\epsilon - \mu = 0,-U$.
In contrast, the required value $p_{z \text{L}}|_{I = 0}$ is obtained by evaluating the explicit expression \eq{eq_parity_expectation} of $p_{z,\text{eq}} = p_{z\text{R}}$ at temperature $T+\Delta T$, and at a chemical potential shifted compared to $\muR = \mu$ by the nonlinear thermovoltage $V|_{I=0}= S_\text{nl}\Delta T$,
which depends nontrivially on the parameters [\Eq{eq_thermovoltage_nl}]: 
\begin{align}
p_{z\text{L}} & = p_{z,\text{eq}}\sqbrack{\mu \rightarrow \mu - S_\text{nl}(\epsilon-\mu,U,T)\Delta T\,\,\, ,\,\,\, T \rightarrow T+\Delta T}.
	   \label{eq_parity_nonlinear}
\end{align}
In the following, more detailed analysis of \Eq{eq_fourier_nonlineara}, we can exploit \eq{eq_parity_nonlinear} in the regime of moderate temperature gradients $0 < \Delta T < T$.
Namely, an inspection of Figs.~\ref{fig_fourier_nl}(c) and (d) shows that the parity with respect to the left lead, Eq.~\eq{eq_parity_nonlinear}, is expected to take a constant value, equal to the value of $p_{z\text{R}} = \pzeq$ at the electron-hole symmetric point but at a different temperature $T+\Delta T$, in a $\Delta T$-dependent range around $\epsilon-\mu=-U/2$.
%%%%%%%%%%%%%%%%%%%%%%%%%%%%%%%%%%%%%%%%%%%%%%%%%%%%%%%
%%%%%%%%%%%%%%%%%%%%%%%%%%%%%%%%%%%%%%%%%%%%%%%%%%%%%%%%
And indeed, explicitly evaluating the thermovoltage \eq{eq_thermovoltage_nl} in the piecewise-linear approximation \eq{eq_thermovoltage_nl_lines} and for $U\Delta T/(T(T+\Delta T)) \gg 1$, we find
\begin{align}
p_{z\text{L}}
\approx
\begin{cases}
p_{z\text{R}}
&,\quad \epsilon \approx \mu
\\
p_{z,\text{eq}}\left(\epsilon - \mu = -U/2,U,T+\Delta T\right)
&,\quad \epsilon+U/2 \approx \mu
\\  
p_{z\text{R}}
&,\quad \epsilon + U \approx \mu\ .
\end{cases}
\label{eq_parity_nl_lines}
\end{align}
With the help of equations \eq{eq_fourier_nonlineara} to \eq{eq_parity_nl_lines}, we can now fully understand the plots of $\kappa^{\textrm{L}}_\text{nl}$ as a function of the dot level $\epsilon$ as shown in \Fig{fig_fourier_nl}(a) and (b) for different values of the Coulomb interaction. For increasing thermal bias $\Delta T$, the linear response peak at $\epsilon - \mueq = - U / 2$ [\Fig{fig_fourier}] increases in both height and width, depending on the ratio $U/T$. Namely, the
broadened peak around $\epsilon-\mu=-U/2$ for $U/T=10$ clearly assumes a plateau shape at large\footnote{With decreasing temperature relative to $U$, it is interesting how cotunneling corrections modify the plateau.} values of $U/T$, such as $U/T=30$ in Fig.~\ref{fig_fourier_nl}(b).
Now, the approximate expressions for the Fourier coefficient in Eq.~(\ref{eq_parity_nl_lines}) tell us that there is a difference between the parities $p_{z\text{L}}$ and $p_{z\text{R}}$ only  between the two resonances in the regime delimited by condition \eq{eq_regime}. The width of this regime, $U\frac{\Delta T}{T+\Delta T}$,  
first increases linearly with $\Delta T \ll T$ and starts to become of order $U$ once $\Delta T\gtrsim T$, as indicated by thin, dashed vertical lines in panels (a) and (b).
The difference in parities assumes a constant value in the regime given by $U\Delta T/(T(T+\Delta T)) \gg 1$, for which the second line in \Eq{eq_parity_nl_lines} is valid. As visible in panel (b) of Fig.~\ref{fig_fourier_nl}, this results in the plateau at the maximum value  
\begin{equation}\label{eq_plateau}
\kappa^{\textrm{L}}_\text{nl}
= -
	  \frac{\GamL \GamR}{\Gamma}
	  \frac{U}{4 \Delta T}
	  \left[\tanh \frac{U}{4(T+\Delta T)}-\tanh \frac{U}{4T} \right]
	  \qquad \text{for $|\epsilon + \tfrac{1}{2}U -\mu | \lesssim \frac{U}{2} \frac{\Delta T}{T+\Delta T}$},
\end{equation}
increasing with $\Delta T$ as indicated by the different dashed horizontal lines in panel (a) and (b) of Fig.~\ref{fig_fourier_nl}.
Taking the limit $\Delta T\to 0$, we obtain the linear response formula \Eq{eq_plateau_linearized} reported in \Sec{sec_fourier}.

%%%%%%%%%%%%%%%%%%%%%%%%%%%%%%%%%%%%%%%%%%
\section{Conclusion and outlook}\label{sec_conclusion}
%%%%%%%%%%%%%%%%%%%%%%%%%%%%%%%%%%%%%%%%%%

We have shown that besides time-reversal symmetry, our duality involving 
fermion-parity superselection is a crucial general principle for understanding both the linear and nonlinear thermoelectric transport through strongly interacting electronic nanoscale systems.
If one is unaware of this, many computed results seem to reflect model-specific features
whereas in reality they are fixed from the beginning by the physical restrictions imposed by duality.

Our study illustrates that thermoelectrics is not only interesting for future applications, but also provides new ways in which fundamental effects can be studied: thermal transport properties are particularly susceptible to effects tied to the new duality, since the Coulomb interaction is an additional channel for storing and transporting energy. This naturally brings the fermion-parity operator into the problem, which is an open-system evolution mode that is fundamentally ``protected''  by the duality.

Concretely, the duality explains why the thermoelectric response of a strongly repulsive system shows features characteristic for attractive interaction. This is particularly visible in the characteristic energy scale of stationary linear transport of heat and charge, which is governed by a resonance occurring at an anomalous energy with unexpected thermal broadening. Both were shown to derive from the occupation number of a quantum dot with inverted Coulomb interaction. In addition, the fluctuations of this dual occupation number were shown to play a similarly important role for the linear heat conductance as the usual occupation fluctuations have for the charge conductance.

The technical advantage offered by the new duality relation is most obvious in the more complicated nonlinear regime. We provided very compact analytical formulas for all transport coefficients, namely the nonlinear thermopower, Peltier coefficient and Fourier heat, allowing for intuitive predictions and analyses of their characteristic features. Strikingly, in most cases, the duality allowed us to express the nonlinear thermoelectric properties in terms of \textit{equilibrium} variables.

%%%% Outlook %%%
The definite advantages for the analysis and understanding of the thermoelectric response of an interacting single-level quantum dot are expected to extend also to more complex systems with, e.g., multiple levels and contacts.
Indeed, the duality applies quite generally beyond the limitations~\cite{Schulenborg16} of the present paper, also to low-temperature, strongly-coupled quantum dot systems, which are of current interest~\cite{Dorda2016Dec,Sierra2017Aug,Karki2017Sep} as there are still many interesting thermoelectric properties yet to be explored.

%%%%%%%%%%%%%%%%%%%%%%%%%%%%%%%%%%%%%%%%%%
\acknowledgments{We thank Rafael S\'anchez for useful comments on the manuscript. We acknowledge funding from the Knut and Alice Wallenberg foundation through their Academy Fellows program (J. Sp. and A. dM.), from the Swedish VR (J. Sp. and J. Sc.), from the Erasmus Mundus program (J. V.), and from the DFG project SCHO 641/7-1 (M. R. W.).}
%%%%%%%%%%%%%%%%%%%%%%%%%%%%%%%%%%%%%%%%%%

%%%%%%%%%%%%%%%%%%%%%%%%%%%%%%%%%%%%%%%%%%
\authorcontributions{%For research articles with several authors, a short paragraph specifying their individual contributions must be provided. The following statements should be used ``X.X. and Y.Y. conceived and designed the experiments; X.X. performed the experiments; X.X. and Y.Y. analyzed the data; W.W. contributed reagents/materials/analysis tools; Y.Y. wrote the paper.'' Authorship must be limited to those who have contributed substantially to the work reported.
J. Sp. conceived the idea.
J. Sc., A. dM., J. Sp., and J. V. performed the calculations. All authors analyzed the results and wrote the paper.
}
%%%%%%%%%%%%%%%%%%%%%%%%%%%%%%%%%%%%%%%%%%

%%%%%%%%%%%%%%%%%%%%%%%%%%%%%%%%%%%%%%%%%%
\conflictsofinterest{The authors declare no interest in conflicts.} 
%%%%%%%%%%%%%%%%%%%%%%%%%%%%%%%%%%%%%%%%%%

%%%%%%%%%%%%%%%%%%%%%%%%%%%%%%%%%%%%%%%%%%
\appendixtitles{yes} %Leave argument "no" if all appendix headings stay EMPTY (then no dot is printed after "Appendix A"). If the appendix sections contain a heading then change the argument to "yes".
\appendixsections{multiple} %Leave argument "multiple" if there are multiple sections. Then a counter is printed ("Appendix A"). If there is only one appendix section then change the argument to "one" and no counter is printed ("Appendix").
\appendix
%%%%%%%%%%%%%%%%%%%%%%%%%%%%%%%%%%%%%%%%%%

%%%%%%%%%%%%%%%%%%%%%%%%%%%%%%%%%%%%%%%%%%
\section{Non-equilibrium master equation kernel from physical principles and symmetries}\label{app:kernel}
%%%%%%%%%%%%%%%%%%%%%%%%%%%%%%%%%%%%%%%%%%

In this appendix, we derive the non-equilibrium master equation kernel by appealing only to its physicality (dissipativity), the validity of the fermion-parity duality, and the fact that each lead individually is in a local equilibrium state. Moreover, we use that we only consider the first order in the tunnel coupling. This means that only sequential tunneling processes are possible and, due to the full spin symmetry, that the coherent dynamics decouple from the time evolution of the energy eigenstate probabilities. In the following, we can thus restrict the treatment to these probabilities only. 

%%%%%%%%%%%%%%%%%%%%%%%%%
\subsection{Kernel for a single lead}
\label{app:kernel_lead}
%%%%%%%%%%%%%%%%%%%%%%%%%

Before we turn to the full, non-equilibrium kernel $W$, we start by arguing only for any individual kernel $W_\alpha$ in the reservoir sum
\begin{equation}
W = \sum_\alpha W_\alpha \ .  \label{eq_derive_step1}
\end{equation}
Note that this decomposition can be obtained in the sequential tunneling approximation, in which simultaneous transitions from two reservoirs are neglected. In particular, any $W_\alpha$ contains only the dot variables and quantities with respect to the lead $\alpha$, meaning $\mu_\alpha,\beta_\alpha$ and $\Gamma_\alpha$.\par

Now, by construction, each kernel $W_\alpha$ must be linear in $\Gamma_\alpha$. This means that if we turn off the couplings $\Gamma_{\alpha'}$ to all other reservoirs $\alpha'\neq\alpha$, the complete dynamics of the dot are governed exclusively by $W_\alpha$. Each $W_\alpha$ is a physically valid, probability conserving and dissipative kernel in its own right. This implies
\begin{equation}
\Bra{\one}W_\alpha = 0 \lrsepa \Bra{x}\frac{W_\alpha + (W_\alpha)^\dagger}{2}\Ket{x} \leq 0 \quad\text{for any } \Ket{x},\label{eq_dissipative} 
\end{equation}
and furthermore dictates real non-negative transition rates\footnote{Otherwise, it is always possible to find an initial state $\Ket{\rho_0}$ for which the master equation predicts energy eigenstate projections $\Braket{i}{\rho(t)} < 0$ or even $\Im\sqbrack{\Braket{i}{\rho(t)}} \neq 0$ at some time $t$, which obviously forbids a probability interpretation.}, $W_\alpha (f\leftarrow i) /\Braket{f}{f}= \Bra{f}W_\alpha\Ket{i}/\Braket{f}{f} \geq 0$, for transitions from an initial dot energy eigenstate $\Ket{i}$ to a \emph{different} final dot state $\Ket{f}$. 
Additionally, for reasons that become clear in the next paragraph, we make the -- not explicitly required yet plausible\footnote{A hypothetical way to violate this condition in the wide-band description of the single-level dot would be to take the zero temperature limit in the Coulomb blockade regime, leading to completely blocked transitions out of the singly occupied state. This is, however, neither physical nor practical, since the zero temperature limit cannot be appropriately described by the Markovian sequential tunneling approximation, and since it is of course impossible to realize $T_\alpha = 0$ \emph{exactly} in practice.} -- assumption that a sequence of single electron tunneling events to and from lead $\alpha$ eventually connects any possible initial dot energy eigenstate $\Ket{i}$ to any other final dot state $\Ket{f}\neq\Ket{i}$:
\begin{equation}
\Exists n\in\mathds{N} \quad:\quad \Bra{f}(W_\alpha)^n\Ket{i} \neq 0 \quad\text{for any } \Ket{i},\Ket{f} \ . \label{eq_connectiveness}
\end{equation}
Note now that the first statement in \Eq{eq_dissipative} means that $\Bra{\one}$ is a nontrivial left eigenvector of $W_\alpha$ to the eigenvalue $0$, $\Bra{\one}W_\alpha = 0\cdot\Bra{\one}$. Importantly, the property \eq{eq_connectiveness} then ensures~\cite{Schulenborg16} that this eigenvalue $0$ is \textit{non-degenerate}, and that the only trace-normalized right zero-eigenvector $\Ket{z_\alpha} \neq 0$ is the unique, stationary mixed state of the dot, with truly positive probabilities $0 < P_\chi < 1$ for all dot energy eigenstates $\Ket{\chi}$. Since the lead itself is assumed to be in equilibrium, and since the grand-canonical ensemble $\sim e^{-\beta_\alpha(\hat{\Hdot} - \mu_\alpha \hat{N})}$ is a possible stationary state of the dot if the latter is in equilibrium with the lead in terms of particle and energy exchange, the unique stationary state $\Ket{z_\alpha}$ must be the grand-canonical ensemble:
\begin{align}
\Ket{z_\alpha} &= \frac{\expfn{-\beta_\alpha\nbrack{\hat{\Hdot} - \mu_\alpha \hat{N}}}}{\Tr\sqbrack{\expfn{-\beta_\alpha\nbrack{\hat{\Hdot} - \mu_\alpha \hat{N}}}}}\notag\\
&= \frac{1}{Z(\epsilon,U,\mu_\alpha,\beta_\alpha)}\Ket{0}  + \frac{2e^{-\beta_\alpha(\epsilon - \mu_\alpha)}}{Z(\epsilon,U,\mu_\alpha,\beta_\alpha)}\Ket{1} + \frac{2e^{-\beta_\alpha(2\epsilon + U - 2\mu_\alpha)}}{Z(\epsilon,U,\mu_\alpha,\beta_\alpha)}\Ket{2}
\label{eq_derive_step3}
\end{align}
with the partition function $Z(\epsilon,U,\mu_\alpha,\beta_\alpha) = 1 + 2e^{-\beta_\alpha(\epsilon - \mu_\alpha)} +  e^{-\beta_\alpha(2\epsilon + U - 2\mu_\alpha)}$. We have used that, due to the spin-degeneracy, the occupation number states and energy eigenstates \eq{eq_states} of $\hat{\Hdot}$, as well as their dual counterparts, form an orthogonal\footnote{but not orthonormal, since $\Braket{1}{1} = 1/2$!} set of Liouville space vectors spanning the relevant Liouville space:
\begin{equation}
 \mathcal{I} = \Ket{0}\Bra{0} + 2\Ket{1}\Bra{1} + \Ket{2}\Bra{2}.\label{eq_liouville_identity_occuapation}
\end{equation}

In the wide-band limit, $W_\alpha$ belongs to the class for which the fermion-parity duality \eq{eq_duality} holds:
\begin{equation}
\sqbrack{W_{\alpha}(\hat{\Hdot},\mu_\alpha)}^\dagger = -\Gamma_\alpha -\hat{\P} W_\alpha(-\hat{\Hdot},-\mu_\alpha)\hat{\P} \lrsepa \hat{\P}\Ket{\bullet} = \Ket{\fpOp \bullet}\label{eq_derive_step4}
\end{equation}
As in \Ref{Schulenborg16}, we assume that neither the physicality \eq{eq_dissipative} nor the connectiveness of all states \eq{eq_connectiveness} are spoiled by the energy inversion in the dual kernel $W_\alpha(-\hat{\Hdot}, -\mu_\alpha)$. In the wide-band limit\footnote{For energy-dependent bare couplings $\Gamma(E)$, this situation would change if $\Gamma(E)$ had roots on the real axis!}, this is reasonable because both local energies and electrochemical potential are inverted, and because only energy \emph{differences} to the electrochemical potential matter for the tunneling rates!
The important consequence of this assumption is that the parity rate
$-\Gamma_\alpha$ is a non-degenerate eigenvalue
of the kernel $W_\alpha$  with the left and right eigenvectors $\Bra{\zia\fpOp}$ and $\Ket{\fpOp}$, as reported in Eq.~(\ref{eq_table}).
Here, we have introduced the inverted or \emph{dual} stationary state $\Ket{\zia} = \Ket{z(-H,-\mu_\alpha)}$ as defined in \Eq{eq_def_zi} as the stationary state of the dual kernel; it derives from the stationary state $\Ket{z_\alpha}$ for lead $\alpha$ by inverting the sign of all local energy scales and of the electrochemical potential. Since the stationary state \eq{eq_derive_step3} is the grand-canonical ensemble, this inverted stationary state can also explicitly be written as
\begin{align}
\Ket{\zia} 
&\equalbyeqn{eq_derive_step3} \frac{1}{Z(-\epsilon,-U,-\mu_\alpha,\beta_\alpha)}\Ket{0}  + \frac{2e^{\beta_\alpha(\epsilon - \mu_\alpha)}}{Z(-\epsilon,-U,-\mu_\alpha,\beta_\alpha)}\Ket{1} + \frac{2e^{\beta_\alpha(2\epsilon + U - 2\mu_\alpha)}}{Z(-\epsilon,-U,-\mu_\alpha,\beta_\alpha)}\Ket{2}.\label{eq_derive_step6}
\end{align}
By now we have already found two left and right eigenvectors of $W_\alpha$ that must be biorthogonal to each other. Since the  relevant Liouville subspace and its dual are spanned only by the 3 spin-symmetric physical states written in \eq{eq_states}, we span the entire relevant Liouville [dual] space by only one other spin-symmetric [left] right vector [$\Bra{c'_\alpha}$] $\Ket{c_\alpha}$ that is orthogonal to the previously found two [right] left eigenvectors. Up to normalization factors, both $\Ket{c_\alpha}$ and $\Bra{c'_\alpha}$ are hence uniquely determined by the Gram-Schmidt algorithm.
To efficiently carry out this orthogonalization procedure, we use the fact that also $\Ket{\one},\Ket{N-\one}$, and $\Ket{\fpOp}$ form an orthogonal basis of the Liouville space of interest:
\begin{equation}
 \mathcal{I} = \frac{1}{4}\Ket{\one}\Bra{\one} + \frac{1}{2}\Ket{N - \one}\Bra{N - \one} + \frac{1}{4}\Ket{\fpOp}\Bra{\fpOp}.\label{eq_liouville_identity_alt}
\end{equation}
This shows that in order to be orthogonal to the parity mode $\Ket{\fpOp}$, $\Bra{c'_\alpha}$ must be a linear combination of $\Bra{N}$ and $\Bra{\one}$. Enforcing also orthogonality to the stationary state $\Ket{z_\alpha}$, applying an analogous procedure for the right vector $\Ket{c_\alpha}$, and finally choosing an appropriate normalization factor, such that $\Braket{c'_\alpha}{c_\alpha} = 1$, we obtain
\begin{equation}
\Bra{c'_\alpha} = \Bra{N} - \nza\Bra{\one} \lrsepa \Ket{c_\alpha} = \frac{1}{2}\fpOpHat\sqbrack{\Ket{N} - \nzia\Ket{\one}}.\label{eq_derive_step8}
\end{equation}
Here, $\nza = \Braket{N}{z_\alpha}$ is the average charge number in the stationary state, and $\nzia = \Braket{N}{\zia}$ the corresponding average in the \emph{dual} stationary state:
\begin{align}
\nza &= \Braket{N}{z_\alpha} \equalbyeqn{eq_states} 2\sqbrack{\Braket{1}{z_\alpha} + \Braket{2}{z_\alpha}} \equalbyeqn{eq_derive_step3} \frac{2f_\alpha(\epsilon)}{1+f_\alpha(\epsilon) - f_\alpha(\epsilon+U)} \notag\\
\nzia &= \Braket{N}{\zia}  \equalbyeqn{eq_states} 2\sqbrack{\Braket{1}{\zia} + \Braket{2}{\zia}} \equalbyeqn{eq_derive_step6} \frac{2(1-f_\alpha(\epsilon))}{1-f_\alpha(\epsilon) + f_\alpha(\epsilon+U)},\label{eq_derive_step9}
\end{align}
where $f_{\alpha}(x) = \sqbrack{\expfn{\beta_\alpha\sqbrack{x - \mu_\alpha}} + 1}^{-1}$ is the Fermi function. The left vector is thus interpreted as the charge deviation from the stationary average, and the right vector is called charge \emph{mode}.
Since the zero eigenvalue
and the parity rate are by construction non-degenerate\footnote{
	As noted before, $0$ is a non-degenerate eigenvalue of $W_\alpha$. Hence, the vectors \eq{eq_derive_step8} can only be eigenvectors or generalized eigenvectors in a Jordan block of $W_\alpha$ to the eigenvalue $\Gamma_\alpha$, or true eigenvectors to an eigenvalue that differs from $0,-\Gamma_\alpha$. Remembering moreover that we have assumed full state connectiveness \BrackEq{eq_connectiveness} and a non-degenerate eigenvalue $0$ for the \emph{dual} kernel $W_\alpha(-\hat{\Hdot}, -\mu_\alpha)$, the duality \eq{eq_derive_step4} dictates $-\Gamma_\alpha$ to also be a non-degenerate eigenvalue of $W_\alpha$. Thus, using finally that $W_\alpha$ is real, we can conclude that $\Bra{c'_\alpha}$ and $\Ket{c_\alpha}$ must be amplitudes and modes of $W_\alpha$ to an eigenvalue $-\gamca$ obeying $0 > -\gamca > -\Gamma_\alpha$.},
both vectors are eigenvectors of $W_\alpha$ that can be associated with
the decay of the average charge number to its stationary value $\nza$. The timescale of this decay -- the charge rate -- is set by the eigenvalue $-\gamca$ to be determined.
 Collecting all previous results, we arrive at the set of left and right eigenvectors that diagonalizes the kernel, and that is stated in \Eq{eq_table}:
\begin{equation}
\begin{array}{c|c|c}
          \text{Amplitude}                &                            \text{$-$ Eigenvalue $=$ decay rate}                             &                             \text{Mode}                                \\\hline\hline
         \Bra{z'_\alpha} = \Bra{\one}          &                            \gamma_{z \alpha}                             &                            \Ket{z_\alpha}                             \\\hline
 \Bra{{c_\alpha}'} = \Bra{N} - \nza \Bra{\one} & \gamca & \Ket{c_\alpha} = \frac12 \fpOpHat \Big[\Ket{N}-\nzia \Ket{\one}\Big]           \\\hline
     \Bra{p'_\alpha} = \Bra{\zia \fpOp}       &                          \gampa                           &                     \Ket{p_\alpha} = \Ket{\fpOp}                          \\\hline
\end{array}
\label{eq_kernel_singlelead_eigenvectors}
\end{equation}
with eigenvalues $- \gamma_{z \alpha}  =0,-\gamca,- \gamma_{p \alpha} = -\Gamma_\alpha$. The eigenmode expansion of the kernel reads
\begin{equation}
W_\alpha = -\gamca\sqbrack{\frac{1}{2}\fpOpHat\sqbrack{\Ket{N} - \nzia\Ket{\one}}}\sqbrack{\vphantom{\frac{1}{2}}\Bra{N} - \nza\Bra{\one}} -\Gamma_\alpha\Ket{\fpOp}\Bra{\zia\fpOp}.\label{eq_derive_kernel}
\end{equation}
The explicit expression of $\gamca$ finally follows from the fact that in sequential tunneling, direct transitions between the empty and doubly occupied state are not possible. Using $\fpOpHat\Ket{0/2} = \Ket{0/2}$ and $\hat{N}\Ket{0} = 0,\hat{N}\Ket{2} = 2\Ket{2}$, this leads to
\begin{equation}
\Bra{0}W_\alpha\Ket{2} \equalby{!} 0 \overset{\eq{eq_derive_kernel}}{\Rightarrow} \gamca = 2\Gamma_\alpha\frac{\Braket{2}{\zia}}{\nzia(2-\nza)} \equalbyeqns{eq_derive_step6}{eq_derive_step9} \frac{\Gamma_\alpha}{2}\sqbrack{1 + f_\alpha(\epsilon) - f_{\alpha}(\epsilon + U)}.\label{eq_charge_rate}
\end{equation}
In summary, the equations \eq{eq_derive_step9} to \eq{eq_charge_rate} show that instead of calculating explicit transition rates using, e.g., Fermi's golden rule, an equivalent result for the master equation kernel $W_\alpha$ of the single-level dot with one lead can be obtained by appealing to general principles, reasonable assumptions and symmetries. The latter includes in particular the fermion-parity duality.

%%%%%%%%%%%%%%%%%%%%%%%%%
\subsection{Full multi-lead kernel}
%%%%%%%%%%%%%%%%%%%%%%%%%

The remaining question is what happens in case of a non-equilibrium situation with several leads, at different electrochemical potentials and temperatures. Since the fermion-parity duality holds for the full non-equilibrium kernel $W$, we can \emph{formally} obtain its eigenmode expansion by repeating the same procedure as for $W_\alpha$ in \Sec{app:kernel_lead}, making the same assumptions about the connectiveness and physicality of the (dual) kernel\footnote{If each kernel $W^\alpha$ is dissipative \BrackEq{eq_dissipative} and eventually connects all states \BrackEq{eq_connectiveness}, this is also true for the sum of all kernels, since their off-diagonal elements in the probability sector are non-negative!}. Similarly to \Eq{eq_kernel_singlelead_eigenvectors}, this yields the three biorthonormal left and right eigenvectors
\begin{gather}
\begin{array}{c|c|c}
          \text{Amplitude}                &                            \text{$-$ Eigenvalue $=$ decay rate}                             &                             \text{Mode}                                \\\hline\hline
         \Bra{z'} = \Bra{\one}          &                            \gamma_{z}                             &                            \Ket{z}                             \\\hline
 \Bra{{c}'} = \Bra{N} - \nz \Bra{\one} & \gamc & \Ket{c} = \frac12 \fpOpHat \Big[\Ket{N}-\nzi \Ket{\one}\Big]           \\\hline
     \Bra{p'} = \Bra{z_\text{i} \fpOp}       &                           \gamp                           &                     \Ket{p} = \Ket{\fpOp}                          \\\hline
\end{array}\label{eq_kernel_eigenvectors_formal}
\end{gather}
to the eigenvalues $-\gamma_z=0,-\gamc,-\gamma_p=-\Gamma = -\sum_\alpha\Gamma_\alpha$, and
\begin{equation}
W = -\gamc\sqbrack{\frac{1}{2}\fpOpHat\sqbrack{\Ket{N} - \nzi\Ket{\one}}}\sqbrack{\Bra{N} - \nz\Bra{\one}} -\Gamma\Ket{\fpOp}\Bra{\zin\fpOp},\label{eq_kernel_formal}
\end{equation}
where $0 < \gamc < \Gamma$ restricts the total charge rate $\gamc$.\par

The problem of \Eq{eq_kernel_formal} is that it is not yet obvious what exactly the \emph{non-equilibrium} stationary state $\Ket{z}$ and all derived quantities including $\Ket{\zin}$ are, since we cannot simply assume the grand-canonical ensemble as for the equilibrium case \eq{eq_derive_step3}. Starting from \eq{eq_derive_step1}, a straightforward but cumbersome approach to this problem would be to calculate the full kernel explicitly from all single-lead kernels $W_\alpha$, and then explicitly solve $W\Ket{z} = 0$ with $\Braket{\one}{z} = 1$ to obtain the non-equilibrium stationary state $\Ket{z}$. However, our goal here -- and in general in this article -- is to instead show that for \emph{everything we are interested in}, we never once need the full expression of $\Ket{z}$.

As a first step, we derive $n_z$ and $\gamma_\text{c}$. Since $\Bra{\one}W = \Bra{\one}W_\alpha = 0$, we can always write
\begin{align}
\Bra{N}W &= \Bra{c'}W \equalbyeqn{eq_kernel_eigenvectors_formal} -\gamc\sqbrack{\Bra{N} - \nz\Bra{\one}} = -\gamc\Bra{N-\one} + \gamc (\nz-1)\Bra{\one}\notag\\
&\equalbyeqn{eq_derive_step1} \sum_\alpha\Bra{N}W_\alpha = \sum_\alpha\Bra{c'_\alpha}W_\alpha = -\sum_\alpha\gamca\sqbrack{\Bra{N} - \nza\Bra{\one}}\notag\\
&=\sqbrack{-\sum_\alpha\gamca}\Bra{N-\one} + \sqbrack{\sum_\alpha\gamca\nbrack{\nza-1}}\Bra{\one} \ . \label{eq_derive_step10}
\end{align}
Acting from the right with $\Ket{N-\one}$ yields the total charge rate
\begin{equation}
\gamc = \sum_{\alpha}\gamca \equalbyeqn{eq_charge_rate}\sum_{\alpha}\frac{\Gamma_\alpha}{2}\sqbrack{1 + f_\alpha(\epsilon) - f_{\alpha}(\epsilon + U)} \ . \label{eq_charge_rate_full}
\end{equation}
Likewise, acting from the right with the unit operator $\Ket{\one}$ gives
\begin{equation}
\nz = \sum_\alpha\frac{\gamca}{\gamc}\nza \equalbyeqns{eq_charge_rate_full}{eq_derive_step9} \frac{1}{\gamc}\sum_{\alpha}\Gamma_\alpha \cdot f_{\alpha}(\epsilon) \ . \label{eq_charge_full}
\end{equation}
Second, when evaluated at equilibrium, $\mu_{\alpha'} = \mu_\alpha = \mueq\,,\,\beta_{\alpha'} = \beta_\alpha = \beta \text{ for all reservoirs } \alpha',\alpha$, denoted by $\evalAtEqui{\dotsc}$, the expressions \eq{eq_charge_rate_full} and \eq{eq_charge_full} for the charge rate and average particle number simplify to
\begin{equation}
\gamceq = \evalAtEqui{\gamc} = \frac{\Gamma}{\Gamma_\alpha}\evalAtEqui{\gamca} \lrsepa \evalAtEqui{\nz} = \evalAtEqui{\nza} \qquad\forall \alpha.\label{eq_charge_rate_equilibrium}
\end{equation}
Moreover, $\Ket{z}$ must also become the grand-canonical ensemble with respect to the common electrochemical potential $\mueq$ and inverse temperature $\beta$ under these conditions, and equivalently for  the \emph{dual} stationary state:\footnote{While this is intuitively clear, it now follows explicitly from our construction. Namely, we realize from the equations \eq{eq_derive_step9}, \eq{eq_derive_kernel}, and \eq{eq_charge_rate} that at equilibrium, the kernels $W_\alpha$ differ only by the coupling prefactor $\Gamma_\alpha$, such that $W_\alpha = \Gamma_\alpha W^0$ with some common superoperator $W^0$. More explicitly, $\evalAtEqui{W_{\alpha}} = \Gamma_\alpha W^0 \overset{\eq{eq_derive_step1}}{\Rightarrow} \evalAtEqui{W} = \sum_{\alpha'}\Gamma_{\alpha'}W^0 = \frac{\Gamma}{\Gamma_\alpha}\evalAtEqui{W_{\alpha}}$. Since $W\Ket{z} = 0$ is a homogeneous equation, and since the stationary state is unique by construction, this implies Eq.~(\ref{eq_stationary_state_equilibrium}).}
\begin{equation}
\Ket{\zeq} := \evalAtEqui{\Ket{z}} = \evalAtEqui{\Ket{z_\alpha}}  \lrsepa \Ket{\zieq} := \evalAtEqui{\Ket{\zin}} = \evalAtEqui{\Ket{\zia}} \ ,\label{eq_stationary_state_equilibrium}
\end{equation}
with $\Braket{\one}{\zeq} = 1$ and $\Braket{\one}{\zieq} = 1$. This then finally also implies
\begin{equation}
\Ket{c_{\text{eq}}} = \evalAtEqui{\Ket{c}} \equalbyeqn{eq_stationary_state_equilibrium} \evalAtEqui{\Ket{c_\alpha}} \lrsepa \Bra{c'_{\text{eq}}} = \evalAtEqui{\Bra{c'}} \equalbyeqn{eq_stationary_state_equilibrium} \evalAtEqui{\Bra{c'_\alpha}}.\label{eq_charge_mode_equilibrium}
\end{equation}
As it turns out, the relations \eq{eq_stationary_state_equilibrium} are almost everything we need to know about the full non-equilibrium stationary state and its dual, apart from the stationarity $W\Ket{z} = 0$ and the orthogonality relation $\Braket{\zin\fpOp}{z} = 0$ that also hold far away from equilibrium.

%%%%%%%%%%%%%%%%%%%%%%%%%%%%%%%%%%%%%%%%%%
\section{Explicit expressions for the kernel}\label{app:explicit}
%%%%%%%%%%%%%%%%%%%%%%%%%%%%%%%%%%%%%%%%%%
In this appendix, we provide explicit expressions for the full non-equilibrium Kernel $W$ and its stationary state for completeness. We emphasize that these expressions are not needed except for producing plots of the nonlinear Peltier coefficient as shown in Fig.~\ref{fig_peltier_nl}, requiring the non-equilibrium parity $p$.\par

As mentioned before, the transition rates $W_{f,i} = \Bra{f}W\Ket{i}/\Braket{f}{f}$ of the full kernel $W$ can be derived from the sequential tunneling decomposition \Eq{eq_kernel_decomposition} of $W$ combined with explicitly known form of the lead resolved kernel $W_\alpha$ obtained using duality, see \Eq{eq_derive_kernel} combined with \Eq{eq_derive_step6}, \Eq{eq_derive_step9} and \Eq{eq_charge_rate}. More traditionally, one finds $(W_{f,i})_\alpha$ by explicitly using Fermi's Golden rule or first order diagrammatic perturbation theory. In any case, the full matrix of transition rates in the occupation number basis $\Ket{0},\Ket{1},\Ket{2}$ reads 
\begin{equation}
\left(\begin{array}{ccc}
-W_{1,0} & W_{0,1} & 0\\
W_{1,0} & -W_{0,1}-W_{2,1} & W_{1,2}\\
0 & W_{2,1} & -W_{1,2}
\end{array}\right) = \sum_\alpha \left(\begin{array}{ccc}
-(W_{1,0})_\alpha & (W_{0,1})_\alpha & 0\\
(W_{1,0})_\alpha & -(W_{0,1})_\alpha-(W_{2,1})_\alpha & (W_{1,2})_\alpha\\
0 & (W_{2,1})_\alpha & -(W_{1,2})_\alpha
\end{array}\right)\label{eq_W_matrix}
\end{equation}
with the transition rates $(W_{f,i})_\alpha = \Bra{f}W_\alpha\Ket{i}/\Braket{f}{f}$ given by
\begin{eqnarray}
(W_{1,0})_\alpha = \Gamma_\alpha f^+_\alpha(\epsilon) && (W_{2,1})_\alpha = (\Gamma_\alpha/2) f^+_\alpha(\epsilon+U)\nonumber \\
(W_{0,1})_\alpha = (\Gamma_\alpha/2) f^-_\alpha(\epsilon) && (W_{1,2})_\alpha = \Gamma_\alpha f^-_\alpha(\epsilon+U).
\end{eqnarray}
The stationary state $\Ket{z}$ -- being the right eigenvector of $W$ to the eigenvalue $\gamma_z=0$ is found to be represented by
\begin{equation}
\Ket{z} \equalby{\cdot} \frac{1}{\frac{W_{0,1}}{W_{1,0}}+\frac{W_{2,1}}{W_{1,2}}+1}\left(\begin{array}{c}
\frac{W_{0,1}}{W_{1,0}}\\
1\\
\frac{W_{2,1}}{W_{1,2}}
\end{array}\right)
\end{equation}
in the basis $\Ket{0},\Ket{1},\Ket{2}$.
The inverted stationary state, $\Ket{\zin}$, is then simply found by replacing all $f_\alpha^+\leftrightarrow f_\alpha^-$.
From this expression for the stationary state of the original and the inverted model, the following quantities required for the plot of the nonlinear Peltier coefficient can be obtained from a standard trace operation: $\nz=\Tr\{\hat{N}\hat{z}\}$, $\nzi=\Tr\{\hat{N}\hat{z}_\text{i}\}$, $p_z=\Tr\{(-\one)^{\hat{N}}\hat{z}\}$, and $p_\text{i}=\Tr\{(-\one)^{\hat{N}}\hat{z}_\text{i}\}$.

%%%%%%%%%%%%%%%%%%%%%%%%%%%%%%%%%%%%%%%%%%
\section{Derivation of the linear response coefficients}
\label{app:linear_response}
%%%%%%%%%%%%%%%%%%%%%%%%%%%%%%%%%%%%%%%%%%

Here, we show how to derive the well-known linear response coefficients in their new shape \eq{eq_G}-\eq{eq_K} by combining the linear state response \Eq{eq_linear_response_state} with the eigenmode expansion \eq{eq_table} induced by the duality \eq{eq_duality}.

%%%%%%%%%%%%%%%%%%%%%%%%%%%%%%%%%%%%%%%%%%
\subsection{First derivatives at equilibrium}
\label{app:near_equi}
%%%%%%%%%%%%%%%%%%%%%%%%%%%%%%%%%%%%%%%%%%

The most important prerequisite to calculate the linear response coefficients is the state linearization relation \eq{eq_linear_response_state}, which we here derive using the decay eigenmode expansion \eq{eq_table}. In principle, the final result stated in \Eq{eq_linear_response_state} is well-known from earlier works on Coulomb blockade oscillations \cite{Beenakker1991Jul,Beenakker1992Oct} in more general quantum dot systems. However, we stress that due to the simplicity, the symmetries and due to the resulting conservation laws of the model considered here, we can obtain the result \emph{without} ever assuming that detailed balance~\cite{Beenakker1991Jul,Beenakker1992Oct} or more general \emph{linear balance} relations~\cite{Erdman2017Jun} continue to hold for small deviations away from equilibrium.

Due to current conservation under stationary conditions we find for all leads $\alpha$,
\begin{gather}
\sum_\alpha \INa \equalbyeqn{eq_INz} -\sum_\alpha\gamca\Braket{c'_\alpha}{z} \equalbyeqn{eq_IN} \sum_\alpha \Bra{N}W_\alpha\Ket{z} \equalbyeqn{eq_derive_step1} \Bra{N}W\Ket{z} \equalbyeqn{eq_kernel_eigenvectors_formal}  0\notag\\
\sum_\alpha \IEa \equalbyeqn{eq_IEz} \sum_\alpha\cbrack{\sqbrack{\epsilon + \nbrack{2 - \nzia}\frac{U}{2}}\INa - U\Gamma_\alpha\Braket{\zia\fpOp}{z}} \equalbyeqn{eq_IE} \sum_\alpha \Bra{\Hdot}W_\alpha\Ket{z} \equalbyeqn{eq_derive_step1} \Bra{\Hdot}W\Ket{z} \equalbyeqn{eq_kernel_eigenvectors_formal}  0.\label{eq_derive_current_conservation}
\end{gather}
As intuitively clear, at equilibrium, charge and energy current to every lead $\alpha$ vanish separately:
\begin{gather}
\evalAtEqui{\INa} \equalbyeqn{eq_INz} -\evalAtEqui{\gamca\Braket{c'_\alpha}{z}} \equalbyeqn{eq_charge_mode_equilibrium} -\evalAtEqui{\gamca\Braket{c'}{z}} \equalbyeqn{eq_kernel_eigenvectors_formal} 0\label{eq_derive_equi_charge_current}\\
\evalAtEqui{\IEa} \equalbyeqn{eq_IEz} \evalAtEqui{\sqbrack{\epsilon + \nbrack{2 - \nzia}\frac{U}{2}}\INa} -U\Gamma_\alpha\evalAtEqui{\Braket{\zia\fpOp}{z}} \equalbyeqns{eq_derive_equi_charge_current}{eq_stationary_state_equilibrium} -U\Gamma_\alpha\evalAtEqui{\Braket{\zin\fpOp}{z}} \equalbyeqn{eq_kernel_eigenvectors_formal} 0.\label{eq_derive_equi_energy_current}
\end{gather}
Now the main trick to derive the desired result is to cleverly represent the non-equilibrium stationary state $\Ket{z}$ in terms of all local equilibrium states $\Ket{z_\alpha}$. Using the complete basis \eq{eq_kernel_singlelead_eigenvectors}, one can write
\begin{equation}
\Ket{z} = \underbrace{\Braket{\one}{z}}_{=1}\cdot \Ket{z_\alpha} + \Braket{c'_\alpha}{z} \cdot\Ket{c_\alpha} + \Braket{\zia\fpOp}{z} \cdot \Ket{\fpOp}\label{eq_derive_step11}
\end{equation}
for any lead $\alpha$. Multiplying by $\gamca$, summing over $\alpha$, and dividing\footnote{This is allowed since we know that all $\gamca > 0$.} the result by $\sum_\alpha\gamca = \gamc$ \BrackEq{eq_charge_rate_full} yields
\begin{align}
\Ket{z} &= \sum_{\alpha}\frac{\gamca}{\gamc}\Ket{z_\alpha} + \sum_{\alpha}\gamca\Braket{c'_\alpha}{z} \cdot\frac{1}{\gamc}\Ket{c_\alpha} + \sqbrack{\sum_{\alpha}\frac{\gamca}{\gamc}\cdot\Braket{\zia\fpOp}{z}}\Ket{\fpOp}\notag\\
&\equalbyeqn{eq_INz} \sum_{\alpha}\frac{\gamca}{\gamc}\Ket{z_\alpha} - \sum_{\alpha}\INa \cdot\frac{1}{\gamc}\Ket{c_\alpha} + \sqbrack{\sum_{\alpha}\frac{\gamca}{\gamc\cdot U\Gamma_\alpha}\cdot U\Gamma_\alpha\Braket{\zia\fpOp}{z}}\Ket{\fpOp}\notag\\
&\equalbyeqn{eq_IEz} \sum_{\alpha}\frac{\gamca}{\gamc}\Ket{z_\alpha} - \sum_{\alpha}\INa \cdot\frac{1}{\gamc}\Ket{c_\alpha} - \cbrack{\sum_{\alpha}\frac{\gamca}{\gamc\cdot U\Gamma_\alpha}\cdot\sqbrack{\IEa - \nbrack{\epsilon + \nbrack{2 - \nzia}\frac{U}{2}}\INa}}\Ket{\fpOp}.\label{eq_derive_step12}
\end{align}
In the second and third step, we have identified the charge current $\INa$ and energy current $\IEa$ from lead $\alpha$ as part of the expansion coefficients in order to use the conservation laws \eq{eq_derive_current_conservation} and the equilibrium conditions \Eq{eq_derive_equi_charge_current} and \Eq{eq_derive_equi_energy_current} in the following steps.\par

When taking equilibrium derivatives of $\Ket{z}$ with respect to $x = \mu_\alpha$ or $x = \beta_\alpha$, we are always free to take the equilibrium limit separately for all terms outside of derivatives (when using the product rule), and to interchange sums with derivatives or the equilibrium limit, since every term in \Eq{eq_derive_step12} is separately continuously differentiable. This finally justifies the following computation:
\begin{align}
\evalAtEqui{\frac{d}{dx}\Ket{z}} &\equalbyeqn{eq_derive_step12} \sum_\alpha\evalAtEqui{\frac{d}{dx}\nbrack{\frac{\gamca}{\gamc}}}\cdot\underbrace{\evalAtEqui{\sqbrack{\Ket{z_\alpha}}}}_{\equalbyeqn{eq_stationary_state_equilibrium}\Ket{\zeq}} + \sum_{\alpha}\underbrace{\evalAtEqui{\frac{\gamca}{\gamc}}}_{\equalbyeqn{eq_charge_rate_equilibrium}\Gamma_\alpha/\Gamma} \cdot\evalAtEqui{\frac{d}{dx}\Ket{z_\alpha}} - \sum_\alpha\evalAtEqui{\frac{d}{dx}\INa}\cdot\underbrace{\evalAtEqui{\frac{1}{\gamc}\Ket{c_\alpha}}}_{\equalbyeqn{eq_charge_mode_equilibrium}\evalAtEqui{\Ket{c}/\gamc}}\notag\\
&\phantom{=}-\sum_{\alpha}\underbrace{\evalAtEqui{\INa}}_{\equalbyeqn{eq_derive_equi_charge_current}0}\cdot\evalAtEqui{\frac{d}{dx}\nbrack{\frac{1}{\gamc}\Ket{c_\alpha}}} -\clbrack{\sum_{\alpha}\evalAtEqui{\frac{d}{dx}\nbrack{\frac{\gamca}{\gamc U\Gamma_\alpha}}}}\cdot\underbrace{\evalAtEqui{\sqbrack{\IEa - \nbrack{\epsilon + \frac{2-\nzia}{2}U}\INa}}}_{\equalbyeqns{eq_derive_equi_charge_current}{eq_derive_equi_energy_current}0}\notag\\
&\phantom{=} + \sum_\alpha\underbrace{\evalAtEqui{\frac{\gamca}{\gamc U\Gamma_\alpha}}}_{\equalbyeqn{eq_charge_rate_equilibrium}1/(U\Gamma)}\cdot\sqbrack{\evalAtEqui{\frac{d}{dx}\IEa} - \evalAtEqui{\frac{d}{dx}\nbrack{\epsilon + \frac{2 - \nzia}{2}U}}\cdot \smash{ \underbrace{\evalAtEqui{\INa}}_{  \equalbyeqn{eq_derive_equi_charge_current}0}}  }\notag\\
&\phantom{=} - \sum_\alpha\underbrace{\evalAtEqui{\frac{\gamca}{\gamc U\Gamma_\alpha}}}_{\equalbyeqn{eq_charge_rate_equilibrium}1/(U\Gamma)}\cdot\sqbrack{ \smash{ \underbrace{\evalAtEqui{\nbrack{\epsilon + \frac{2-\nzia}{2}U}}}_{\equalbyeqn{eq_stationary_state_equilibrium}\epsilon + \frac{2-\nzieq}{2}U}} \cdot\evalAtEqui{\frac{d}{dx}\INa}}\crbrack{\vphantom{\sum_\alpha}}\Ket{\fpOp}\notag\\
&= \sum_\alpha\frac{\Gamma_\alpha}{\Gamma}\evalAtEqui{\frac{d}{dx}\Ket{z_\alpha}} + \vphantom{\underbrace{\frac{\sum_\alpha\gamca}{\gamc}}_{\equalbyeqn{eq_charge_rate_full}1}} \sqbrack{\frac{d}{dx}\evalAtEqui{\vphantom{\frac{\sum_\alpha\gamca}{\gamc}} \smash{ \underbrace{\frac{\sum_\alpha\gamca}{\gamc}}_{\equalbyeqn{eq_charge_rate_full}1} } }}\cdot\Ket{\zeq} - 
\sqbrack{\evalAtEqui{ \vphantom{\sum_\alpha \INa}\smash{\frac{d}{dx}\underbrace{\sum_\alpha \INa}_{\equalbyeqn{eq_derive_current_conservation}0} } }}\cdot\evalAtEqui{\frac{1}{\gamc}\Ket{c}}\notag\\
&\phantom{=} \vphantom{\frac{d}{dx}\underbrace{\sum_\alpha \IEa}_{\equalbyeqn{eq_derive_current_conservation}0}} -\frac{1}{U\Gamma}\cdot\sqbrack{\vphantom{ \frac{d}{dx}\sum_\alpha \IEa }  \evalAtEqui{ \vphantom{\sum_\alpha \IEa} \smash{\frac{d}{dx}\underbrace{\sum_\alpha \IEa}_{\equalbyeqn{eq_derive_current_conservation}0} } } - \evalAtEqui{\nbrack{\epsilon + \frac{2-\nzieq}{2}U}} \cdot \evalAtEqui{ \vphantom{\sum_\alpha \INa} \smash{\frac{d}{dx}\underbrace{\sum_\alpha \INa}_{\equalbyeqn{eq_derive_current_conservation}0} } }}\Ket{\fpOp}.\label{eq_derive_side_formula_one_step}
\end{align}
Equation \eq{eq_derive_side_formula_one_step} yields the central relation
\begin{equation}
\evalAtEqui{\frac{d}{dx}\Ket{z}} = \sum_\alpha\frac{\Gamma_\alpha}{\Gamma}\evalAtEqui{\frac{d}{dx}\Ket{z_\alpha}}.\label{eq_derive_side_formula_one}
\end{equation}
It reduces equilibrium derivatives of the non-equilibrium stationary state $\Ket{z}$ to derivatives of the local equilibrium states $\Ket{z_\alpha}$. The latter are in turn straightforward to perform for $x=\mu_\alpha$ or $x = \beta_\alpha$. Namely, defining the affinities $\hat{A}_\alpha = \beta_\alpha\nbrack{\hat{\Hdot} - \mu_\alpha \hat{N}}$, using the exponential form \eq{eq_derive_step3} of $\Ket{z_\alpha}$ and the fact that the local dot Hamiltonian conserves the local particle number, $[\hat{\Hdot},\hat{N}] = 0$, the basic rules of taking derivatives yield
\begin{equation}
\evalAtEqui{\frac{d}{dx}\Ket{z_\alpha}} = -\evalAtEqui{\sqbrack{\nbrack{\frac{d}{dx}\hat{A}_\alpha} - \Braket{\nbrack{\frac{d}{dx}A_{\alpha}}}{z_\alpha}\cdot\one}}\cdot\evalAtEqui{\Ket{z_\alpha}} .\label{eq_derive_side_formula_two}
\end{equation}
The relations \eq{eq_derive_side_formula_one} and \eq{eq_derive_side_formula_two} together prove the wanted result
\begin{gather}
\evalAtEqui{\frac{d}{dx}\Ket{z}} = \sum_\alpha\frac{\Gamma_\alpha}{\Gamma}\evalAtEqui{\frac{d}{dx}\Ket{z_\alpha}} = -\sum_\alpha\frac{\Gamma_\alpha}{\Gamma}\evalAtEqui{\sqbrack{\nbrack{\frac{d}{dx}\hat{A}_\alpha} - \Braket{\nbrack{\frac{d}{dx}A_{\alpha}}}{z_\alpha}\cdot\one}}\cdot\evalAtEqui{\Ket{z_\alpha}} \ .\label{eq_derive_central_formula}
\end{gather}
This is the key ingredient which allows us, in the following steps, to express and relate the linear response coefficients to the additional insights provided by the duality relation. This appeals in particular to the role of the fermion parity and the significance of the \emph{dual} stationary state $\Ket{\zin}$. 

\subsection{Linearized charge current and conductance}
\label{app:linear_response_conductance}
We start with the linear response of the particle current $\INa$ to a variation $\delta x$ of the variable $x=\mu_{\alpha'},T_{\alpha'}$.
\begin{align}
\evalAtEqui{\frac{\partial \INa}{\partial x}}
 & \overset{\eq{eq_INz}}= \evalAtEqui{\frac{\partial}{\partial x}\sqbrack{\gamca\nbrack{\nza - n_z}} }
 \equalbyeqn{eq_charge_rate_equilibrium} \sqbrack{\evalAtEqui{\frac{\partial \gamca}{\partial x}} \cdot \underbrace{\nbrack{n_{z} - n_z}}_{=0} + \gamca\cdot\evalAtEqui{ \nbrack{\frac{\partial \nza }{\partial x} - \frac{\partial \nz}{\partial x}} }  } \notag\\&= \frac{\Gamma_\alpha\cdot\gamceq}{\Gamma}\cdot \evalAtEqui{\sqbrack{\Bra{N}\frac{\partial}{\partial x}\Ket{z_\alpha} - \Bra{N}\frac{\partial}{\partial x}\Ket{z}}} \equalbyeqn{eq_linear_response_state}\frac{\Gamma_\alpha\cdot\gamceq}{\Gamma} \sum_{\alpha''}\frac{\Gamma_{\alpha''}}{\Gamma}\evalAtEqui{\Bra{N}\frac{\partial}{\partial x}\sqbrack{\Ket{z_{\alpha}} - \Ket{z_{\alpha''}}}}\notag\\
&\equalbyeqns{eq_linear_response_state}{eq_stationary_state_equilibrium} -\frac{\Gamma_\alpha\cdot\gamceq}{\Gamma} \sum_{\alpha''}\frac{\Gamma_{\alpha''}}{\Gamma}\clbrack{\evalAtEqui{\frac{d}{dx}\nbrack{\beta_\alpha - \beta_{\alpha''}} }\Bra{N}\sqbrack{\hat{\Hdot}- \Braket{H}{\zeq}\cdot\one}\cdot\Ket{\zeq},}\notag\\
&\phantom{\equalbyeqns{eq_linear_response_state}{eq_stationary_state_equilibrium} -\frac{\Gamma_\alpha\cdot\gamceq}{\Gamma}\sum_{\alpha''}\frac{\Gamma_{\alpha''}}{\Gamma}\clbrack{}}-\crbrack{\evalAtEqui{\frac{d}{dx}\nbrack{\mu_\alpha\beta_\alpha - \mu_{\alpha''}\beta_{\alpha''}} } \Bra{N}\sqbrack{\hat{N} - \Braket{N}{\zeq}\cdot\one}\cdot\Ket{\zeq} }\notag\\
&= \frac{\Gamma_\alpha\cdot\gamceq}{\Gamma} \sum_{\alpha''}\frac{\Gamma_{\alpha''}}{\Gamma}\cbrack{\evalAtEqui{ \frac{d}{dx}\nbrack{\mu_\alpha\beta_\alpha - \mu_{\alpha''}\beta_{\alpha''}}}\delnsqzeq - \evalAtEqui{\frac{d}{dx}\nbrack{\beta_\alpha - \beta_{\alpha''}} }\cdot\sqbrack{\left\langle \hat{N}\hat{\Hdot}\right\rangle_{\text{eq}} - \left\langle \hat{N}\right\rangle_{\text{eq}}\left\langle \hat{\Hdot}\right\rangle_{\text{eq}}  } },\label{eq_linear_response_step1}
\end{align}
where
\begin{equation}
 \delnsqzeq = \left\langle \hat{N}^2\right\rangle_{\text{eq}} - \left\langle \hat{N}\right\rangle_{\text{eq}}^2 \lrsepa \left\langle \bullet \right\rangle_{\text{eq}} = \Braket{\bullet}{\zeq} = \Tr\sqbrack{\bullet\cdot\hatzeq}\label{eq_linear_response_step2}
\end{equation}
for any dot observable $\bullet = \bullet^\dagger$.

To compute the linear response of the particle current $\INa$ in lead $\alpha$ to a variation of the electrochemical potential $\delta\mu_{\alpha'}$ from the equilibrium, we set $x = \mu_{\alpha'}$ and realize that only the first term proportional to the average charge fluctuations survives in \Eq{eq_linear_response_step1}. More precisely, we obtain
\begin{equation}
 \evalAtEqui{\frac{d\INa}{d\mu_{\alpha'}}} \equalbyeqn{eq_linear_response_step1} \frac{\gamceq}{T}\cdot\frac{\Gamma_\alpha}{\Gamma}\cdot\sqbrack{\sum_{\alpha''}\frac{\Gamma_{\alpha''}}{\Gamma}\nbrack{\delta_{\alpha\alpha'} - \delta_{\alpha''\alpha'}}}\cdot\delnsqzeq = \frac{\gamceq}{T}\cdot\frac{\Gamma_\alpha(\Gamma\delta_{\alpha\alpha'} - \Gamma_{\alpha'})}{\Gamma^2}\cdot\delnsqzeq.\label{eq_linear_response_charge_mu}
\end{equation}
Restricting our considerations to the two-contact case as presented in the main paper the result simplifies to
\begin{align}
 L_{11} &= -\frac{T}{2}\sqbrack{\evalAtEqui{\frac{dI_N^\text{L}}{dV}} - \evalAtEqui{\frac{dI_{N}^\text{R}}{dV}} } \equalbyeqn{eq_linear_response_charge_mu} \gamceq\cdot\frac{ \Gamma_\text{L}\Gamma_\text{R}}{\Gamma^2}\cdot\delnsqzeq.\label{eq_linear_response_charge_mu_symmetrized}
\end{align}

To finally plot the conductance, one needs the explicit expressions for the equilibrium charge fluctuation $\delnsqzeq$ as a function of the system parameters, including the equilibrium expectation value of $\hat{N}^2$. One way to obtain this expectation value is by using the Liouville space identity \eq{eq_liouville_identity_alt}
and the  scalar products\footnote{This can be shown using the definition of the states $\Ket{0},\Ket{1},\Ket{2}$ and observables $\hat{N},\fpOpHat$ in terms of creation and annihilation operators, their anti-commutation relations, and the fact that an annihilation operator acting on the vacuum gives $0$.}
$\Braket{\one}{N^2} = 6,\ \Braket{N-\one}{N^2} = 4,\ \Braket{\fpOp}{N^2} = 2.$
This allows us to reduce $\hat{N}^2$ to
\begin{align}
\hat{N}^2 &\equalbyeqn{eq_liouville_identity_alt} \frac{1}{4}\Braket{\one}{N^2}\cdot \one + \frac{1}{2}\Braket{N-\one}{N^2}\cdot (N-\one) + \frac{1}{4}\Braket{\fpOp}{N^2}\cdot \fpOpHat = -\frac{1}{2}\cdot \one + 2\cdot \hat{N} + \frac{1}{2}\cdot \fpOpHat,\label{eq_fluctuation_expansion}
\end{align}
and to write
\begin{equation}
 \left\langle \hat{N}^2\right\rangle_{\text{eq}} = \Braket{N^2}{\zeq} \equalbyeqn{eq_fluctuation_expansion} 2\Braket{N}{\zeq} + \frac{1}{2}\sqbrack{\Braket{\fpOp}{\zeq} - \Braket{\one}{\zeq}} = 2\nzeq + \frac{1}{2}\nbrack{\pzeq - 1}.\label{eq_fluctuation_expectation}
\end{equation}
The parity $\pzeq$ is, due to \Eq{eq_stationary_state_equilibrium}, calculated by taking the equilibrium limit of the lead resolved parity:
\begin{align}
 \pzeq = \evalAtEqui{p_{z\alpha}} \lrsepa p_{z\alpha} = \Braket{0}{z_\alpha} + \Braket{2}{z_\alpha} - 2\Braket{1}{z_\alpha} \equalbyeqns{eq_states}{eq_derive_step3} 1 - 4\frac{f_\alpha(\epsilon)[1-f_\alpha(\epsilon + U)]}{1+f_\alpha(\epsilon) - f_\alpha(\epsilon+U)}.\label{eq_parity_expectation}
\end{align}
With this, we explicitly find
\begin{align}
 \delnsqzeq &= \left\langle \hat{N}^2\right\rangle_{\text{eq}} - (\nzeq)^2 \equalbyeqn{eq_fluctuation_expectation} \nzeq(2-\nzeq) + \frac{1}{2}\nbrack{\pzeq - 1}\notag\\
 &\equalbys{\eq{eq_derive_step9},\eq{eq_charge_rate_equilibrium}}{\eq{eq_parity_expectation}} 2\frac{f(\epsilon)\sqbrack{1-f(\epsilon+U)}\sqbrack{1 - f(\epsilon) + f(\epsilon + U)}}{\sqbrack{1 + f(\epsilon) - f(\epsilon + U)}^2},\label{eq_fluctuation_expectation_explicit} 
\end{align}
where $f(x) = f_\text{R}(x)$ is the Fermi function with respect to the equilibrium potential $\mu$ and temperature $T$. As stated in the main paper, an alternative way to arrive at the explicit expression \Eq{eq_fluctuation_expectation_explicit} is to take an $\epsilon$-derivative of the equilibrium occupation number. Namely, this follows from the relation
\begin{align}
 \frac{d\nzeq}{d\epsilon} &= \Bra{N}\frac{d}{d\epsilon}\Ket{\zeq} = \Bra{N}\evalAtEqui{\frac{d}{d\epsilon}\Ket{z_\alpha}} \equalbyeqn{eq_derive_side_formula_two} -\Bra{N}\evalAtEqui{\sqbrack{\nbrack{\frac{d}{d\epsilon}\hat{A}_\alpha} - \Braket{\nbrack{\frac{d}{d\epsilon}A_{\alpha}}}{z_\alpha}\cdot\one}}\cdot\evalAtEqui{\Ket{z_\alpha}}\notag\\
 &\equalbyeqn{eq_stationary_state_equilibrium} -\beta\Bra{N}\sqbrack{\frac{d\hat{\Hdot}}{d\epsilon} - \Braket{\nbrack{\frac{d\Hdot}{d\epsilon}}}{\zeq}}\cdot\Ket{\zeq} \equalbyeqn{eq_hamiltonian_dot} -\beta\sqbrack{\Braket{N^2}{\zeq} - \Braket{N}{\zeq}^2} = -\frac{1}{T}\delnsqzeq.\label{eq_fluctuation_expectation_derivative}
\end{align}
In summary, the above analysis shows that we can explicitly calculate every ingredient in \Eq{eq_linear_response_charge_mu_symmetrized}: $\gamceq$ and $\nzeq$ using \Eq{eq_charge_rate} and \Eq{eq_derive_step9} combined with \Eq{eq_charge_rate_equilibrium}, and finally $\delnsqzeq$ according to \Eq{eq_fluctuation_expectation_explicit}.

%%%%%%%%%%%%%%%%%%%%%%%%%%%%%%%%%%%%%%%%%%%%%%%%%%%%%%%%%%%%%%%%%%%%%%%%%%%%%%%
\subsection{Charge-Energy correlation and Seebeck effect}
\label{app:linear_response_seebeck}
%%%%%%%%%%%%%%%%%%%%%%%%%%%%%%%%%%%%%%%%%%%%%%%%%%%%%%%%%%%%%%%%%%%%%%%%%%%%%%%
We continue by calculating the response of the charge current to a temperature gradient. The first crucial step is to evaluate the equilibrium correlation function $\left\langle \hat{N}\hat{\Hdot}\right\rangle_{\text{eq}}$ by expanding it in terms of the decay eigenmodes and amplitudes \eq{eq_table} taken at equilibrium. Using that all operators, i.e., $\hat{N}$, $\hat{\Hdot}$ and $\hatzeq$, are hermitian and mutually commute\footnote{The particle number is locally conserved, $[\hat{\Hdot},\hat{N}] = 0$ and the equilibrium state is just the Boltzmann factor containing $\hat{\Hdot}$ and $\hat{N}$ in the exponential.}, we can insert the Liouville space identity 
\begin{equation}
\mathcal{I} \equalbyeqn{eq_table} \Ket{z}\Bra{\one} + \Ket{c}\Bra{c'} + \Ket{\fpOp}\Bra{\zin\fpOp} \equalbyeqn{eq_stationary_state_equilibrium} \Ket{\zeq}\Bra{\one} + \Ket{c_{\mathrm{eq}}}\Bra{c'_{\mathrm{eq}}} + \Ket{\fpOp}\Bra{\zieq\fpOp}\label{eq_liouville_identity} 
\end{equation}
to write
\begin{align}
 \left\langle \hat{N}\hat{\Hdot}\right\rangle_{\text{eq}} - \left\langle \hat{N}\right\rangle_{\text{eq}}\left\langle \hat{\Hdot}\right\rangle_{\text{eq}}&= \Braket{H}{N\zeq} - \Braket{H}{\zeq}\Braket{N}{\zeq}\notag\\
 &\equalbyeqn{eq_liouville_identity} \Braket{H}{c_{\mathrm{eq}}}\Braket{c'_{\mathrm{eq}}}{N\cdot\zeq} + \Braket{H}{\fpOp}\Braket{\zieq\fpOp}{N\cdot\zeq}\notag\\
 &\equalbyeqns{eq_table}{eq_linear_response_step2}\Braket{H}{c_{\mathrm{eq}}}\cdot\delnsqzeq + \Braket{H}{\fpOp}\Braket{\zieq\fpOp}{N\cdot\zeq}.\label{eq_linear_response_step4}
\end{align}
To further simplify the remaining traces in \Eq{eq_linear_response_step4}, we realize that the scalar products
\begin{gather}
\Braket{\one}{\fpOp} = \Braket{N}{\fpOp} = 0 \lrsepa \Braket{2}{N} = \Braket{2}{\fpOp N} = 2 \lrsepa \Braket{2}{\fpOp} = \Braket{2}{\one} = 1,\label{eq_orthogonality_relation_one}
\end{gather}
and the important orthogonality 
\begin{align}
 \evalAtEqui{\Braket{\zia\fpOp}{N\cdot z}} = 0 \label{eq_orthogonality_relation_two}
\end{align}
pointed out in the main text \BrackEq{eq_orthogonality_relations} causes the second term of \eq{eq_linear_response_step4} to vanish. 
The charge-energy correlator of the dot at equilibrium with the bath is thus simply proportional to the average equilibrium charge fluctuation:
\begin{align}
 \left\langle \hat{N}\hat{\Hdot}\right\rangle_{\text{eq}} - \left\langle \hat{N}\right\rangle_{\text{eq}}\left\langle \hat{\Hdot}\right\rangle_{\text{eq}}&\equalbyeqns{eq_linear_response_step4}{eq_orthogonality_relation_two} \Braket{H}{c_{\mathrm{eq}}}\cdot\delnsqzeq.\label{eq_linear_response_step5}
\end{align}
The proportionality factor is obtained from the decay mode expansion of the Hamiltonian~\cite{Schulenborg16} 
\begin{align}
 \Braket{H}{c_{\mathrm{eq}}} &\equalbyeqns{eq_hamiltonian_dot}{eq_table} \frac{\epsilon}{2}\sqbrack{\Braket{N}{\fpOp N} - \nzieq\Braket{N}{\fpOp}} + \frac{U}{2}\sqbrack{\Braket{2}{\fpOp N} - \nzieq\Braket{2}{\fpOp}}\notag\\ &\equalbyeqn{eq_orthogonality_relation_one} \epsilon+U\frac{2-\nzieq}{2} \equalbyeqn{eq_seebeck} \mueq + \Eeq  .\label{eq_linear_response_step6}
\end{align}
Let us finally calculate the linear response of the particle current to temperature gradients. Evaluating \Eq{eq_linear_response_step1} with \Eq{eq_linear_response_step5} and \Eq{eq_linear_response_step6} gives for $x = \mu_{\alpha'},T_{\alpha'}$
\begin{align}
\evalAtEqui{\frac{d\INa}{dx}} &\equalbyeqns{eq_linear_response_step5}{eq_linear_response_step6} \frac{\Gamma_\alpha\cdot\gamceq}{\Gamma} \sum_{\alpha''}\frac{\Gamma_{\alpha''}}{\Gamma}\evalAtEqui{ \frac{d}{dx}\nbrack{\nbrack{\mu_\alpha - \mueq - \Eeq}\beta_\alpha - \nbrack{\mu_{\alpha''} - \mueq - \Eeq}\beta_{\alpha''}}}\delnsqzeq.\label{eq_linear_response_step7}
\end{align}
With $x = T_{\alpha'}$ and $d\beta_{\alpha}/dT_{\alpha'} = -\delta_{\alpha\alpha'}\beta_\alpha^2$, we furthermore find
\begin{equation}
 \evalAtEqui{\frac{d\INa}{dT_{\alpha'}}} \equalbyeqn{eq_linear_response_step7} \frac{\Eeq}{T}\cdot \evalAtEqui{\frac{d\INa}{d\mu_{\alpha'}}}.\label{eq_linear_response_step8}
\end{equation}
For the symmetrized charge current response between two contacts to a temperature bias $\Delta T = T_\text{L} - T$, we then find the relation $S = -L_{12}/(TL_{11}) = \Eeq/T$ used in the main paper:
 \begin{align}
 L_{12} &=  -\frac{T^2}{2}\evalAtEqui{\sqbrack{\frac{d(I_N^\text{L} - I_N^\text{R})}{dT_\text{L}}}} \equalbys{\eq{eq_linear_response_step8}}{\eq{eq_linear_response_charge_mu_symmetrized}} -\Eeq\cdot L_{11}\label{eq_linear_response_charge_T_symmetrized} \ .
\end{align}

%%%%%%%%%%%%%%%%%%%%%%%%%%%%%%%%%%%%%%%%%%%%%%%%%%%%%%%%%%%%%%%%%%%%%%%%%%%%%%%%%%%
\subsection{Peltier effect and Fourier heat}
\label{app:linear_response_peltier_fourier}
%%%%%%%%%%%%%%%%%%%%%%%%%%%%%%%%%%%%%%%%%%%%%%%%%%%%%%%%%%%%%%%%%%%%%%%%%%%%%%%%%%%
We now compute the linear response of the heat current to a potential or temperature gradient
\begin{align}
 \evalAtEqui{\frac{\partial J^\alpha}{\partial x}} 
 &\overset{\eq{eq_charge_heat}}=  \evalAtEqui{\frac{\partial \IEa}{\partial x}} - \evalAtEqui{\frac{\partial (\mu_\alpha \INa)}{\partial x}}\notag\\
 &\equalbyeqns{eq_derive_equi_energy_current}{eq_derive_equi_charge_current}  \sqbrack{\evalAtEqui{\frac{\partial \IEa}{\partial x}} - \mueq\evalAtEqui{\frac{\partial \INa}{\partial x}}} \notag\\
 &\equalbyeqn{eq_IEz}  \cbrack{\evalAtEqui{\frac{d}{dx}\nbrack{\left[\epsilon-\mueq+\frac{U}{2}\left(2-\nzia\right)\right]\INa}} - \Gamma_\alpha \, U \, \evalAtEqui{\frac{d}{dx}\Braket{\zia\fpOp}{z}}}\notag\\
 &\equalbyeqn{eq_seebeck} \sqbrack{\Eeq\cdot\evalAtEqui{\frac{d\INa}{dx}} - \Gamma_\alpha \, U \, \evalAtEqui{\frac{d}{dx}\Braket{\zia\fpOp}{z}}}.\label{eq_linear_response_step9}
\end{align}
The first part in \Eq{eq_linear_response_step9} can be fully understood from the charge current response discussed in the previous subsections \ref{app:linear_response_conductance} and \ref{app:linear_response_seebeck}.
To further simplify \Eq{eq_linear_response_step9}, we rewrite the second term as
\begin{align}
 \evalAtEqui{\frac{d}{dx}\Braket{\zia\fpOp}{z}} &= \Braket{\evalAtEqui{\frac{d}{dx}\zia}\fpOp}{\zeq} + \Braket{\zieq\fpOp}{\evalAtEqui{\frac{d}{dx}z}}\notag\\
 &\equalbyeqns{eq_stationary_state_equilibrium}{eq_derive_side_formula_two} \Bra{\zieq}\nbrack{\evalAtEqui{\frac{d\hat{A}_\alpha}{dx}}}\fpOpHat\Ket{\zeq} + \Braket{\zieq\fpOp}{\evalAtEqui{\frac{d}{dx}z}}\notag\\
 &=\evalAtEqui{\frac{d\beta_\alpha}{dx}}\Bra{\zieq}\hat{\Hdot}\fpOpHat\Ket{\zeq} - \evalAtEqui{\frac{d(\mu_\alpha\beta_\alpha)}{dx}}\Bra{\zieq}\hat{N}\fpOpHat\Ket{\zeq}
 + \Braket{\zieq\fpOp}{\evalAtEqui{\frac{d}{dx}z}}\notag\\
 &\equalbyeqn{eq_orthogonality_relation_two}\evalAtEqui{\frac{d\beta_\alpha}{dx}}\Bra{\zieq}\hat{\Hdot}\fpOpHat\Ket{\zeq} + \Bra{\zieq\fpOp}\evalAtEqui{\frac{d}{dx}\Ket{z}}.\label{eq_linear_response_step10}
\end{align}
Expanding the dot Hamiltonian as
\begin{align}
 \hat{\Hdot} &\equalbyeqns{eq_liouville_identity_alt}{eq_orthogonality_relation_one} \nbrack{\epsilon + \frac{U}{2}}\cdot \hat{N} + \frac{U}{4}\fpOpHat - \frac{U}{4}\one,\label{eq_hamiltonian_dot_expansion}
\end{align}
and using the orthogonality relations \eq{eq_orthogonality_relation_two} as well as $\Braket{\zieq\fpOp}{\zeq} = 0$ \BrackEq{eq_kernel_eigenvectors_formal}, Eq.~\eq{eq_linear_response_step10} reduces to
\begin{align}
\evalAtEqui{\frac{d}{dx}\Braket{\zia\fpOp}{z}} &\equalbys{\eq{eq_linear_response_step10}}{\eq{eq_hamiltonian_dot_expansion},\eq{eq_orthogonality_relation_two}}\evalAtEqui{\frac{d\beta_\alpha}{dx}}\cdot\frac{U}{4}\Braket{\zieq}{\zeq} + \Bra{\zieq\fpOp}\evalAtEqui{\frac{d}{dx}\Ket{z}}\notag\\
&\,\,\,\quad\equalbyeqn{eq_linear_response_state}\,\,\,\quad \evalAtEqui{\frac{d\beta_\alpha}{dx}}\cdot\frac{U}{4}\Braket{\zieq}{\zeq}\notag\\
&\phantom{\equalbys{\eq{eq_linear_response_step10}}{\eq{eq_hamiltonian_dot_expansion},\eq{eq_orthogonality_relation_two}}}- \sum_{\alpha'}\frac{\Gamma_{\alpha'}}{\Gamma}\evalAtEqui{\frac{d\beta_{\alpha'}}{dx}} \vphantom{\underbrace{\Braket{\zieq\fpOp}{\zeq}}_{\equalbyeqn{eq_kernel_eigenvectors_formal}0}} \sqbrack{  \Bra{\zieq}\fpOpHat \hat{\Hdot}\Ket{\zeq} - \smash{\underbrace{\Braket{\zieq\fpOp}{\zeq}}_{\equalbyeqn{eq_kernel_eigenvectors_formal}0}}\Braket{\zieq\cdot H}{\zeq}  }\notag\\
&\phantom{\equalbys{\eq{eq_linear_response_step10}}{\eq{eq_hamiltonian_dot_expansion},\eq{eq_orthogonality_relation_two}}}+ \sum_{\alpha'}\frac{\Gamma_{\alpha'}}{\Gamma}\evalAtEqui{\frac{d(\beta_{\alpha'}\mu_{\alpha'})}{dx}} \vphantom{\underbrace{\Braket{\zieq\fpOp}{\zeq}}_{\equalbyeqn{eq_kernel_eigenvectors_formal}0}} \sqbrack{  \Bra{\zieq}\fpOpHat \hat{N} \Ket{\zeq} - \smash{\underbrace{\Braket{\zieq\fpOp}{\zeq}}_{\equalbyeqn{eq_kernel_eigenvectors_formal}0}}\Braket{\zieq\cdot N}{\zeq}  }\notag\\
&\equalbys{\eq{eq_kernel_eigenvectors_formal}}{\eq{eq_hamiltonian_dot_expansion},\eq{eq_orthogonality_relation_two}}\frac{U}{4}\cdot\sum_{\alpha'}\frac{\Gamma_{\alpha'}}{\Gamma}\evalAtEqui{\frac{d(\beta_\alpha - \beta_{\alpha'})}{dx}}\cdot \Braket{\zieq}{\zeq}.\label{eq_linear_response_step11}
\end{align}
Consequently, the equilibrium derivative of the heat current \eq{eq_linear_response_step9} can be expressed as
\begin{equation}
 \evalAtEqui{\frac{dJ^\alpha}{dx}} \equalbyeqns{eq_linear_response_step9}{eq_linear_response_step11} \Eeq\cdot\evalAtEqui{\frac{d\INa}{dx}} + \Gamma_\alpha \nbrack{\frac{U}{2T}}^2\cdot\sum_{\alpha''}\frac{\Gamma_{\alpha''}}{\Gamma}\evalAtEqui{\frac{d(T_\alpha - T_{\alpha''})}{dx}}\cdot \Braket{\zieq}{\zeq} \qquad \text{for } x = \mu_{\alpha'},T_{\alpha'}.\label{eq_linear_response_step12}
\end{equation}
Let us summarize what we learn from \Eq{eq_linear_response_step11} and \Eq{eq_linear_response_step12}. The first important message is that for potential gradients, $x = \mu_{\alpha'}$, only the first part of the heat current that is tightly coupled to the particle current contributes, since
\begin{equation}
  \evalAtEqui{\frac{d}{d\mu_{\alpha'}}\Braket{\zia\fpOp}{z}} \equalbyeqn{eq_linear_response_step11} 0.\label{eq_vanishing_response}
\end{equation}
As claimed in the main text, this ensures compliance with Onsager's reciprocity relation. Namely, the linear response of the heat current with respect to electrochemical potential variations is given by
\begin{equation}
 \evalAtEqui{\frac{dJ^\alpha}{d\mu_{\alpha'}}} \equalbyeqn{eq_linear_response_step12} \Eeq\cdot\evalAtEqui{\frac{d\INa}{d\mu_{\alpha'}}}.\label{eq_linear_response_step13}
\end{equation}
For two contacts, this yields
 \begin{align}
 L_{21} &= \frac{T}{2}\evalAtEqui{\frac{d(J^\text{L}-J^\text{R}}{dV}} \equalbyeqn{eq_linear_response_step13} -\Eeq\cdot \frac{T}{2}\cdot \evalAtEqui{\frac{d(I_N^\text{L}-I_N^\text{R})}{dV}} \equalbyeqn{eq_linear_response_charge_mu_symmetrized} -\Eeq\cdot L_{11} \equalbyeqn{eq_linear_response_charge_T_symmetrized} L_{12}.\label{eq_linear_response_heat_mu_symmetrized}
\end{align}

The final task is to determine the linear response of the heat current to a temperature gradient, involving in particular the second term in \Eq{eq_linear_response_step12}. Motivated by the general fluctuation-dissipation theorem at equilibrium, we first aim to express the state overlap $\Braket{\zieq}{\zeq}$ entering the second term in terms of particle number fluctuations. With respect to the stationary and dual stationary state, these fluctuations can be conveniently written as
\begin{align}
\delnsqzeq &= \Braket{N^2}{\zeq} - \sqbrack{\Braket{N}{\zeq}}^2 \equalbyeqn{eq_kernel_eigenvectors_formal}\Braket{c'_{\mathrm{eq}}}{N\cdot \zeq} \notag\\
\delnsqzieq &= \Braket{N^2}{\zieq} - \sqbrack{\Braket{N}{\zieq}}^2 \equalbyeqn{eq_kernel_eigenvectors_formal}2\Braket{\zieq\cdot\fpOp\cdot N}{c_{\mathrm{eq}}}.\label{eq_stationary_fluctuations}
\end{align}
Forming the product of both fluctuations, we can rewrite the result by solving the \emph{Liouville space} identity \eq{eq_liouville_identity} for $\Ket{c_{\mathrm{eq}}}\Bra{c'_{\mathrm{eq}}}$, inserting the result and using the nontrivial orthogonality relation \eq{eq_orthogonality_relation_two}:
\begin{align}
\delnsqzieq\cdot \delnsqzeq &\equalbyeqn{eq_stationary_fluctuations} 2\Braket{\zieq\cdot\fpOp N}{c_{\mathrm{eq}}}\Braket{c'_{\mathrm{eq}}}{N\cdot\zeq}\notag\\
&\equalbyeqn{eq_liouville_identity} 2\Braket{\zieq\fpOp N^2}{\zeq} - 2\underbrace{\Braket{\zieq\fpOp N}{\zeq}}_{\equalbyeqn{eq_orthogonality_relation_two}0}\cdot \Braket{\one}{N\cdot\zeq}\notag\\
&\phantom{\equalbyeqn{eq_liouville_identity}} - 2\Braket{\zieq\cdot N\fpOp}{\fpOp} \cdot \underbrace{\Braket{\zieq\fpOp N}{\zeq}}_{\equalbyeqn{eq_orthogonality_relation_two}0}\notag\\
&\equalbyeqns{eq_orthogonality_relation_two}{eq_fluctuation_expansion}-\underbrace{\Braket{\zieq\fpOp}{\zeq}}_{\equalbyeqn{eq_kernel_eigenvectors_formal}0} + 4\underbrace{\Braket{\zieq\cdot \fpOp N}{\zeq}}_{\equalbyeqn{eq_orthogonality_relation_two}0} + \Braket{\zieq\cdot\fpOp\fpOp}{\zeq}\notag\\
&= \Braket{\zieq}{\zeq}.\label{eq_fluctuation_relation_step}
\end{align}
This leads to an explicit expression for the heat current response to a temperature gradient:
\begin{align}
 \evalAtEqui{\frac{dJ^\alpha}{dT_{\alpha'}}} &\equalbyeqns{eq_linear_response_step12}{eq_fluctuation_relation_step} \Eeq\cdot\evalAtEqui{\frac{d\INa}{dT_{\alpha'}}}  + \Gamma_\alpha \nbrack{\frac{U}{2T}}^2\cdot\sum_{\alpha''}\frac{\Gamma_{\alpha''}}{\Gamma}\nbrack{\delta_{\alpha\alpha'} - \delta_{\alpha''\alpha'}}\cdot\delnsqzieq\cdot \delnsqzeq\notag\\
 &\equalbyeqn{eq_linear_response_step8}\frac{\Eeq^2}{T}\cdot\evalAtEqui{\frac{d\INa}{d\mu_{\alpha'}}} + \nbrack{\frac{U}{2T}}^2\cdot\frac{\Gamma_\alpha(\Gamma\delta_{\alpha\alpha'} - \Gamma_{\alpha'})}{\Gamma}\cdot\delnsqzieq\cdot \delnsqzeq.\label{eq_linear_response_step14}
\end{align}
The corresponding result for the symmetrized heat current and temperature bias $\Delta T$ for the two-contact system of the main paper reads
 \begin{align}
 L_{22} & = \frac{T^2}{2}\sqbrack{\evalAtEqui{\frac{dJ^\text{L}}{dT_\text{L}}} - \evalAtEqui{\frac{dJ^\text{R}}{dT_\text{L}}} }\equalbyeqns{eq_linear_response_step14}{eq_linear_response_charge_mu_symmetrized} \Eeq^2\cdot L_{11} + \nbrack{\frac{U}{2}}^2\cdot\frac{\Gamma_\text{L} \Gamma_\text{R}}{\Gamma}\cdot\delnsqzieq\cdot \delnsqzeq.\label{eq_linear_response_step15}
\end{align}
Note that in order to explicitly calculate  $\delnsqzieq$, we use analogously to \Eq{eq_fluctuation_expectation} and \Eq{eq_fluctuation_expectation_explicit} that
\begin{equation}
 \Braket{N^2}{\zieq} \equalbyeqn{eq_fluctuation_expansion} 2\Braket{N}{\zieq} + \frac{1}{2}\sqbrack{\Braket{\fpOp}{\zieq} - \Braket{\one}{\zieq}} \equalbyeqn{eq_stationary_state_equilibrium} 2\nzieq + \frac{1}{2}\nbrack{\pzieq - 1}.\label{eq_fluctuation_dual_expectation}
\end{equation}
The dual parity $\pzieq$ is obtained in the same way as the equilibrium parity $\pzeq$:
\begin{align}
 \pzieq = \evalAtEqui{\pzia} \lrsepa \pzia = \Braket{0}{\zia} + \Braket{2}{\zia} - 2\Braket{1}{\zia} \equalbyeqns{eq_states}{eq_derive_step6} 1 - 4\frac{f_\alpha(\epsilon + U)[1-f_\alpha(\epsilon)]}{1 - f_\alpha(\epsilon) + f_\alpha(\epsilon+U)}.\label{eq_dual_parity_expectation}
\end{align}

The last remaining step is to confirm that \Eq{eq_linear_response_step15} indeed contains all contributions of the energy fluctuation \Eq{eq_E_fluct}, i.e., when the single-particle energy is measured with respect to the electrochemical potential:
$ \hat{\mathcal{H}} = \hat{\Hdot}- \mueq\cdot \hat{N}$. The fact that $\hat{N},\hat{\Hdot},\hatzeq$ mutually commute leads to
\begin{align}
 \delta \hat{\mathcal{H}}_{\textrm{eq}}^2 &= \Braket{\mathcal{H}^2}{\zeq}  - \sqbrack{\Braket{\mathcal{H}}{\zeq}}^2\notag\\
 &= \Braket{H^2}{\zeq} - \sqbrack{\Braket{H}{\zeq}}^2 - 2\mueq\sqbrack{\Braket{NH}{\zeq} - \Braket{N}{\zeq}\Braket{H}{\zeq}} + \mueq^2\cbrack{\Braket{N^2}{\zeq} -\sqbrack{\Braket{N}{\zeq}}^2}\notag\\
 &\equalbyeqns{eq_linear_response_step5}{eq_linear_response_step6} \Braket{H^2}{\zeq} - \sqbrack{\Braket{H}{\zeq}}^2 - \mueq\nbrack{\mueq + 2\Eeq}\delnsqzeq.\label{eq_linear_response_step16}
\end{align}
The energy fluctuation can be efficiently calculated with the decay mode expansion of the Hamiltonian,
\begin{align}
 \Bra{H} &\equalbyeqn{eq_liouville_identity} \Braket{H}{\zeq}\Bra{\one} + \Braket{H}{c_{\text{eq}}}\sqbrack{\Bra{N} - \Braket{N}{\zeq}\Bra{\one}} + \Braket{H}{\fpOp}\Bra{\zieq\fpOp}\notag\\
 &\equalbys{\eq{eq_linear_response_step6}}{\eq{eq_hamiltonian_dot_expansion},\eq{eq_orthogonality_relation_one}} \Braket{H}{\zeq}\Bra{\one} + (\mueq + \Eeq)\sqbrack{\Bra{N} - \Braket{N}{\zeq}\Bra{\one}} + U\Bra{\zieq\fpOp}.\label{eq_hamiltonian_dot_expansion_in_modes}
\end{align}
Using $\Braket{H^2}{\zeq} = \Braket{H}{H\zeq}$ together with \eq{eq_hamiltonian_dot_expansion_in_modes} in \Eq{eq_linear_response_step16} yields
\begin{align}
 \delta \hat{\mathcal{H}}_{\textrm{eq}}^2 &\equalbyeqns{eq_linear_response_step16}{eq_hamiltonian_dot_expansion_in_modes} (\mueq + \Eeq)\sqbrack{\Braket{NH}{\zeq} - \Braket{N}{\zeq}\Braket{H}{\zeq}} - \mueq\nbrack{\mueq + 2\Eeq}\delnsqzeq + U\Braket{\zieq\fpOp}{H\zeq}\notag\\
 &\equalbys{\eq{eq_linear_response_step5},\eq{eq_linear_response_step6}}{\eq{eq_hamiltonian_dot_expansion}} \Eeq^2\delnsqzeq - \nbrack{\frac{U}{2}}^2\underbrace{\Braket{\zieq\fpOp}{\zeq}}_{\equalbyeqn{eq_kernel_eigenvectors_formal}0} + U\nbrack{\epsilon + \frac{U}{2}}\underbrace{\Braket{\zieq\fpOp}{N\zeq}}_{\equalbyeqn{eq_orthogonality_relation_two}0} + \nbrack{\frac{U}{2}}^2\underbrace{\Braket{\zieq}{\zeq}}_{\equalbyeqn{eq_fluctuation_relation_step}\delnsqzeq\delnsqzieq}\notag\\
 &=  \Eeq^2\delnsqzeq +  \nbrack{\frac{U}{2}}^2\delnsqzeq\delnsqzieq = \text{\Eq{eq_E_fluct}}.\label{eq_linear_response_step17}
\end{align}

\section{Non-equilibrium relations for current-balanced system}
\label{app:cur_bal}
%%%%%%%%%%%%%%%%%%%%%%%%%%%%%%%%%%%%%%%%%%

In this final appendix, we derive the relations used in section \ref{sec_nonlinear} to characterize energy transport in a current-balanced non-equilibrium situation, that is, for
\begin{equation}
\INa = 0 \quad\forall \alpha.\label{eq_current_balanced}
\end{equation}
Since \Eq{eq_current_balanced} means $0 = \INa \equalbyeqn{eq_INz} \gamca\sqbrack{\nza - \nz}$ for all $\alpha$, and since $\gamca > 0$ (see appendix \ref{app:kernel}), we have
\begin{equation}
\nza = \nz \Rightarrow \nza = \nzap \quad\forall \alpha,\alpha'.\label{eq_current_balanced_one}
\end{equation}
Clearly, the converse also holds, i.e., it follows from \Eq{eq_charge_full} that $\nz = \nza$ when assuming $\nza = \nzap$, which leads to $\INa = 0$ when using \Eq{eq_INz}. Let us in the following exploit this equivalence to derive the expansion \eq{eq_bal_state} of the stationary state $\Ket{z}$ used in the main text.

We start by expanding $\Ket{z}$ in the eigenbasis \eq{eq_kernel_singlelead_eigenvectors} of a given reservoir kernel $W_\alpha$: 
\begin{align}
\evalAtBal{\Ket{z}} &\equalbyeqns{eq_kernel_singlelead_eigenvectors}{eq_derive_step11} \evalAtBal{\Ket{z_\alpha}} + \evalAtBal{\nbrack{\nz - \nza}} \cdot\evalAtBal{\Ket{c_\alpha}} + \evalAtBal{\Braket{\zia\fpOp}{z}} \cdot \Ket{\fpOp}\notag\\
&\equalbyeqn{eq_current_balanced_one} \evalAtBal{\Ket{z_\alpha}} + \evalAtBal{\Braket{\zia\fpOp}{z}} \cdot \Ket{\fpOp},\label{eq_current_balanced_two}
\end{align}
where $\evalAtBal{\dots}$ denotes evaluation under the condition \eq{eq_current_balanced_one}. Analogously to \Eq{eq_derive_step12}, we multiply by $\Gamma_\alpha$, sum over $\alpha$ and divide by $\Gamma = \sum_\alpha\Gamma_\alpha$:
\begin{align}
\evalAtBal{\Ket{z}} &\equalbyeqn{eq_current_balanced_two} \sum_\alpha\frac{\Gamma_\alpha}{\Gamma}\evalAtBal{\Ket{z_\alpha}} + \frac{1}{U\Gamma}\sqbrack{\sum_\alpha\evalAtBal{U\Gamma_\alpha\Braket{\zia\fpOp}{z}} }\cdot\Ket{\fpOp}\notag\\
&\equalbyeqn{eq_IEz}\sum_\alpha\frac{\Gamma_\alpha}{\Gamma}\evalAtBal{\Ket{z_\alpha}} - \frac{1}{U\Gamma}\evalAtBal{\sum_\alpha\sqbrack{\IEa - \nbrack{\epsilon + \frac{2-\nzia}{2}U}\INa} }\cdot\Ket{\fpOp}\notag\\
&\equalbyeqn{eq_current_balanced}\sum_\alpha\frac{\Gamma_\alpha}{\Gamma}\evalAtBal{\Ket{z_\alpha}} - \frac{1}{U\Gamma}\evalAtBal{\sum_\alpha \IEa}\cdot\Ket{\fpOp}\notag\\
&\equalbyeqn{eq_derive_current_conservation}\sum_\alpha\frac{\Gamma_\alpha}{\Gamma}\evalAtBal{\Ket{z_\alpha}} = \text{\Eq{eq_bal_state}} . \label{eq_current_balanced_state}
\end{align}

To find the explicit expression of the non-equilibrium Fourier heat coefficient \eq{eq_fourier_overlap}, we apply $\Bra{\fpOp}$ to \Eq{eq_current_balanced_two} from the left, use $\Braket{\fpOp}{\fpOp} = 4$, and combine the result with \Eq{eq_current_balanced_state}. This gives
\begin{equation}
\evalAtBal{\Braket{\zia\fpOp}{z}} \equalbyeqn{eq_current_balanced_two} \frac{1}{4}\evalAtBal{\sqbrack{\Braket{\fpOp}{z} - \Braket{\fpOp}{z_\alpha}}} \equalbyeqn{eq_current_balanced_state} \frac{1}{4}\sum_{\alpha'\neq\alpha}\frac{\Gamma_{\alpha'}}{\Gamma}\evalAtBal{\nbrack{p_{\alpha'} - p_{\alpha}}}\label{eq_fourier_noneq}
\end{equation}
with the average parities $p_{z\alpha} = \Braket{\fpOp}{z_\alpha}$. In the case of only two leads, $\alpha = \text{L,R}$, the relation \eq{eq_fourier_noneq} leads to \Eq{eq_fourier_nonlineara}. \par

Finally, since we have proven equivalence between \Eq{eq_current_balanced} and the right side of \Eq{eq_current_balanced_one}, the prescription $\evalAtBal{\dotsc}$ reduces the calculation of the thermopower to solving
\begin{equation}
\nza = \nzap \overset{\eq{eq_derive_step9}}{\Leftrightarrow} \frac{f_\alpha(\epsilon)}{1+f_\alpha(\epsilon) - f_\alpha(\epsilon+U)} = \frac{f_{\alpha'}(\epsilon)}{1+f_{\alpha'}(\epsilon) - f_{\alpha'}(\epsilon+U)}\label{eq_current_balanced_tosolve}
\end{equation}
for all $\alpha,\alpha'$. Given one fixed potential $\mu_\alpha$, a fixed level position $\epsilon - \mu_\alpha$ and fixed inverse temperatures $T_\alpha,T_{\alpha'}$, this condition leads to an analytic expression for all other $\mu_{\alpha'\neq\alpha}$:
\begin{align}
\evalAtBal{\frac{\mu_{\alpha'} - \mu_{\alpha}}{T_{\alpha'}}} &= \nbrack{\epsilon - \mu_\alpha + U}\nbrack{\frac{1}{T_{\alpha'}} - \frac{1}{T_{\alpha}}} + \ln\sqbrack{\frac{1 - \nzia + \sqrt{\nbrack{1 - \nzia}^2 + \expfn{\frac{U(T_{\alpha'} - T_\alpha)}{T_{\alpha'}T_{\alpha}}}\nzia(2-\nzia) }}{2 - \nzia} } \ , \label{eq_current_balanced_solved}
\end{align}
For two reservoirs $\alpha = \text{L,R}$, this leads to expression \eq{eq_thermovoltage_nl} from the main text.

%%%%%%%%%%%%%%%%%%%%%%%%%%%%%%%%%%%%%%%%%%
% Citations and References in Supplementary files are permitted provided that they also appear in the reference list here. 

%%=====================================
%% References, variant A: internal bibliography
%%=====================================

%% The following MDPI journals use author-date citation: Arts, Econometrics, Economies, Genealogy, Humanities, IJFS, JRFM, Laws, Religions, Risks, Social Sciences. For those journals, please follow the formatting guidelines on http://www.mdpi.com/authors/references

%=====================================
% References, variant B: external bibliography
%=====================================
% \externalbibliography{no}
% \bibliography{cite}

%%%%%%%%%%%%%%%%%%%%%%%%%%%%%%%%%%%%%%%%%%
%%%%%%%%%%%%%%%%%%%%%%%%%%%%%%%%%%%%%%%%%%
\end{document}